\documentclass[11pt,a4paper]{article}

\usepackage[T1]{fontenc}
\usepackage[utf8]{inputenc}
\usepackage{lmodern}
\usepackage{microtype}

\usepackage[
  a4paper,
  left=22mm,
  right=22mm,
  top=25mm,
  bottom=25mm
]{geometry}

\usepackage{amsmath}
\usepackage{amssymb}
\usepackage{amsfonts}
\usepackage{amsthm}
\usepackage{mathtools}
\usepackage{bm}
\usepackage{mathrsfs}

\allowdisplaybreaks[2]

\newtheorem{theorem}{Theorem}[section]
\newtheorem{proposition}[theorem]{Proposition}
\newtheorem{lemma}[theorem]{Lemma}
\newtheorem{corollary}[theorem]{Corollary}

\theoremstyle{definition}
\newtheorem{definition}[theorem]{Definition}
\newtheorem{example}[theorem]{Example}

\theoremstyle{remark}

\usepackage{graphicx}
\usepackage{booktabs}
\usepackage{array}
\usepackage{caption}
\usepackage{cellspace}
\makeatletter
\let\@startpbox@action\@startpbox
\makeatother
\usepackage{multirow}
\usepackage{longtable}
\usepackage{enumitem}
\setlist[enumerate]{topsep=6pt, itemsep=3pt, parsep=0pt}

\setlist{nosep}

\usepackage{xcolor}

\usepackage[
  authoryear,
  round
]{natbib}

\usepackage{xurl}

\usepackage[
  unicode=true,
  colorlinks=true,
  linkcolor=blue,
  citecolor=blue,
  urlcolor=blue
]{hyperref}

\usepackage[
  nameinlink,
  capitalise,
  noabbrev
]{cleveref}

\newcommand{\bbp}{\mathbb P}
\newcommand{\bbe}{\mathbb E}

\newcommand{\bbr}{\mathbb R}

\newcommand{\dcal}{\mathcal D}
\newcommand{\fcal}{\mathcal F}

\newcommand{\mcal}{\mathcal M}

\newcommand{\ycal}{\mathcal Y}
\newcommand{\xcal}{\mathcal X}

\title{Risk Adequacy and Anti-Discrimination in Insurance Pricing:\\
Classification, Aggregation, and Transfers}

\author{
\setlength{\tabcolsep}{20pt}
\begin{tabular}{ccc}
Wiebke Hansen
&
Stefan Weber\thanks{Corresponding author: Stefan Weber, House of Insurance \& Institute of Actuarial and Financial Mathematics, Leibniz Universit\"at Hannover, Welfengarten 1, 30167 Hannover, Germany. Email: stefan.weber@insurance.uni-hannover.de}
&
Mario V. W\"uthrich
\\[0.5em]
\multicolumn{2}{c}{
 Leibniz Universit\"at Hannover}
&
{ ETH Z\"urich}
\end{tabular}
}

\date{\today}

\begin{document}

\maketitle


\begin{abstract}
Insurance tariffs must reconcile risk-based pricing, portfolio coverage, and anti-discrimination requirements. We introduce a framework in which tariffs are constructed through classification, aggregation of benchmark risk prices within classes, and balanced transfers, and are evaluated by several distinct criteria. This yields a multi-objective tariff-design problem whose attainable objective set and Pareto frontier describe the resulting trade-offs.

Risk adequacy is measured by the discrepancy between charged premiums and individual benchmark risk prices. Portfolio coverage is assessed through the capital required to make aggregate claims net of premium income acceptable. Anti-discrimination is examined through differences in protected-group composition across tariff classes and through premium effects experienced by protected groups. We analyze the trade-offs among these criteria and prove mathematical
results characterizing their relationships:
(i) the inferential threshold for perfect risk adequacy with finitely
many benchmark-price levels;
(ii) lower bounds on risk inadequacy under stricter inferential constraints;
(iii) exact changes in risk inadequacy and inferential disparity when
classes are merged, and dynamic programming for scalarized optimization
over consecutive benchmark-price classes;
(iv) attainable protected-group total premium effects and their minimum
risk inadequacy costs under balanced transfers;
(v) a sharp lower bound linking inferential disparity to the risk
inadequacy required under a bound on protected-group premium-effect
dispersion; and
(vi) conditions for portfolio-coverage improvement and its additional
risk inadequacy cost at fixed protected-group total premium effects.  A numerical motor-insurance study illustrates the results.
\end{abstract}

\medskip



\section{Introduction}

Insurance pricing combines two features that do not naturally collapse into a
single objective. Policyholders differ in the risks they bring to an insurance
portfolio, while insurance operates by pooling risks and financing losses
collectively. A risk model and a premium principle can therefore associate an
individual benchmark risk price with each policyholder type, but an insurance
tariff does not generally charge these benchmark risk prices individually.
Instead, policyholders are assigned to tariff classes, benchmark risk prices
are aggregated within those classes, and the resulting class premiums may be
modified by transfers between classes.

Anti-discrimination requirements introduce further constraints on insurance
pricing. Protected groups are groups defined by one or more characteristics
designated for protection against discrimination, such as sex or ethnic
origin. Insurance pools risks and therefore treats policyholders
collectively, but this collective treatment can interact with protected-group
membership in different ways. Tariff classes may differ in their
protected-group composition, so one may make the classification coarser in
order to reduce these differences. Doing so, however, can require pooling
policyholders with different benchmark risk prices and thereby move charged
premiums away from individual benchmark risk prices. Pooling within tariff
classes can also generate different average premium effects across protected
groups. Transfers between tariff classes can be used to modify these
protected-group-level effects, but doing so can move individual premiums still
further away from their benchmark risk prices. The resulting tariff-design
problem therefore involves trade-offs between protected-group requirements and
risk adequacy.

These mechanisms concern different properties of a tariff and should therefore
be distinguished. We separate tariff construction into three components: a
classification, an aggregation component, and a transfer. The classification
determines which policyholders receive the same tariff treatment. The
aggregation component converts the benchmark risk prices within each class
into an aggregate benchmark price. The transfer modifies the charged class
premium relative to that aggregate benchmark price. We evaluate the resulting
tariff along several dimensions. Portfolio coverage concerns the aggregate
financial consequences of the charged premiums. Risk adequacy
compares charged premiums with individual benchmark risk prices. The inferential
criterion measures how strongly the protected-group composition varies across
tariff classes. Protected-group premium effects describe how pooling and
transfers change premiums relative to benchmark risk prices when these changes
are averaged within protected groups.

Protected-group premium effects can themselves be evaluated in different ways.
A deviation measure captures how much the mean premium effects differ across
protected groups. A monetary risk measure instead evaluates the distribution
of these mean effects across protected groups and can focus on its adverse
upper tail, where the largest premium increases occur. Deviation and monetary risk measures therefore capture different aspects of
protected-group premium effects.

The relevant criteria also need not be evaluated under the same population
law. Risk adequacy and portfolio coverage concern the insurer's
current portfolio. By contrast, regulatory evaluation may concern the
population of prospective policyholders to whom the tariff will be applied,
whose composition need not coincide with that of the current portfolio.
The inferential criterion, protected-group premium effects, and transfer balance
may therefore be evaluated under a regulatory or other reference population.
We distinguish these population laws while keeping the structural risk model
fixed. This distinction is material: changing the reference population can
change both the protected-group premium effects being evaluated and the
transfers whose surcharges and rebates offset one another in expectation
under that reference population.

At its core, the framework defines a multi-objective tariff-design problem.
Ideally, one would characterize the attainable objective set and its Pareto
frontier, and then use additional criteria to select among the resulting
efficient tariffs. Our analysis addresses this problem through a set of structural results for
specific trade-offs. We ask three concrete questions. First, how similar can tariff classes be in
their protected-group composition while still allowing every policyholder to
be charged exactly the benchmark risk price? If the tariff classes are made even more
similar, how far must charged premiums move away from benchmark risk prices?

Second, once the tariff classes and the pooling of benchmark risk prices within
them are fixed, what average premium changes can still be created for the
different protected groups by shifting premiums between classes without
changing the average under the reference population? Moreover, even when an average premium change has been specified for each protected group, different transfers between tariff classes
may still be admissible. Among these possibilities,
which one keeps individual premiums closest to their benchmark risk prices?

Third, how much do these possibilities depend on how the protected groups are
distributed across tariff classes? And how much of this can be learned from the
inferential criterion alone?

\subsection{Main contributions}
\label{subsec:intro-contributions}

The paper develops a framework for answering these questions and establishes
three groups of results.

\begin{enumerate}
\item We formulate tariff construction in terms of three controls:
classification, aggregation, and balanced transfers.
We evaluate the resulting tariffs in terms of portfolio coverage,
risk adequacy, the inferential criterion, and protected-group premium
effects. Different population laws can be used for the different
criteria. The resulting attainable objective set provides the natural
basis for Pareto comparisons without collapsing the criteria into a
single notion of tariff quality.

\item We characterize the trade-off between classification, risk adequacy, and the inferential criterion. For finitely many benchmark-price levels,
we identify the smallest value of the inferential criterion compatible with perfect
risk adequacy. This threshold is attained by the classification
that distinguishes exactly the benchmark-price levels. Requiring a value of the inferential
criterion below the threshold necessarily entails a positive loss of
risk adequacy. Under squared Euclidean dissimilarity, we obtain a
quantitative lower bound for this loss, derive exact formulas for the effects
of merging tariff classes, and show how optimal coarsenings with consecutive
benchmark-price levels can be computed by dynamic programming.

\item For fixed classification and aggregation components, we characterize,
within the family of balanced transfers with otherwise unrestricted
real-valued class premiums, which protected-group total premium effects
can be attained. For each attainable target, we determine the minimum
risk inadequacy among all such balanced transfers that attain it
and describe the remaining transfer directions that leave the target unchanged.
Attainability and minimum risk inadequacy depend on how the
protected groups are distributed across tariff classes. The inferential
criterion constrains this relation but does not determine it: classifications
with the same value of the inferential criterion can require very different
levels of risk inadequacy to attain the same protected-group
target. Finally, when a protected-group target is fixed, the remaining
transfer directions can generate a further trade-off between risk adequacy and portfolio coverage.
\end{enumerate}

\subsection{Literature}
\label{subsec:intro-literature}

Premium principles provide methods for assigning prices to risk
distributions, while credibility theory studies the combination of
individual and collective information
\citep{Young2014PremiumPrinciples,Buhlmann1967,BuhlmannGisler2005}.
The economic and legal literature examines the efficiency, welfare, and
redistributive consequences of risk classification and restrictions on
its use \citep{CrockerSnow1986,Hoy2006,ThieryVanSchoubroeck2006}.
Recent actuarial research examines fairness and discrimination in
insurance pricing, including the use of rating information and the
construction of premiums under fairness requirements
\citep{FreesHuang2023,Charpentier2024,FahrenwaldtEtAl2024,
LindholmEtAl2022,XinHuang2024,AraizaIturriaEtAl2024}.
This research distinguishes proxy discrimination from demographic
group fairness \citep{LindholmEtAl2024,LindholmEtAl2026} and also
investigates constrained cross-subsidization between actuarial and
solidarity-based pricing \citep{ZhangLiuShi2026}.

The broader statistical fairness literature considers different
requirements, including individual fairness, demographic parity,
equalized odds, and equal opportunity
\citep{DworkEtAl2012,ZemelEtAl2013,HardtPriceSrebro2016};
see \citet{BarocasHardtNarayanan2023}.
It establishes incompatibilities among certain fairness criteria
and investigates their trade-offs with predictive accuracy
\citep{Chouldechova2017,KleinbergMullainathanRaghavan2017,
MenonWilliamson2018,ChzhenSchreuder2022}.
In insurance pricing, DPTree restricts protected-group proportions
during the construction of regression trees
\citep{SimonDenuitTrufin2026}.
Our inferential criterion measures dissimilarity between the
protected-group distributions within tariff classes, independently
of the premiums charged.
We study how restrictions on this criterion interact with risk
adequacy. For finitely many benchmark risk-price levels, we identify
the smallest inferential-criterion value compatible with perfect
risk adequacy and derive quantitative lower bounds on risk inadequacy
under stricter requirements.

Another connection concerns adjustments intended to satisfy
group-level requirements. In health-insurance risk adjustment,
\citet{ZinkRose2020} study group-level prediction errors and
undercompensation. Fair regression methods impose constraints on
group means and characterize optimal adjustments and their accuracy
costs \citep{FitzsimonsEtAl2019,WeiEtAl2023}.
Related actuarial research studies balance, autocalibration, and
multicalibration as relationships between premiums and claims
\citep{DenuitCharpentierTrufin2021,LindholmWuthrich2026,
DenuitMichaelidesTrufin2026}.
In our framework, the relevant group effects are average differences
between charged premiums and individual benchmark risk prices.
For fixed tariff classes, we characterize which combinations of
these effects can be achieved through balanced transfers and determine
the minimum risk inadequacy for each attainable target.
We show that these costs depend on the distribution of protected
groups across tariff classes and cannot be determined from the
inferential criterion alone. We also characterize the transfer
directions that leave group effects unchanged.

The choice of population law is another important aspect of fairness
evaluation. An insurer's portfolio may differ from the population
for which fairness is to be assessed.
\citet{CoteEtAl2026} study the consequences of this distinction
for fairness adjustments under selection bias.
Our framework allows portfolio coverage, risk inadequacy,
protected-group criteria, and transfer balance to be evaluated
under different population laws.
We study the resulting consequences for protected-group effects
and the feasibility of balanced transfers.

Monetary risk measures and deviation measures provide complementary
ways to evaluate adverse protected-group effects and their dispersion
\citep{ArtznerEtAl1999,FollmerSchied2025,NendelRiedelSchmeck2021}.
More generally, attainable objective sets and Pareto efficiency
provide a framework for analyzing multiple criteria \citep{Ehrgott2005}.
Recent insurance-pricing research studies multiobjective optimization
over predictive pricing models \citep{BoonenFanQuan2026} and the
effects of pricing constraints on demand, insurer outcomes, and
consumer welfare \citep{HuangShimao2026}.
Within this broader literature, we study how the construction of
tariffs from given individual benchmark risk prices constrains
the attainable combinations of actuarial and protected-group
objectives. We establish limits on the compatibility of risk
adequacy and protected-group composition across tariff classes,
characterize attainable protected-group premium effects and their
minimum risk inadequacy costs under balanced transfers, and
analyze how the remaining transfer freedom can be used to improve
portfolio coverage.

\subsection{Organization of the paper}
\label{subsec:intro-organization}

Section~\ref{sec:tariff-framework} introduces the tariff framework, the
evaluation criteria, the relevant population laws, and the multi-objective
formulation. Section~\ref{sec:main-result-I} isolates the role of
classification by allowing class premiums to vary freely subject only to being
constant within tariff classes. It identifies the inferential threshold for
zero risk inadequacy and the cost of imposing stricter inferential
constraints. Section~\ref{sec:balanced-transfers} then fixes the classification
and aggregation components and varies the transfer subject to balance,
characterizing attainable protected-group total premium effects, their
minimum risk inadequacy costs, and the remaining trade-off with
portfolio coverage. Section~\ref{sec:finite-model} illustrates these
structural results in a finite insurance case study with neighboring
benchmark-price levels and overlapping protected groups.
Section~\ref{sec:conclusion} summarizes the results and discusses further
questions on structured and implementable tariff design. Proofs,
supplementary results, and detailed calculations are collected in the
appendices.


\section{Tariff construction, objectives, and reference populations}
\label{sec:tariff-framework}


This section specifies the probabilistic structure underlying the analysis and
then develops the corresponding tariff model. We first describe a
B\"uhlmann-type model \citep{Buhlmann1967,BuhlmannGisler2005}
for a heterogeneous insurance portfolio, linking the
portfolio background, policyholder types, observable information, claims, and
protected characteristics. Within this structural model, a premium principle
assigns a benchmark risk price to each policyholder type. Tariffs are then
constructed from these benchmark risk prices through classification, within-class
aggregation, and transfers. Finally, we introduce the criteria by which the
resulting tariffs are evaluated: portfolio coverage, risk adequacy, and protected-group considerations, with the latter distinguishing inferential
properties of the classification from protected-group premium effects.
Portfolio and regulatory criteria may be evaluated under different
population laws while the underlying structural risk model remains fixed.

\subsection{Policyholder risks and structural kernels}
\label{subsec:structural-kernels}

The tariff analysis requires a model connecting individual risks,
portfolio losses, and the populations under which tariffs are evaluated.
At the individual level, we need a conditional claim distribution from
which to assign a benchmark risk price to each policyholder type.
For portfolio coverage, we also need a joint distribution of policyholders
and their claims. Finally, we want to distinguish changes in population
composition from changes in the conditional risk structure.
We construct a B\"uhlmann-type hierarchical model that provides these
three ingredients.

We begin with the individual risk structure. Let
$(\widetilde\Theta,\fcal_{\widetilde\Theta})$ and
$(\xcal,\fcal_\xcal)$ be standard Borel spaces representing policyholder
types and observable information. Let $\ycal\subseteq\bbr$ be a Borel
set of monetary claim amounts, equipped with its Borel $\sigma$-field
$\fcal_\ycal$. A \emph{type} $\vartheta\in\widetilde\Theta$ describes an
individual's risk characteristics at the finest level of differentiation
retained in the model. It need not be observable to the insurer.
Observable information may reveal only part of this modeled heterogeneity.

A probability kernel $\kappa_{X,Y\mid\Theta}$ from
$(\widetilde\Theta,\fcal_{\widetilde\Theta})$ to
$(\xcal\times\ycal,\fcal_\xcal\otimes\fcal_\ycal)$ specifies the joint
distribution of observables and claims for each type.
Thus $\kappa_{X,Y\mid\Theta}(\vartheta,\cdot)$ describes both the
information associated with type $\vartheta$ and the uncertainty in its
claim. Keeping this kernel fixed allows us to change the distribution
of types in a population without changing the conditional risk model
of any given type. Related distinctions between changes in population
composition and changes in conditional claim distributions arise in
insurance model monitoring; see \citet{BrauerMenzelWuthrich2025}.
Here the distinction is made at the level of policyholder types.

Protected-group membership is represented by a measurable map
$D:(\widetilde\Theta,\fcal_{\widetilde\Theta})\to
(\dcal,\fcal_\dcal)$, where $(\dcal,\fcal_\dcal)$ is a standard Borel
space. The map selects a \emph{protected characteristic} from the modeled type.
The type is taken to include protected-group membership, so this membership
is determined once the type is fixed. Observable information may retain
randomness conditional on type, as specified by the kernel
$\kappa_{X,Y\mid\Theta}$.
The tariff classification introduced in the next subsection will instead
be a decision about which types receive the same tariff treatment.
The specification of individual risks therefore precedes the choice of
tariff classes.

To obtain an individual benchmark risk price, we use the claim marginal
of the risk kernel. Define the probability kernel $\kappa_{Y\mid\Theta}$
from $(\widetilde\Theta,\fcal_{\widetilde\Theta})$ to
$(\ycal,\fcal_\ycal)$ by
$\kappa_{Y\mid\Theta}(\vartheta,B)
=\kappa_{X,Y\mid\Theta}(\vartheta,\xcal\times B)$ for
$B\in\fcal_\ycal$. A premium principle assigns a price to this
conditional claim distribution. For a measurable space
$(E,\mathcal E)$, write $\mcal_1(E,\mathcal E)$ for the set of
probability measures on that space.

\begin{definition}[Benchmark risk price]
Let $\Pi$ be a real-valued premium principle defined on a subset of
$\mcal_1(\ycal,\fcal_\ycal)$ containing
$\kappa_{Y\mid\Theta}(\vartheta,\cdot)$ for every
$\vartheta\in\widetilde\Theta$. Assume that the map
$\vartheta\mapsto\Pi(\kappa_{Y\mid\Theta}(\vartheta,\cdot))$
is measurable. The \emph{benchmark risk price} of type $\vartheta$ is
\begin{equation}
 \pi(\vartheta)
 =
 \Pi\!\left(\kappa_{Y\mid\Theta}(\vartheta,\cdot)\right),
 \qquad \vartheta\in\widetilde\Theta.
 \label{eq:benchmark-risk-price}
\end{equation}
\end{definition}

For example, expectation pricing gives
$\pi(\vartheta)=\int_\ycal y\,\kappa_{Y\mid\Theta}(\vartheta,dy)$
when the conditional claim law is integrable.
Other premium principles may include a prescribed risk loading.
In each case, the benchmark risk price depends on the conditional claim
law of the type. It is defined before tariff construction and need not
equal the premium eventually charged after classification, aggregation,
and transfers. We impose throughout the integrability conditions needed
for the quantities under consideration.

The individual risk kernel does not yet specify which types occur in a
portfolio or how risks across policyholders are related. To provide this
joint structure, let
$(\widetilde\Psi,\fcal_{\widetilde\Psi})$ be a standard Borel space
representing a \emph{portfolio background}. A probability kernel
$\lambda_{\Theta\mid\Psi}$ from
$(\widetilde\Psi,\fcal_{\widetilde\Psi})$ to
$(\widetilde\Theta,\fcal_{\widetilde\Theta})$ specifies the distribution
of types for each background $\psi$.
For example, different backgrounds may assign different proportions to
types with higher or lower benchmark risk prices.
The background is shared across the portfolio, so uncertainty about it
can induce dependence between policyholders.

Let $\mu_\Psi$ be the \emph{background law}, a probability measure on
$(\widetilde\Psi,\fcal_{\widetilde\Psi})$, specifying the distribution
of the portfolio background. The B\"uhlmann-type construction proceeds
in three steps. First, a background $\psi$ is sampled from $\mu_\Psi$.
Second, conditional on $\psi$, policyholder types are sampled
independently from $\lambda_{\Theta\mid\Psi}(\psi,\cdot)$.
Third, conditional on the background and the sampled types,
observables and claims are sampled independently across policyholders
from their respective kernels
$\kappa_{X,Y\mid\Theta}(\vartheta_i,\cdot)$.
Thus, the background governs the distribution of portfolio
composition, while each type determines the conditional distribution
of the corresponding observables and claims.

We realize this structure on one probability space for all portfolio
sizes. Set
$\Omega=\widetilde\Psi\times
(\widetilde\Theta\times\xcal\times\ycal)^{\mathbb N}$
and equip it with the product $\sigma$-field
$\mathcal F=\fcal_{\widetilde\Psi}\otimes
(\fcal_{\widetilde\Theta}\otimes\fcal_\xcal\otimes
\fcal_\ycal)^{\otimes\mathbb N}$.
Denote the coordinate maps by $\Psi,\Theta_i,X_i,Y_i$,
$i\in\mathbb N$, and set $D_i=D(\Theta_i)$.
A portfolio of size $n$ consists of the first $n$ policyholders
in this sequence.

For any probability measure $\mu_\Psi$ on
$(\widetilde\Psi,\fcal_{\widetilde\Psi})$, the Ionescu--Tulcea theorem
yields a unique probability measure $P^{\mu_\Psi}$ on
$(\Omega,\mathcal F)$ with background law $\mu_\Psi$ and the specified
conditional structure. More explicitly, writing
$\vartheta_{1:n}=(\vartheta_1,\ldots,\vartheta_n)$ and analogously for
$x_{1:n}$ and $y_{1:n}$, its finite-dimensional marginals satisfy,
for every $n\in\mathbb N$,
\begin{equation}
 P^{\mu_\Psi}
 (d\psi,d\vartheta_{1:n},dx_{1:n},dy_{1:n})
 =
 \mu_\Psi(d\psi)
 \prod_{i=1}^n
 \lambda_{\Theta\mid\Psi}(\psi,d\vartheta_i)
 \kappa_{X,Y\mid\Theta}(\vartheta_i,dx_i,dy_i).
 \label{eq:portfolio-joint-law}
\end{equation}
Here the left-hand side denotes the joint law of
$\Psi,\Theta_{1:n},X_{1:n},Y_{1:n}$.
Conditional on $\Psi$, the triples $(\Theta_i,X_i,Y_i)$ are
independent and identically distributed. Conditional on $\Psi$ and
the entire type sequence, the pairs $(X_i,Y_i)$ are independent,
with respective conditional laws
$\kappa_{X,Y\mid\Theta}(\Theta_i,\cdot)$.
In particular, given $\Theta_i$, the background $\Psi$ carries no
further information about $(X_i,Y_i)$.

We can now distinguish the populations used for tariff evaluation.
Fix the two kernels $\lambda_{\Theta\mid\Psi}$ and
$\kappa_{X,Y\mid\Theta}$ and the premium principle $\Pi$.
Let $\mu_\Psi$ describe the background of the insurer's portfolio,
and call $\bbp=P^{\mu_\Psi}$ the \emph{portfolio law}.
This law determines both the distribution of individual benchmark risk
prices in the portfolio and the joint distribution needed to evaluate
aggregate claims and premium income.

A \emph{regulatory reference population} is specified by another background
law $\mu_\Psi^{\mathrm{reg}}$, inducing
$\bbp^{\mathrm{reg}}=P^{\mu_\Psi^{\mathrm{reg}}}$ on the same canonical
space. The two populations therefore share the conditional distribution
of types given the background and the conditional distribution of
observables and claims given a type. Their background distributions,
and consequently their unconditional population compositions, may differ.
Since $\Pi$ and the claim kernel remain fixed, the benchmark risk-price
map $\pi$ is common to both populations.

One interpretation of $\bbp^{\mathrm{reg}}$ is a population of prospective
policyholders to whom the tariff may be applied. Evaluating a tariff under
this law describes its effects in that reference population, whose
composition need not coincide with the insurer's current portfolio.
This distinction allows portfolio performance and risk adequacy to be evaluated under $\bbp$, while regulatory criteria may
be evaluated under $\bbp^{\mathrm{reg}}$.

For individual-level quantities, we write
$(\Theta,X,Y)=(\Theta_1,X_1,Y_1)$ and regard this triple
as representing a generic policyholder. Under each hierarchical
portfolio law, its distribution is the common marginal law of
$(\Theta_i,X_i,Y_i)$, $i\in\mathbb N$.
We also write $D=D(\Theta)$ for the protected characteristic
and $R=\pi(\Theta)$ for the benchmark risk price. For subsequent evaluations, we use $\bbp^\star$ to denote a
\emph{reference law} on $(\Omega,\mathcal F)$, with
$\bbe_{\bbp^\star}$ denoting the corresponding expectation.
For the protected-group criteria and transfer-balance conditions
considered below, only its type marginal
$\bbp_\Theta^\star=\bbp^\star\circ\Theta^{-1}$
is relevant.

\subsection{Classification, aggregation, and transfers}
\label{subsec:classification-aggregation-transfers}

A type-level tariff transforms benchmark risk prices into charged premiums
through three components: classification, within-class aggregation, and
transfers. The classification
$C:\widetilde\Theta\to\{1,\ldots,m\}$ is measurable and assigns each type to
one of finitely many tariff classes; we take the class labels so that every
class is nonempty. For a generic policyholder, $C(\Theta)$ is the resulting
tariff class. In practice, however, the insurer has access to
observable information $X$. An implementable class assignment must
therefore be measurable with respect to $\sigma(X)$.

This restriction has two distinct consequences. First, observable
information may reveal only part of the policyholder type.
Even if $X$ is completely determined by $\Theta$, different
types may give rise to the same observable information and
must therefore receive the same class assignment. In this case,
the available classifications form a restricted family of
type-level classifications.

Second, $X$ may contain additional randomness conditional on
$\Theta$. Different observations of the same type may then
lead to different class assignments. Although the assignment
is determined once $X$ is observed, class membership may
remain random conditional on $\Theta$. An observable
classification therefore need not induce a sharp partition
of the type space: each type may instead have positive
probabilities of being assigned to several tariff classes.

If $X$ is interpreted as an imperfect observation of the
underlying risk type, one may additionally assume that
$X$ and $Y$ are conditionally independent given $\Theta$.
Under this assumption, $X$ provides no additional information
about the claim once the type is known, but can still provide
information about the claim through what it reveals about
the unobserved type. This conditional independence is not
assumed in the present framework and is not needed for its
results.

An extension to observable classifications would account for
both the restriction on available information and the possible
randomness of class assignments conditional on type.
The resulting attainable objective set could be determined
and compared with that of the present type-level model,
which serves as a theoretical benchmark rather than an
implementable tariff. In particular, one could quantify
how incomplete and noisy observations change the attainable
values of risk inadequacy, portfolio coverage, and
protected-group criteria, as well as the trade-offs between
them.

Because the protected characteristic $D$ is a measurable function
of $\Theta$, the present
type-level formulation does not by itself exclude classifications that depend
on $D$. Restrictions on the information that may be used for classification,
including restrictions concerning protected characteristics, are therefore
additional admissibility conditions on the tariff family.

The second component of the tariff is a family of within-class aggregation
functionals $\mathcal A=(\mathcal A_c)_{c=1}^m$. Together with the transfer
function $\Lambda$ introduced below, a general type-level tariff can therefore
be represented as $\mathbf C=(C,\mathcal A,\Lambda)$. For each class $c$, its
aggregate benchmark price is
\[
 A^{\mathbf C}(c)
 =
 \mathcal A_c\bigl(\pi|_{C^{-1}(\{c\})}\bigr),
\]
where $\mathcal A_c$ is defined on an appropriate class of real-valued
benchmark-price maps on $C^{-1}(\{c\})$. Natural requirements are monotonicity
and translation equivariance:
\[
 r\leq r'
 \quad\Longrightarrow\quad
 \mathcal A_c(r)\leq\mathcal A_c(r'),
 \qquad
 \mathcal A_c(r+a)=\mathcal A_c(r)+a.
\]
These properties parallel two of the basic axioms of monetary risk measures
and allow nonlinear aggregation rules, including Choquet-type specifications.

An important special case is linear aggregation, in which
the aggregation component $\mathcal A$ is represented by a probability kernel
$\xi_{\Theta\mid C}$ from $\{1,\ldots,m\}$ to $\widetilde\Theta$. In this
case we write $\mathbf C=(C,\xi_{\Theta\mid C},\Lambda)$, with
$\xi_{\Theta\mid C}(c,C^{-1}(\{c\}))=1$. Assuming that $\pi$ is integrable
with respect to the corresponding probability measures, we set
\begin{equation}
 A^{\mathbf C}(c)
 =
 \int_{\widetilde\Theta}
 \pi(\vartheta)\,
 \xi_{\Theta\mid C}(c,d\vartheta).
 \label{eq:class-aggregate}
\end{equation}
The aggregation kernel is part of the tariff specification. In particular,
it need not coincide with the conditional distribution of $\Theta$ given
$C(\Theta)=c$ under either $\bbp$ or $\bbp^{\mathrm{reg}}$.

Finally, $\Lambda:\{1,\ldots,m\}\to\bbr$ assigns a transfer to each class.
The amount $\Lambda(c)$ is the transfer per policyholder assigned to class
$c$, measured in the same monetary units as the premium. A positive value is
a surcharge and a negative value a rebate. The premium charged to a type $\vartheta$ is therefore
\begin{equation}
 \pi^{\mathbf C}(\vartheta)
 =
 A^{\mathbf C}(C(\vartheta))+\Lambda(C(\vartheta)),
 \qquad \vartheta\in\widetilde\Theta.
 \label{eq:charged-premiums}
\end{equation}

We call a transfer $\Lambda$ $\bbp^\star$-balanced if
\begin{equation}
 \bbe_{\bbp^\star}\!\left[\Lambda(C(\Theta))\right]=0.
 \label{eq:transfer-balance}
\end{equation}
Thus a $\bbp^\star$-balanced transfer has mean zero when weighted
by the class frequencies under $\bbp^\star$. In particular,
transfer balance implies
$\bbe_{\bbp^\star}[\pi^{\mathbf C}(\Theta)]
=\bbe_{\bbp^\star}[A^{\mathbf C}(C(\Theta))]$.
The mean charged premium therefore equals the mean benchmark risk
price under $\bbp^\star$ if and only if the aggregation component
satisfies
$\bbe_{\bbp^\star}[A^{\mathbf C}(C(\Theta))]
=\bbe_{\bbp^\star}[\pi(\Theta)]$.
Transfer balance alone does not impose this condition.

The deviation of the charged premium from the benchmark can be written as
\begin{equation}
 \pi^{\mathbf C}(\vartheta)-\pi(\vartheta)
 =
 \bigl[A^{\mathbf C}(C(\vartheta))-\pi(\vartheta)\bigr]
 +
 \Lambda(C(\vartheta)),
 \qquad \vartheta\in\widetilde\Theta.
 \label{eq:premium-decomposition}
\end{equation}
This decomposition separates the effects of classification and aggregation
from the transfer component. Classification determines which policyholder types belong to the same
class, while aggregation determines the common benchmark price
assigned to that class. The first term records the resulting deviation from the
individual benchmark risk price, whereas the second records the class-level transfer.
We refer to the first term as the pooling effect, to the second as the
transfer effect, and to their sum as the total premium effect.

\subsection{Coverage and risk adequacy}
\label{subsec:coverage-risk-adequacy}

We evaluate a tariff along three distinct dimensions. The first is portfolio
coverage (O1), which concerns aggregate financial adequacy. The second is
risk adequacy (O2), which compares charged premiums with benchmark
risk prices. The third comprises protected-group considerations (O3), considered in the next
subsection, which concern both the protected-group information carried by the
classification and the protected-group premium effects of the tariff.

We first define coverage and risk adequacy. For a portfolio of size $n$, let
$S_n=\sum_{i=1}^n Y_i$ denote the aggregate claim. After collection of tariff
premiums, the residual loss of the insurer is
\begin{equation}
 Z^{\mathbf C}
 =
 S_n-\sum_{i=1}^n\pi^{\mathbf C}(\Theta_i).
 \label{eq:residual-loss}
\end{equation}

Under the portfolio law $\bbp$, let $\mathfrak A_{\mathrm{acc}}$ denote the
set of residual losses regarded as acceptable. We assume that it is
monotone in the loss convention,
\[
 Z\leq Z',\quad Z'\in\mathfrak A_{\mathrm{acc}}
 \quad\Longrightarrow\quad
 Z\in\mathfrak A_{\mathrm{acc}},
\]
and that, for every residual loss considered below,
$\{a\in\bbr:Z-a\in\mathfrak A_{\mathrm{acc}}\}$ is nonempty and bounded
from below. The associated monetary risk measure is
\[
 \varrho(Z)
 =
 \inf\{a\in\bbr:Z-a\in\mathfrak A_{\mathrm{acc}}\}.
\]
It is monotone and cash additive in the loss convention:
$Z\leq Z'$ implies $\varrho(Z)\leq\varrho(Z')$, and
$\varrho(Z+a)=\varrho(Z)+a$ for $a\in\bbr$. Thus $\varrho(Z)$ is the
infimal additional capital amount required to make the residual loss $Z$
acceptable. We define the portfolio-coverage criterion (O1) by
\begin{equation}
 f_{\mathrm{cov}}(\mathbf C)=\varrho(Z^{\mathbf C}).
 \label{eq:coverage-criterion}
\end{equation}
Smaller values of $f_{\mathrm{cov}}$ correspond to better portfolio coverage.
Both the aggregate claim and the premium income are evaluated under the
portfolio law $\bbp$.

Risk adequacy (O2) compares each policyholder's charged premium with
the benchmark risk price and evaluates the resulting deviations on
average under the portfolio law $\bbp$. Let
$\operatorname{dist}_{\bbr}:\bbr\times\bbr\to[0,\infty)$ be a measurable
loss function such that
$\operatorname{dist}_{\bbr}(x,y)=0$ if and only if $x=y$. We define
\begin{equation}
 f_{\mathrm{RA}}(\mathbf C)
 =
 \bbe_{\bbp}\!\left[
 \operatorname{dist}_{\bbr}
 \bigl(\pi^{\mathbf C}(\Theta),\pi(\Theta)\bigr)
 \right].
 \label{eq:risk-adequacy-general}
\end{equation}
The main specification used below is squared premium deviation, for which
$\operatorname{dist}_{\bbr}(x,y)=(x-y)^2$ and hence
$f_{\mathrm{RA}}(\mathbf C)
=\bbe_{\bbp}[\,(\pi^{\mathbf C}(\Theta)-\pi(\Theta))^2\,]$.
We use risk adequacy for the desirable property and risk inadequacy for the loss measured by $f_{\mathrm{RA}}$. Thus smaller
values of $f_{\mathrm{RA}}$ correspond to greater risk adequacy.
We call the tariff perfectly risk adequate if
$f_{\mathrm{RA}}(\mathbf C)=0$, equivalently if
$\pi^{\mathbf C}(\Theta)=\pi(\Theta)$ $\bbp$-a.s.

\subsection{Inferential and premium-effect criteria}
\label{subsec:inferential-premium-effect-criteria}

Protected-group considerations (O3) concern two questions. First,
does the tariff classification $C$ carry information about the protected
characteristic $D$, so that protected-group composition varies across tariff
classes? Second, do the individual premium effects relative to benchmark risk
prices differ across the groups defined by $D$? We evaluate these two aspects
separately.
\begin{enumerate}
\item[\textnormal{(O3-I)}] We measure how much the protected-group composition
varies across the tariff classes generated by $C$. Fix a reference law
$\bbp^\star$. For every class $c$ with
$\bbp^\star(C(\Theta)=c)>0$, let
$\eta^\star_{D\mid C}(c)=\bbp^\star(D\in\cdot\mid C(\Theta)=c)$ denote its
protected-group profile. We compare these profiles using a nonnegative
dissimilarity $\operatorname{dist}_{\dcal}$ on
$\mcal_1(\dcal,\fcal_\dcal)\times
\mcal_1(\dcal,\fcal_\dcal)$.

A second reference law $\bbp^\diamond$ determines how the comparisons between
classes are weighted. We assume that
$\bbp^\diamond(C(\Theta)=c)>0$ implies
$\bbp^\star(C(\Theta)=c)>0$, so that every profile receiving positive weight
is defined. The inferential criterion is
\begin{equation}
 g_{\mathrm{inf}}^{\star,\diamond}(C)
 =
 \sum_{\substack{c,c':\,
 \bbp^\diamond(C(\Theta)=c)>0,\,
 \bbp^\diamond(C(\Theta)=c')>0}}
 \bbp^\diamond(C(\Theta)=c)\bbp^\diamond(C(\Theta)=c')\,
 \operatorname{dist}_{\dcal}
 \bigl(\eta^\star_{D\mid C}(c),\eta^\star_{D\mid C}(c')\bigr).
 \label{eq:inferential-criterion}
\end{equation}
The criterion depends on the tariff only through the classification $C$.
Thus $\bbp^\star$ determines the protected-group composition within each
class, while $\bbp^\diamond$ determines how class pairs are weighted.
Smaller values of $g_{\mathrm{inf}}^{\star,\diamond}(C)$ correspond to less
variation in protected-group composition across tariff classes. The
criterion measures this statistical association and does not by itself impose
a restriction on which variables may be used to construct $C$. Any
information-use or protected-characteristic restriction is instead part of
the admissible tariff family.

Our main specification takes
$\bbp^\star=\bbp^\diamond=\bbp^{\mathrm{reg}}$. Under the
prospective-policyholder interpretation,
$\eta^{\mathrm{reg}}_{D\mid C}(c)$ is the protected-group distribution among
policyholders from the regulatory reference population who are assigned to
tariff class $c$. For the main specification, we use the shorthand
$\Gamma(C)=g_{\mathrm{inf}}^{\mathrm{reg},\mathrm{reg}}(C)$.
The quantitative results in Section~\ref{sec:main-result-I}
and the finite case study in Section~\ref{sec:finite-model}
use squared Euclidean distance between finite-group probability vectors.

\item[\textnormal{(O3-P)}] Premium effects are described relative to the benchmark risk price. For a generic
policyholder, define the pooling effect, transfer effect, and total premium
effect by
\begin{equation}
 W^{\mathbf C}
 =
 A^{\mathbf C}(C(\Theta))-\pi(\Theta),
 \qquad
 B^{\mathbf C}
 =
 \Lambda(C(\Theta)),
 \qquad
 \Delta^{\mathbf C}
 =
 W^{\mathbf C}+B^{\mathbf C}
 =
 \pi^{\mathbf C}(\Theta)-\pi(\Theta).
 \label{eq:premium-effects} 
\end{equation}
For any $E^{\mathbf C}\in
\{W^{\mathbf C},B^{\mathbf C},\Delta^{\mathbf C}\}$ that is integrable under
$\bbp^\star$, we first average individual effects within protected groups.
The resulting protected-group effect is the $\sigma(D)$-measurable
random variable
\begin{equation}
 M_E^{\mathbf C,\star}
 =
 \bbe_{\bbp^\star}
 \left[E^{\mathbf C}\mid\sigma(D)\right].
 \label{eq:group-conditional-mean}
\end{equation}
We call $M_W^{\mathbf C,\star}$, $M_B^{\mathbf C,\star}$, and
$M_\Delta^{\mathbf C,\star}$ the protected-group pooling, transfer,
and total premium effects, respectively. Thus ``protected-group''
indicates that individual effects have been averaged within each group.
If $D\in\{1,\ldots,k\}$ and $\bbp^\star(D=d)>0$ for every $d$, we identify
$M_E^{\mathbf C,\star}$ with the vector
$\bigl(\bbe_{\bbp^\star}[E^{\mathbf C}\mid D=d]\bigr)_{d=1}^k\in\bbr^k$.
For $\bbp^\star=\bbp^{\mathrm{reg}}$, the $d$th component can be interpreted
as the average tariff effect experienced by policyholders in protected group
$d$ when the tariff is applied to the regulatory reference population.

We evaluate the protected-group effects in two complementary ways. A
deviation measure quantifies how much these effects vary across protected
groups, independently of their common level. A monetary risk measure instead
evaluates their adverse level and therefore changes when all protected-group
effects are shifted by the same amount.

Following the distinction between deviation measures and monetary risk
measures in \citet{NendelRiedelSchmeck2021}, let $\mathfrak D^\star$ be a
deviation measure on the relevant space of $\sigma(D)$-measurable random
variables, satisfying
\[
 \mathfrak D^\star(Z+a)=\mathfrak D^\star(Z),
 \qquad
 \mathfrak D^\star(0)=0,
 \qquad
 \mathfrak D^\star(Z)\geq0.
\]
Thus $\mathfrak D^\star$ measures variation in protected-group effects
without responding to a common shift across all groups. Examples include
variance, standard deviation, and mean absolute deviation under the
protected-group distribution.

To evaluate the adverse level of the same effects, let $\mathfrak R^\star$
be a monetary risk measure on the same space. We use the loss
convention of \citet{McNeilFreyEmbrechts2015}, under which positive values
represent adverse outcomes; for the general theory of monetary risk measures,
see, for example, \citet{FollmerSchied2025}. Thus
\[
 \mathfrak R^\star(Z+a)=\mathfrak R^\star(Z)+a,
 \qquad
 Z\leq Z' \ \Longrightarrow\
 \mathfrak R^\star(Z)\leq\mathfrak R^\star(Z').
\]
Here positive values of $Z$ represent adverse premium effects on
policyholders. Hence a common increase of all protected-group effects by
$a$ increases $\mathfrak R^\star$ by $a$, and making every protected-group
effect weakly more adverse cannot decrease $\mathfrak R^\star$. Depending on
the chosen risk measure, different parts of the distribution across protected
groups may receive different emphasis. Examples include expectation, AVaR,
and more general convex or coherent monetary risk measures.

For any premium effect $E^{\mathbf C}\in
\{W^{\mathbf C},B^{\mathbf C},\Delta^{\mathbf C}\}$, we define
\begin{equation}
 D_E^\star(\mathbf C)
 =
 \mathfrak D^\star(M_E^{\mathbf C,\star}),
 \qquad
 R_E^\star(\mathbf C)
 =
 \mathfrak R^\star(M_E^{\mathbf C,\star}).
 \label{eq:group-effect-criteria}
\end{equation}
Smaller values of $D_E^\star(\mathbf C)$ correspond to less variation in
mean premium effects across protected groups, while smaller values of
$R_E^\star(\mathbf C)$ correspond to a smaller adverse level according to
the prescribed risk measure.

For the total premium effect, we call the tariff group neutral if
$M_\Delta^{\mathbf C,\star}$ is constant: the average deviation from the
benchmark risk price is then the same across protected groups. We say that
the tariff achieves exact compensation if
$M_\Delta^{\mathbf C,\star}=0$ almost surely, so that this average deviation
vanishes in every protected group. The benchmark risk price $\pi$ is taken as given throughout
the analysis. Exact compensation therefore refers to the
absence of protected-group premium effects relative to this benchmark.
\end{enumerate}

The two protected-group perspectives describe different properties of
a tariff that are relevant to anti-discrimination in insurance pricing.
The inferential criterion measures how protected-group composition
varies across tariff classes, independently of the premiums charged.
The premium-effect criteria measure the consequences of pooling,
aggregation, and transfers relative to the given benchmark risk prices.
Deviation measures quantify differences in average premium effects
across protected groups, while monetary risk measures evaluate their
adverse level.

For illustration, consider the total premium effect $\Delta^{\mathbf C}$
and squared premium deviation, as in
Sections~\ref{sec:balanced-transfers} and~\ref{sec:finite-model}.
In this specification, risk inadequacy is
$f_{\mathrm{RA}}(\mathbf C)=\bbe_{\bbp}[(\Delta^{\mathbf C})^2]$.
The protected-group criteria for the same effect first form
$M_\Delta^{\mathbf C,\star}
=\bbe_{\bbp^\star}[\Delta^{\mathbf C}\mid\sigma(D)]$.
Choosing standard deviation for $\mathfrak D^\star$ then gives
$D_\Delta^\star(\mathbf C)
=\operatorname{SD}_{\bbp^\star}(M_\Delta^{\mathbf C,\star})$:
the standard deviation of the group-average total premium effects,
weighted by the protected-group probabilities under $\bbp^\star$.
Even when the population laws coincide, positive and negative individual
premium deviations can cancel within each protected group.
Consequently, all group-average total premium effects may vanish while
risk inadequacy remains positive.

\subsection{Multi-objective tariff design}
\label{subsec:pareto-formulation}

The evaluation dimensions introduced above concern different but interacting
aspects of tariff design. Coverage concerns the aggregate financial
consequences of the charged premiums. Risk adequacy compares those
premiums with individual benchmark risk prices. Protected-group
considerations concern both the information about $D$ carried by the
classification and the premium effects experienced across the groups defined
by $D$. These objectives cannot in general be optimized simultaneously. Changes in classification granularity can trade risk adequacy against the inferential criterion: finer classification may improve the former while increasing the value of the latter. Transfers may reduce differences in mean premium effects across
protected groups, but can likewise increase individual deviations from the
benchmark. Tariff design therefore naturally gives rise to a multi-objective
optimization problem.

Let $\mathfrak C$ denote the admissible family of tariffs, including any
balance, architectural, or implementability restrictions imposed in the
application. In particular, $\mathfrak C$ may restrict the information from
which class assignments may be constructed and may exclude the use of
specified protected characteristics in classification. These admissibility
conditions are logically separate from the inferential criterion, which
evaluates the protected-group composition of the classifications that remain
admissible. One possible objective vector, incorporating the criteria
introduced above, is
\begin{equation}
 F(\mathbf C)
 =
 \left(
 f_{\mathrm{cov}}(\mathbf C),
 f_{\mathrm{RA}}(\mathbf C),
 g_{\mathrm{inf}}^{\star,\diamond}(C),
 R_\Delta^\star(\mathbf C),
 D_\Delta^\star(\mathbf C)
 \right),
 \label{eq:objective-vector}
\end{equation}
where all coordinates are quantities to be minimized. The particular
objective vector depends on the question under consideration: some
coordinates may be omitted, additional criteria may be included, and the
premium-effect criteria may be applied to $W^{\mathbf C}$ or
$B^{\mathbf C}$ instead of $\Delta^{\mathbf C}$.

In general, there is no single tariff that minimizes every coordinate.
The relevant object is therefore the set of attainable trade-offs
\[
 \mathcal V_F
 =
 \{F(\mathbf C):\mathbf C\in\mathfrak C\}.
\]
A tariff $\mathbf C^\ast$ is Pareto efficient if no admissible tariff has
objective values that are no larger in every coordinate and strictly smaller
in at least one. The corresponding points $F(\mathbf C^\ast)$ form the Pareto
frontier of the attainable objective set. The frontier therefore describes
the combinations of objectives for which an improvement in one direction
cannot be achieved without deterioration in at least one other direction.
This is the same organizing principle as the risk--return frontier in
portfolio selection: rather than collapsing competing objectives into a
single criterion from the outset, it makes their attainable trade-offs
explicit.

The multi-objective formulation can be made operational by fixing bounds on
selected criteria and optimizing another. For example, a representative
$\varepsilon$-constraint formulation is
\begin{equation}
\begin{aligned}
 \inf_{\mathbf C\in\mathfrak C}\quad
 & f_{\mathrm{RA}}(\mathbf C)
 \\
 \text{subject to}\quad
 & f_{\mathrm{cov}}(\mathbf C)\leq\varepsilon_{\mathrm{cov}},
 \qquad
 g_{\mathrm{inf}}^{\star,\diamond}(C)\leq\varepsilon_{\mathrm{inf}},
 \\
 & R_\Delta^\star(\mathbf C)\leq\varepsilon_R,
 \qquad
 D_\Delta^\star(\mathbf C)\leq\varepsilon_D.
\end{aligned}
\label{eq:epsilon-constraint-general}
\end{equation}
Further admissibility conditions can be imposed in the same way. Such
formulations make explicit how the different objectives can be combined in a
concrete tariff-design problem. The following sections do not attempt to
solve this general multi-objective problem. Instead, they study two reduced
problems that isolate different components of the tariff architecture.
Section~\ref{sec:main-result-I} varies the classification while allowing
arbitrary class-constant premiums, in order to characterize the trade-off
between risk adequacy and the inferential criterion.
Section~\ref{sec:balanced-transfers} then fixes the classification and
aggregation components and varies balanced transfers, in order to characterize
the attainability and cost of protected-group targets and the remaining scope
for improving portfolio coverage.

\section{Classification under inferential constraints}
\label{sec:main-result-I}

We now study how classification affects risk adequacy (O2) and the
inferential criterion (O3-I). To isolate the role of classification, we
temporarily abstract from the aggregation and transfer components of the
tariff. For a given classification, class premiums may therefore be chosen
freely, subject only to being constant within each tariff class. The resulting
problem determines the smallest risk inadequacy permitted by the
classification itself and separates this information restriction from the
additional requirements imposed by a specified aggregation component and a
balanced transfer.

Within this classification problem, we first identify the smallest value of
the inferential criterion compatible with zero risk inadequacy.
We then show that requiring a value below this threshold entails a strictly
positive loss of risk adequacy and quantify this loss. Next, we
determine how merging tariff classes affects both criteria. Finally, we study
their trade-off over a specified family of admissible classifications using a
scalarized objective. Implementation of the resulting class premiums by a
specified aggregation component and balanced transfer is a separate question,
considered at the end of this section and developed further in
Section~\ref{sec:balanced-transfers}.

\subsection{Perfect risk adequacy and the inferential threshold}
\label{subsec:exact-inferential-threshold}

Write $R=\pi(\Theta)$ for the benchmark risk price and assume
$R\in L^2(\bbp)$. For a finite classification
$C:\widetilde\Theta\to\{1,\ldots,m\}$, let
$p=(p_1,\ldots,p_m)^\top\in\bbr^m$ be a class-premium vector:
$p_c$ is the premium charged to every policyholder in class $c$, and
$p_{C(\Theta)}$ is the premium charged to a policyholder of type $\Theta$.
Define
\begin{equation}
L(C)
=
\inf_{p\in\bbr^m}
\bbe_{\bbp}\!\left[(R-p_{C(\Theta)})^2\right].
\label{eq:MR1-RA}
\end{equation}
Thus $L(C)$ is the smallest squared premium deviation attainable when the
classification $C$ is fixed and class premiums may otherwise be chosen freely.
It is therefore a benchmark for the risk inadequacy generated by
the classification itself; the optimization defining $L(C)$ does not impose
an aggregation rule or a transfer-balance condition. The optimal premium for
a class $c$ with positive portfolio probability is its conditional mean
benchmark risk price,
$\bbe_{\bbp}[R\mid C(\Theta)=c]$, and hence
$
L(C)
=
\bbe_{\bbp}\!\left[
\operatorname{Var}_{\bbp}(R\mid C(\Theta))
\right].
$
In particular,
$L(C)=0$ if and only if $R$ is
$\sigma(C(\Theta))$-measurable $\bbp$-a.s.

We evaluate the inferential criterion under the regulatory law. Write
$s_c=\bbp^{\mathrm{reg}}(C(\Theta)=c)$ and, for $s_c>0$, let
\[
\eta_c
=
\eta^{\mathrm{reg}}_{D\mid C}(c)
=
\bbp^{\mathrm{reg}}(D\in\cdot\mid C(\Theta)=c).
\]
As introduced in Section~\ref{subsec:inferential-premium-effect-criteria},
the corresponding inferential criterion is
\begin{equation}
\Gamma(C)
=
g_{\mathrm{inf}}^{\mathrm{reg},\mathrm{reg}}(C)
=
\sum_{\substack{c,c'\\s_c,s_{c'}>0}}
s_cs_{c'}\,
\operatorname{dist}_{\dcal}(\eta_c,\eta_{c'}).
\label{eq:MR1-Gamma}
\end{equation}
Thus $\Gamma(C)$ is the probability-weighted average dissimilarity between
the protected-group profiles of the tariff classes. We assume that
$\operatorname{dist}_{\dcal}$ is nonnegative, vanishes exactly for identical
distributions, and is jointly convex. Joint convexity means that mixing
pairs of distributions cannot increase their dissimilarity beyond the
corresponding weighted average. This property holds, for example, for total
variation distance and squared Euclidean distance between finite-group
probability vectors.

To relate the portfolio and regulatory populations, we introduce a
simplifying absolute-continuity assumption. Write
$\bbp_\Theta=\bbp\circ\Theta^{-1}$ and, for any reference law
$\bbp^\star$, $\bbp_\Theta^\star=\bbp^\star\circ\Theta^{-1}$;
in particular,
$\bbp_\Theta^{\mathrm{reg}}
=\bbp^{\mathrm{reg}}\circ\Theta^{-1}$.
Throughout this section, we assume
$\bbp_\Theta^{\mathrm{reg}}\ll\bbp_\Theta$.
This assumption is needed, in particular, for the inferential-threshold
and risk-inadequacy-cost results below. It ensures that identities
involving policyholder types that hold almost surely under the
portfolio law also hold under the regulatory law.

To interpret this condition, consider the stronger assumption that
$\bbp_\Theta$ and $\bbp_\Theta^{\mathrm{reg}}$ are equivalent.
Equivalence means that a measurable set of types has positive
probability under one law if and only if it has positive probability
under the other. In a discrete-type model, this means that the
same types have positive probability in both populations, although
their probabilities may differ.

Within the classification problem considered here, zero risk inadequacy requires the classification to distinguish different benchmark
risk prices, but it does not require distinctions between policyholders with
the same benchmark risk price. As shown above, $L(C)=0$ if and only if
$R$ is $\sigma(C(\Theta))$-measurable $\bbp$-a.s. Let $\mathfrak C_{\mathrm{cl}}$ denote the admissible family of
finite classifications. The following lemma
describes how merging tariff classes affects the inferential criterion.

\begin{lemma}[Coarsening]
\label{lem:inferential-coarsening}
Let $C:\widetilde\Theta\to\{1,\ldots,m\}$ and
$\widetilde C:\widetilde\Theta\to\{1,\ldots,\widetilde m\}$ be finite
classifications. If there is a map
$\varphi:\{1,\ldots,m\}\to\{1,\ldots,\widetilde m\}$ such that
$\widetilde C(\Theta)=\varphi(C(\Theta))$ $\bbp^{\mathrm{reg}}$-a.s.,
then $\Gamma(\widetilde C)\leq\Gamma(C)$.
\end{lemma}
The coarsening lemma shows that merging tariff classes cannot increase the value of the
inferential criterion. Its proof is given in
Appendix~\ref{app:proof-inferential-coarsening}.

Suppose now that $R$ has finite essential range
$\{r_1,\ldots,r_K\}$ under $\bbp$, with distinct $r_j$.
Let $C^\pi$ assign one class to each benchmark-price level:
$C^\pi(\vartheta)=j$ when $\pi(\vartheta)=r_j$, with arbitrary assignments
on the remaining $\bbp_\Theta$-null set. Every classification $C$ with $L(C)=0$ refines
$C^\pi$ almost surely under both laws. By the coarsening lemma,
$\Gamma(C^\pi)\leq\Gamma(C)$. Thus, among classifications compatible with
perfect risk adequacy, $C^\pi$ attains the smallest value of the inferential
criterion.

\begin{theorem}[Inferential threshold for perfect risk adequacy]
\label{thm:inferential-threshold}
Suppose that $R$ has finite essential range,
$\bbp_\Theta^{\mathrm{reg}}\ll\bbp_\Theta$, and
$C^\pi\in\mathfrak C_{\mathrm{cl}}$. Then
\begin{equation}
\min\left\{
\Gamma(C):
C\in\mathfrak C_{\mathrm{cl}},\ L(C)=0
\right\}
=
\Gamma(C^\pi)
=:\varepsilon_0.
\label{eq:MR1-threshold-min}
\end{equation}
Within the classification problem, zero risk inadequacy under the
constraint $\Gamma(C)\leq\varepsilon$ is therefore possible exactly when
$\varepsilon\geq\varepsilon_0$.
\end{theorem}
Appendix~\ref{app:proof-inferential-threshold} provides the proof.
The threshold is determined by the joint distribution of the benchmark risk
price $R$ and the protected characteristic $D$ under the regulatory law.
It does not depend on a particular aggregation rule or transfer scheme,
since $L(C)$ evaluates the classification independently of these
components. The theorem establishes that any classification allowing
exact recovery of the benchmark risk price has an inferential-criterion
value of at least $\varepsilon_0$. Whether these benchmark prices can
actually be implemented using a specified aggregation component and
balanced transfer is a separate question.

\begin{corollary}[Vanishing inferential criterion]
\label{cor:exact-inferential-neutrality}
Under the assumptions of Theorem~\ref{thm:inferential-threshold},
$\varepsilon_0=0$ if and only if $R$ and $D$ are independent under
$\bbp^{\mathrm{reg}}$.
\end{corollary}

Appendix~\ref{app:proof-exact-inferential-neutrality} provides the proof.
If $R$ has infinite essential range under $\bbp$, no finite
classification can recover it exactly. Consequently, every finite
classification incurs positive risk inadequacy, even in the absence
of an inferential constraint.

The benchmark-price classification $C^\pi$ attains the threshold
$\varepsilon_0$ in eq.~\eqref{eq:MR1-threshold-min}.
We now quantify, in a special case, how refinements of $C^\pi$
affect the inferential criterion. For this analysis and the remaining
quantitative results, suppose that $R$ has finite essential range
$\{r_1,\ldots,r_K\}$ and that $D\in\{1,\ldots,k\}$.
Identify distributions of $D$ with probability vectors in $\bbr^k$,
and use squared Euclidean dissimilarity,
$\operatorname{dist}_{\dcal}(\mu,\nu)=\|\mu-\nu\|_2^2$,
for the remainder of this section.

Under squared Euclidean dissimilarity, $\Gamma(C)$ measures the
average squared distance between the protected-group profiles of
pairs of tariff classes, weighted by their regulatory class
frequencies. Let $\eta=\bbp^{\mathrm{reg}}(D\in\cdot)$ denote
the protected-group probability vector under the regulatory law.
Since $\eta=\sum_{c:s_c>0}s_c\eta_c$, the criterion can equivalently
be expressed as
\begin{equation}
\Gamma(C)
=
2\sum_{c:s_c>0}s_c\|\eta_c-\eta\|_2^2.
\label{eq:MR1-Gamma-variance}
\end{equation}
Thus $\Gamma(C)$ is twice the probability-weighted squared deviation
of the class profiles from the population profile.
Appendix~\ref{app:proof-Gamma-variance} verifies the identity.

Consider now a classification $C$ with $L(C)=0$.
Within each tariff class, the benchmark risk price $R$ is then
constant $\bbp$-almost surely. By
$\bbp_\Theta^{\mathrm{reg}}\ll\bbp_\Theta$, this also holds
under the regulatory law, so $C$ refines $C^\pi$ up to null sets.
For each benchmark-price level $r$ of positive regulatory probability,
let $\widetilde\eta_r=\bbp^{\mathrm{reg}}(D\in\cdot\mid R=r)$
denote its protected-group distribution.
Applying the variance decomposition within each benchmark-price
level gives
\begin{equation}
\Gamma(C)-\varepsilon_0
=
2\bbe_{\bbp^{\mathrm{reg}}}
\!\left[
\|\eta_{C(\Theta)}-\widetilde\eta_R\|_2^2
\right].
\label{eq:MR1-inferential-excess}
\end{equation}
This identity quantifies the increase in inferential disparity
resulting from the refinement. The increase is zero if and only if
every tariff class of positive regulatory probability has the same
protected-group distribution as its corresponding benchmark-price
level.
Appendix~\ref{app:proof-inferential-excess} provides the proof.

\subsection{The cost of stricter inferential constraints}
\label{subsec:finite-class-approximation}

We now impose an inferential requirement
$\Gamma(C)\leq\varepsilon$ with $0\leq\varepsilon<\varepsilon_0$.
The threshold result excludes perfect risk adequacy under this
requirement. Can risk inadequacy nevertheless be made
arbitrarily small, or must every qualifying classification incur a
uniformly positive cost?
After answering this question, we examine how merging particular
classes changes both criteria and how these changes can be used to
construct an optimal classification within a specified family.
Class premiums remain unrestricted in the definition of $L(C)$;
implementation through aggregation and balanced transfers is considered
at the end of the subsection.

For the cost analysis, suppose that $R$ has $K\geq2$ distinct essential
values. The separation between these values matters because policyholders with
different benchmark risk prices must be charged the same premium when
assigned to the same tariff class. Write
$\delta_R=\min_{i\ne j}|r_i-r_j|>0$
for the smallest such separation.
The comparison between population laws also matters:
$L(C)$ weights premium deviations under $\bbp$, while $\Gamma(C)$
weights protected-group profiles under $\bbp^{\mathrm{reg}}$.
Absolute continuity suffices for a uniform positive cost.
An explicit bound follows when the regulatory type density relative
to the portfolio type density is bounded by a finite constant $M$.
The density bound allows the regulatory expected within-class
variance of benchmark risk prices to be bounded by the product
$M\cdot L(C)$.

\begin{theorem}[Risk inadequacy cost below the threshold]
\label{thm:inferential-cost}
Suppose that $R$ has $K\geq2$ distinct essential values and
$\bbp_\Theta^{\mathrm{reg}}\ll\bbp_\Theta$.
For every $0\leq\varepsilon<\varepsilon_0$, there exists
$c_\varepsilon>0$ such that every finite classification satisfying
$\Gamma(C)\leq\varepsilon$ also satisfies $L(C)\geq c_\varepsilon$.

If, in addition,
$d\bbp_\Theta^{\mathrm{reg}}/d\bbp_\Theta\leq M$
$\bbp_\Theta$-almost surely for a finite constant $M$, then every
finite classification satisfies
\begin{equation}
L(C)
\geq
\frac{\delta_R^2}{4M}
\bigl(\varepsilon_0-\Gamma(C)\bigr)_+,
\label{eq:MR1-inferential-cost}
\end{equation}
where $x_+=\max\{x,0\}$.
The numerical constant $1/4$ is sharp: no larger constant makes
this bound valid under all the stated assumptions, even when
$\bbp^{\mathrm{reg}}=\bbp$.
\end{theorem}

The first assertion rules out a sequence of finite classifications
that satisfies the same requirement $\Gamma(C)\leq\varepsilon$
while approaching perfect risk adequacy.
Under the density bound, the second assertion quantifies the cost:
the improvement over the inferential threshold gives a lower bound
on $L(C)$, scaled by the squared separation of benchmark prices
and the comparison between the two population laws.
This bound applies to all finite classifications, without imposing
a common upper limit on their number of classes.

The proof refines $C$ by additionally recording the benchmark-price
level. The resulting increase in the inferential criterion bounds
any improvement below $\varepsilon_0$ and is controlled by the
within-class variance of benchmark risk prices under the regulatory
law. Absolute continuity then relates this variance to $L(C)$.
Appendix~\ref{app:proof-inferential-cost} gives the argument and
an example attaining the constant $1/4$.

The lower bound does not tell us how to choose a classification
satisfying a given inferential requirement.
One possibility is to merge existing tariff classes.
By the coarsening lemma, this cannot increase $\Gamma(C)$,
but it may increase risk inadequacy.
We therefore calculate both effects of merging two classes:
the reduction in $\Gamma(C)$ and the increase in $L(C)$.

\begin{proposition}[Exact effects of pooling]
\label{prop:pooling-effects}
Let $C^{a,b}$ be obtained from $C$ by merging two classes $a,b$
of positive probability under both laws.
Write $q_c=\bbp(C(\Theta)=c)$ and
$\bar r_c=\bbe_{\bbp}[R\mid C(\Theta)=c]$.
With $s_c=\bbp^{\mathrm{reg}}(C(\Theta)=c)$ and
$\eta_c=\bbp^{\mathrm{reg}}(D\in\cdot\mid C(\Theta)=c)$
as above, we have
\begin{equation}
L(C^{a,b})-L(C)
=\frac{q_aq_b}{q_a+q_b}(\bar r_a-\bar r_b)^2,
\qquad
\Gamma(C)-\Gamma(C^{a,b})
=\frac{2s_as_b}{s_a+s_b}\|\eta_a-\eta_b\|_2^2.
\label{eq:MR1-pooling-effects}
\end{equation}
\end{proposition}

The additional risk inadequacy comes from replacing two
conditional mean benchmark prices by one common premium.
The reduction in inferential disparity comes from replacing two
protected-group profiles by their mixture.
Accordingly, the two changes depend on different features of the
classes and are weighted under different population laws.
Classes with similar conditional mean benchmark prices but substantially
different protected-group profiles can yield a small increase in
$L(C)$ together with a substantial decrease in $\Gamma(C)$.
Appendix~\ref{app:proof-pooling-effects} proves the identities.

These formulas evaluate a particular merge. They do not yet solve
the problem of minimizing $L(C)$ under a prescribed bound on
$\Gamma(C)$. At each step, several merges may be possible,
leading to different classifications and different choices for
subsequent merges. Finding the best resulting classification
therefore remains a combinatorial optimization problem.
To obtain an exact optimization procedure, we next consider a
specified family in which the complete partition can be optimized
jointly.

The starting point is the individual benchmark risk price assigned
by the chosen premium principle for a specified insurance product
and a common unit of insured exposure. This basis is fixed throughout
the classification problem. Individuals with equal or similar
benchmark prices may differ substantially in their types and
observable covariates. Their conditional claim distributions may
also differ, but the premium principle assesses them as equal or
similar for pricing purposes. Since insurance involves collective
treatment based on the pricing of insured risks, individuals with
similar benchmark prices can be grouped for common pricing without
necessarily introducing substantial risk inadequacy, even if they
differ considerably in other respects. A tariff class represents such an operational collective: its members
are assigned a common premium. This common pricing treatment defines
the class for the purposes of our analysis. Our general framework
allows classifications based on other characteristics of policyholder
types, including refinements that distinguish individuals with the
same benchmark risk price. For the analysis that follows, we specialize
to classifications constructed by grouping policyholders according
to their benchmark risk prices.

Order the benchmark-price levels as $r_1<\cdots<r_K$.
Consider coarsenings of $C^\pi$ with at most a prescribed number
$m\geq1$ of tariff classes, each containing consecutive benchmark-price
levels. This restriction pools adjacent parts of the benchmark-price
range: a class containing two levels must also contain all levels
between them.
Covariates may provide information for implementing such a classification,
but they may also support classifications outside this family. Here we optimize only within the consecutive-level
family, evaluating risk inadequacy through $L(C)$ and
inferential disparity through $\Gamma(C)$.

For a fixed weight $\lambda\geq0$, minimize
$L(C)+\lambda\Gamma(C)$ over this family.
The weight specifies the relative contribution of inferential
disparity to the scalarized objective. For each class, $L(C)$
includes the weighted squared deviations of benchmark prices
from their class mean, while $\Gamma(C)$ includes the weighted
squared deviation of its protected-group profile from the
population profile. These contributions depend only on the
benchmark-price levels assigned to each class and add up to
the total objective. Once the final class is fixed, the
preceding classes can therefore be optimized in the same way
on the remaining initial segment. This gives the
dynamic-programming structure.

\begin{proposition}[Optimal classification with consecutive benchmark-price levels]
\label{prop:classification-DP}
For $\lambda\geq0$, minimizing $L(C)+\lambda\Gamma(C)$ over
coarsenings of $C^\pi$ with at most $m$ tariff classes, each consisting
of consecutive benchmark-price levels, is an additive finite
segmentation problem and can be solved by dynamic programming.
\end{proposition}

Appendix~\ref{app:proof-classification-DP} specifies the class
contributions, the recursion, and the recovery of an optimal
classification. This optimality concerns the scalarized objective
within the consecutive-level family, not all admissible classifications.
Moreover, varying $\lambda$ need not recover every efficient
trade-off within this family. In particular, a classification
minimizing $L(C)$ under a prescribed bound on $\Gamma(C)$
need not minimize $L(C)+\lambda\Gamma(C)$ for any $\lambda\geq0$.
Section~\ref{sec:finite-model} illustrates the trade-offs using
consecutive-level classifications.

Finally, we return to the tariff architecture.
The classification analysis permits arbitrary class premiums and
therefore identifies the risk inadequacy that classification
alone imposes. If an aggregation component is prescribed and transfers
must balance, the available premiums face an additional restriction.
For perfect risk adequacy, the resulting implementation question has
a direct answer.

Fix $C$ and $\mathcal A$, and vary only $\Lambda$, writing
$\mathbf C=(C,\mathcal A,\Lambda)$.
The aggregate benchmark prices $A^{\mathbf C}(c)$ are then fixed.
Suppose that $\bbp_\Theta^\star\ll\bbp_\Theta$,
that $R$ and $A^{\mathbf C}(C(\Theta))$ are
$\bbp^\star$-integrable, and that transfers are unrestricted apart
from class dependence and $\bbp^\star$-balance.
If $L(C)=0$, perfect risk adequacy can be implemented if and only if
$\bbe_{\bbp^\star}[A^{\mathbf C}(C(\Theta))]
=\bbe_{\bbp^\star}[R]$.

Indeed, $L(C)=0$ means that $R$ is a function of the tariff class
$\bbp$-almost surely. Charging this benchmark price therefore requires
the class-dependent transfer
$\Lambda(C(\Theta))=R-A^{\mathbf C}(C(\Theta))$.
Absolute continuity makes this identity valid under
$\bbp^\star$ as well, and the stated expectation equality says that
the required transfer has zero reference-law mean.
Thus, once classification permits perfect risk adequacy, implementation
by balanced transfers requires the aggregation component to preserve
the reference-law mean benchmark price.
Appendix~\ref{app:classification-implementation} provides the proof.

Section~\ref{sec:balanced-transfers} studies the broader transfer
problem for fixed classification and aggregation: which protected-group
total premium effects can be attained, what minimum risk inadequacy they require, and what freedom remains to improve
portfolio coverage.

\section{Balanced transfers: attainable objectives and their costs}
\label{sec:balanced-transfers}

Section~\ref{sec:tariff-framework} formulates tariff construction as a
multi-objective problem, with Pareto efficiency describing the trade-offs
among the different objectives. Section~\ref{sec:main-result-I} studies one
part of this problem by varying the classification and analyzing the resulting
trade-off between risk adequacy and the inferential criterion. We
now turn to the role of transfers. To isolate what transfers can achieve, we
fix the classification $C$ and aggregation component $\mathcal A$ and vary
the transfer $\Lambda$ subject to $\bbp^\star$-balance. Apart from this
balance condition, class premiums are allowed to take arbitrary real values.

For a fixed classification and aggregation component, we first characterize
which protected-group total premium effects can be attained and determine
the minimum risk inadequacy required to attain each feasible
target. We also characterize the remaining freedom among transfers that
produce the same prescribed protected-group total premium effects. 

We then relate inferential disparity, risk inadequacy, and
the dispersion of protected-group total premium effects.
When the population laws coincide, the inferential criterion yields
a lower bound on the risk inadequacy required to reduce
this dispersion to a prescribed level. We establish that the bound
can be attained and then examine what the inferential criterion
leaves undetermined. By varying the classification within a fixed
population, we show that the minimum cost of exact compensation
can become arbitrarily large even while the inferential criterion
and the attainable protected-group targets remain unchanged.

Finally, for a fixed classification, aggregation component, and
protected-group target, we study whether the remaining transfer
freedom can improve portfolio coverage and what additional
risk inadequacy this implies. Detailed proofs and
example constructions are collected in
Appendix~\ref{app:balanced-transfers}.

\subsection{Attainable protected-group total premium effects}
\label{subsec:transfer-outcomes-cost}

Fix a classification $C$ with $m$ tariff classes and an aggregation
component $\mathcal A$. As specified above, we vary only the transfer
$\Lambda$, subject to $\bbp^\star$-balance, and otherwise allow arbitrary
real-valued class premiums. Which protected-group total premium effects
can this transfer family generate? Among the transfers attaining a prescribed
target, which minimizes risk inadequacy?

Throughout this section, risk inadequacy is measured by
squared premium deviation. Recall that $R=\pi(\Theta)$, and assume
$R\in L^2(\bbp)\cap L^1(\bbp^\star)$. Let
$D\in\{1,\ldots,k\}$, and write
$q_c=\bbp(C(\Theta)=c)>0$,
$q_c^\star=\bbp^\star(C(\Theta)=c)$, and
$\rho_d^\star=\bbp^\star(D=d)>0$.
Denote the corresponding vectors by $q$, $q^\star$, and $\rho^\star$.
For reference, the probability notation used here and in
Section~\ref{sec:main-result-I} is summarized below.
\begin{center}
\begin{tabular}{c l}
Notation & Meaning\\
\hline
$q_c$ & $\bbp(C(\Theta)=c)$: portfolio class probability\\
$s_c$ & $\bbp^{\mathrm{reg}}(C(\Theta)=c)$: regulatory class probability\\
$q_c^\star$ & $\bbp^\star(C(\Theta)=c)$: reference class probability\\
$\rho_d^\star$ & $\bbp^\star(D=d)$: reference protected-group probability\\
$\eta_c$ & $\bbp^{\mathrm{reg}}(D\in\cdot\mid C(\Theta)=c)$:
regulatory class profile
\end{tabular}
\end{center}
Risk inadequacy and portfolio coverage are evaluated under
the portfolio law $\bbp$, while transfer balance and protected-group
total premium effects are evaluated under the reference law
$\bbp^\star$. These laws remain fixed throughout each optimization.

As in Section~\ref{sec:main-result-I}, let
$p=(p_1,\ldots,p_m)^\top$ denote the class-premium vector. Here
$p_c=A^{\mathbf C}(c)+\Lambda(c)$, with
$\mathbf C=(C,\mathcal A,\Lambda)$. Since $C$ and $\mathcal A$ are
fixed, the aggregate benchmark prices $A^{\mathbf C}(c)$ are fixed.
Every $p\in\bbr^m$ therefore specifies a unique transfer,
$\Lambda(c)=p_c-A^{\mathbf C}(c)$. For this fixed classification and
aggregation component, we write $f_{\mathrm{RA}}(p)$ and
$f_{\mathrm{cov}}(p)$ for the corresponding tariff criteria.

The transfer is $\bbp^\star$-balanced if and only if
$(q^\star)^\top p
=\bbe_{\bbp^\star}[A^{\mathbf C}(C(\Theta))]$.
Thus the available class-premium vectors form the affine hyperplane
$\mathcal H
=\{p\in\bbr^m:(q^\star)^\top p
=\bbe_{\bbp^\star}[A^{\mathbf C}(C(\Theta))]\}$.
A change $v$ in the class-premium vector preserves balance if and only if
$(q^\star)^\top v=0$.

We next express the protected-group total premium effects of
these class premiums. Define
$Q_{dc}=\bbp^\star(C(\Theta)=c\mid D=d)$ and
$\mu_d^\star=\bbe_{\bbp^\star}[R\mid D=d]$, and write $\mu^\star$
for the vector of protected-group conditional mean benchmark risk prices.
The $d$th row of $Q$ gives the conditional distribution of
tariff classes given $D=d$ under $\bbp^\star$. Hence $(Qp)_d$ is its mean charged premium,
and the finite-dimensional representation introduced in
Section~\ref{subsec:inferential-premium-effect-criteria} gives
$M_\Delta^{\mathbf C,\star}=Qp-\mu^\star$.

The attainable protected-group total premium-effect vectors
therefore form the affine set
$\mathcal T_C=\{Qp-\mu^\star:p\in\mathcal H\}
=Q\mathcal H-\mu^\star$. Balance determines the available class-premium
vectors, and the map $p\mapsto Qp-\mu^\star$ determines their
protected-group total premium effects. Balance fixes the mean of these effects. Indeed,
$Q\mathbf1_m=\mathbf1_k$ and
$(\rho^\star)^\top Q=(q^\star)^\top$, where $\mathbf1_j$
denotes the vector of $j$ ones. Consequently, every
$y\in\mathcal T_C$ satisfies $(\rho^\star)^\top y=b$, with
$b=\bbe_{\bbp^\star}[W^{\mathbf C}]
=\bbe_{\bbp^\star}[A^{\mathbf C}(C(\Theta))-R]$.
Balanced transfers can redistribute protected-group total
premium effects around this fixed mean.

Fix an attainable target $y\in\mathcal T_C$.
This determines the protected-group total premium effects,
but not necessarily the premiums charged to individual tariff
classes. The class-premium vectors implementing this target
are the solutions of $Qp-\mu^\star=y$.
These solutions automatically satisfy transfer balance:
the fixed mean of the protected-group effects implies that
the charged premiums have the same mean under $\bbp^\star$
as the aggregate benchmark prices.
Thus the set of possible implementations is
$\mathcal F_y=\{p\in\bbr^m:Qp-\mu^\star=y\}$.
The remaining freedom is exactly $\ker Q$:
class premiums can be changed along these directions without
changing any protected-group total premium effect or violating balance.

To minimize risk inadequacy on $\mathcal F_y$, recall from
Proposition~\ref{prop:pooling-effects} the portfolio
conditional mean benchmark risk price for tariff class $c$,
$\bar r_c=\bbe_{\bbp}[R\mid C(\Theta)=c]$.
Let $\bar r$ denote the vector of these tariff-class-conditional
mean benchmark risk prices.
Equip $\bbr^m$ with the inner product
$\langle x,z\rangle_q=\sum_{c=1}^m q_cx_cz_c$ and the
corresponding squared norm
$\|z\|_q^2=\sum_{c=1}^m q_cz_c^2$.
Conditional-expectation orthogonality gives
$f_{\mathrm{RA}}(p)=L(C)+\|p-\bar r\|_q^2$:
the cross term vanishes because
$\bbe_{\bbp}[R-\bar r_{C(\Theta)}\mid C(\Theta)]=0$.
Since $L(C)$ is fixed, minimizing risk inadequacy amounts
to finding the point of $\mathcal F_y$ closest to $\bar r$ in the
$q$-norm. The portfolio law determines this distance; the reference
law determines the balance and protected-group constraints defining
$\mathcal F_y$.

This is an orthogonal-projection problem in a finite-dimensional
Hilbert space. Denote the unique $q$-orthogonal projection of
$\bar r$ onto $\mathcal F_y$ by $p_y$, and define its squared
distance by
\begin{equation}
c_C(y)
=
\min_{p\in\mathcal F_y}\|p-\bar r\|_q^2
=
\|p_y-\bar r\|_q^2.
\label{eq:transfer-target-cost}
\end{equation}
Thus $c_C(y)$ is the minimum increase in risk inadequacy
above $L(C)$ required to attain $y$. The vector $\bar r$
minimizes risk inadequacy over unrestricted class premiums;
it need not satisfy balance or attain $y$.

For an equivalent algebraic description of attainability, define
$w_C=Q\bar r-\mu^\star$, the protected-group total
premium-effect vector generated by $\bar r$.
The equation for attaining $y$ becomes
$Q(p-\bar r)=y-w_C$. Together with the mean restriction imposed
by balance, this gives the following result.

\begin{theorem}[Attainable protected-group total premium effects and minimum risk inadequacy]
\label{thm:transfer-target-cost}
A target $y\in\bbr^k$ is attainable by a $\bbp^\star$-balanced
transfer if and only if
\begin{equation}
(\rho^\star)^\top y=b,
\qquad
y-w_C\in\operatorname{im}Q.
\label{eq:transfer-target-feasibility}
\end{equation}
For every attainable $y$, the unique class-premium vector minimizing
risk inadequacy is the $q$-orthogonal projection $p_y$
of $\bar r$ onto $\mathcal F_y$, and
\begin{equation}
f_{\mathrm{RA}}(p_y)=L(C)+c_C(y).
\label{eq:transfer-minimum-risk-inadequacy}
\end{equation}
Every class-premium vector attaining $y$ is uniquely of the form
$p_y+v$ with $v\in\ker Q$, and
\begin{equation}
f_{\mathrm{RA}}(p_y+v)
=
L(C)+c_C(y)+\|v\|_q^2.
\label{eq:transfer-target-decomposition}
\end{equation}
\end{theorem}

The last identity is the Pythagorean decomposition:
$p_y-\bar r$ is $q$-orthogonal to the direction space $\ker Q$
of $\mathcal F_y$.
Appendix~\ref{app:transfers-quadratic-formulation} develops the
underlying quadratic formulation, while
Appendix~\ref{app:transfers-minimum-cost} provides the proof
and derives explicit formulas for computing $p_y$ and $c_C(y)$.

The attainability result also identifies limitations of
class-dependent transfers. If the conditional distribution of
tariff classes given protected-group membership is identical
for two protected groups, changes in tariff-class premiums
affect their mean charged premiums equally. The difference
between their protected-group total premium effects
therefore remains unchanged. The theorem also shows that
every target satisfying $(\rho^\star)^\top y=b$ is attainable
if and only if $\operatorname{rank}Q=k$.

Two targets are of particular interest. Group neutrality means
that the average premium effect relative to the benchmark risk
price is the same for every protected group. Exact compensation
requires this effect to be zero for every protected group.
The corresponding targets are $y=b\mathbf1_k$ and $y=0$,
respectively. Since balance fixes the overall mean effect at
$b$, exact compensation is possible only if $b=0$.
For either target, \eqref{eq:transfer-target-feasibility}
determines whether it can be attained, while
\eqref{eq:transfer-minimum-risk-inadequacy} gives the minimum
risk inadequacy required.

Keep the classification, benchmark risk prices, and population laws
fixed, and consider different aggregation components. Each aggregation
component determines the mean premium effect
$b=\bbe_{\bbp^\star}[A^{\mathbf C}(C(\Theta))-R]$.
When class-dependent transfers are unrestricted apart from balance,
this mean is the only way the aggregation component restricts
the available class-premium vectors. Consequently, two aggregation
components with the same value of $b$ allow exactly the same
class-premium vectors, and hence the same attainable protected-group
total premium effects and minimum risk inadequacy costs.
Additional restrictions on premiums or transfers may change
these conclusions.

To recover the corresponding part of the Pareto problem, let $y_D$
denote the random variable taking value $y_d$ on $\{D=d\}$.
For $y\in\mathcal T_C$, the objective vector after minimizing
risk inadequacy is
$(L(C)+c_C(y),\mathfrak R^\star(y_D),\mathfrak D^\star(y_D))$.
Its nondominated values form the Pareto frontier for risk inadequacy and the two protected-group premium-effect criteria,
with $C$ and $\mathcal A$ fixed. The inferential criterion remains
fixed. Portfolio coverage can distinguish other class-premium
vectors in $\mathcal F_y$ and is considered in
Subsection~\ref{subsec:transfer-coverage}.

\subsection{Inferential disparity, risk inadequacy, and protected-group dispersion}
\label{subsec:transfer-inferential-cost}

We now relate three tariff criteria: inferential disparity, risk inadequacy, and the dispersion of protected-group total premium
effects. For fixed classification and aggregation components, we ask:
if balanced transfers must reduce this dispersion to a prescribed level,
what lower bound on risk inadequacy follows from the
inferential criterion $\Gamma(C)$?

The relation arises because class-dependent premium adjustments
affect the mean charged premium of each protected group according
to the conditional distribution of tariff classes given membership
in that group. When these distributions are similar, the same
adjustments change the protected-group means by similar amounts.
Reducing differences between protected-group total premium effects
can require substantial changes in tariff-class premiums.
The following result quantifies this relation.

Retain the notation and assumptions of
Subsection~\ref{subsec:transfer-outcomes-cost}.
For the lower-bound analysis, fix $C$ and $\mathcal A$,
and vary only the balanced transfer, allowing otherwise
unrestricted real-valued class premiums.
The subsequent constructions compare different classifications
to examine the sharpness and limitations of this bound.
For inferential disparity $\Gamma(C)$, use squared Euclidean
dissimilarity and the representation \eqref{eq:MR1-Gamma-variance}.
For protected-group dispersion $D_\Delta^\star(\mathbf C)$ in
\eqref{eq:group-effect-criteria}, choose standard deviation as
the deviation measure $\mathfrak D^\star$.
Throughout this subsection, suppose
$\bbp^\star=\bbp^{\mathrm{reg}}=\bbp$.
Thus risk inadequacy, inferential disparity, protected-group
dispersion, and transfer balance all refer to the same population.
For a prescribed $s\geq0$, the requirement is
$D_\Delta^\star(\mathbf C)\leq s$.

We first identify the balanced tariff that minimizes risk inadequacy.
Recall that $\bar r_c=\bbe_{\bbp}[R\mid C(\Theta)=c]$ is the
portfolio conditional mean benchmark risk price for tariff class $c$,
and that $f_{\mathrm{RA}}(p)=L(C)+\|p-\bar r\|_q^2$.
Transfer balance requires $q^\top(p-\bar r)=b$, where
$b=\bbe_{\bbp}[A^{\mathbf C}(C(\Theta))-R]$.
Among balanced class-premium vectors, risk inadequacy is minimized
by setting $p_c=\bar r_c+b$ for every tariff class $c$, that is,
by adding the same amount $b$ to each tariff-class conditional
mean benchmark risk price. The minimizing class-premium vector
is $p=\bar r+b\mathbf1_m$, and the minimum risk inadequacy
is $L(C)+b^2$.

We take this balanced tariff as our starting point for reducing
protected-group dispersion. Its protected-group total premium-effect
vector is $b\mathbf1_k+w_C$, where $w_C=Q\bar r-\mu^\star$
has zero $\rho^\star$-weighted mean.
Using the protected-group probabilities $\rho_d^\star$, define
$\|z\|_{\rho^\star}^2=\sum_{d=1}^k\rho_d^\star z_d^2$.
Since $w_C$ has zero weighted mean, its norm
$\|w_C\|_{\rho^\star}$ is the standard deviation of the
protected-group total premium effects.
If this baseline dispersion does not exceed $s$, no further
adjustment is needed. Otherwise, reducing dispersion requires
changing the balanced class premiums.

To determine the additional cost, recall from the balance
condition that any balanced class-premium vector can be written
as $p=\bar r+b\mathbf1_m+u$, where $q^\top u=0$.
The vector $u$ represents the adjustment from the balanced baseline.
Since $u$ is $q$-orthogonal to the constant vector, we obtain
\begin{equation}
f_{\mathrm{RA}}(p)=L(C)+b^2+\|u\|_q^2,
\qquad
M_\Delta^{\mathbf C,\star}=b\mathbf1_k+w_C+Qu.
\label{eq:transfer-risk-adequacy-cost}
\end{equation}
Thus the adjustment increases risk inadequacy by $\|u\|_q^2$
and changes the protected-group total premium effects by $Qu$.

The inferential disparity $\Gamma(C)$ measures variation in
the protected-group profiles of tariff classes and yields a
bound on the weighted operator norm of $Q$ restricted to
balanced adjustments $q^\top u=0$.
The $q$-norm measures the size of the premium adjustments,
while the $\rho^\star$-norm measures the resulting changes
in protected-group total premium effects.

More precisely, let
$\rho_{\min}^\star=\min_d\rho_d^\star>0$ denote the
smallest protected-group probability.
A Cauchy--Schwarz argument gives, for every $u$ with
$q^\top u=0$,
$
\|Qu\|_{\rho^\star}^2
\leq
\frac{\Gamma(C)}{2\rho_{\min}^\star}\|u\|_q^2.
$
Thus small inferential disparity limits the changes in
protected-group total premium effects that can be achieved
for a given increase $\|u\|_q^2$ in risk inadequacy.
Appendix~\ref{app:transfers-inferential-cost} provides
the derivation.

We now apply this estimate to the dispersion requirement.
Since $w_C$ and $Qu$ both have zero $\rho^\star$-weighted mean,
the resulting protected-group dispersion is
$D_\Delta^\star(\mathbf C)=\|w_C+Qu\|_{\rho^\star}$.
By the triangle inequality, achieving
$D_\Delta^\star(\mathbf C)\leq s$ requires
$\|Qu\|_{\rho^\star}
\geq(\|w_C\|_{\rho^\star}-s)_+$.

Combining this necessary adjustment with the preceding estimate
gives a lower bound on the additional risk inadequacy,
as stated in the following proposition. Appendix~\ref{app:transfers-inferential-cost} provides the proof. 

\begin{proposition}[Risk inadequacy under a bound on protected-group dispersion]
\label{prop:transfer-inferential-bound}
Let $s\geq0$. If $\Gamma(C)>0$, every balanced class-premium
vector whose tariff satisfies $D_\Delta^\star(\mathbf C)\leq s$
obeys
\begin{equation}
f_{\mathrm{RA}}(p)
\geq
L(C)+b^2+
\frac{2\rho_{\min}^\star}{\Gamma(C)}
\bigl(\|w_C\|_{\rho^\star}-s\bigr)_+^2.
\label{eq:transfer-budget-bound}
\end{equation}
If $\Gamma(C)=0$, every balanced class-premium vector generates
the same protected-group total premium-effect vector
$b\mathbf1_k+w_C$. The requirement
$D_\Delta^\star(\mathbf C)\leq s$ is then feasible if and only if
$\|w_C\|_{\rho^\star}\leq s$.
\end{proposition}

The proposition gives a lower bound on the risk inadequacy
required to meet a prescribed limit on protected-group dispersion.
The first two terms, $L(C)+b^2$, represent the minimum risk
inadequacy under transfer balance alone. Reducing dispersion
below its baseline value requires an additional cost, which
the bound relates to the required reduction and the inferential
criterion $\Gamma(C)$. The bound is necessary but does not
guarantee that the dispersion requirement can be met.

We next show that the bound is sharp by constructing a simple
example in which it is attained.

\begin{example}[Sharpness of the risk-inadequacy bound]
\label{ex:transfer-dispersion-sharpness}
Consider two equally likely protected groups and two equally
likely tariff classes. Let the benchmark risk price $R$ equal $r_0-a$ for every
individual in protected group 1 and $r_0+a$ for every
individual in protected group 2, where $r_0\geq a>0$.

Specify the conditional distribution of tariff classes
within the protected groups by
\[
Q=\frac12
\begin{pmatrix}
1+\theta & 1-\theta\\
1-\theta & 1+\theta
\end{pmatrix},
\qquad 0<\theta\leq1,
\]
where $Q_{dc}=\bbp(C(\Theta)=c\mid D=d)$.
Thus $\theta$ controls the dependence between tariff-class
membership and protected-group membership.
Use conditional-mean aggregation, so $b=0$.
The calculations in
Appendix~\ref{app:transfers-binary-example} give
$\Gamma(C)=\theta^2$,
$L(C)=a^2(1-\theta^2)$, and
$\|w_C\|_{\rho^\star}=a(1-\theta^2)$.

Exact compensation is attainable for every $\theta>0$.
Since $b=0$, zero protected-group dispersion is equivalent
to exact compensation in this example. As defined in Section~\ref{subsec:transfer-outcomes-cost},
$\mathcal F_0$ denotes the set of class-premium vectors
achieving exact compensation. The minimum risk inadequacy is
\begin{equation}
\min_{p\in\mathcal F_0}f_{\mathrm{RA}}(p)
=
a^2\left(\frac{1}{\Gamma(C)}-1\right).
\label{eq:transfer-binary-cost}
\end{equation}
This equals the lower bound
\eqref{eq:transfer-budget-bound} for $s=0$ and diverges
as $\theta\downarrow0$. The appendix verifies these results
and establishes that the bound is also attained for every
$s\geq0$ in this family.
\end{example}

The example shows that exact compensation can become
arbitrarily costly as inferential disparity approaches zero.
This raises a further question: can compensation costs also
become arbitrarily large when $\Gamma(C)$ remains fixed
at a strictly positive value?

The preceding bound controls how much balanced premium adjustments
can change protected-group total premium effects for a given
increase in risk inadequacy. It does not establish how effectively
premium adjustments can eliminate particular differences between
protected groups. The reason is that inferential disparity
$\Gamma(C)$ measures the overall variation in the protected-group
profiles of tariff classes, without determining how strongly
premium adjustments affect each particular difference.

To see the distinction, consider three protected groups.
Premium adjustments may be effective at changing the difference
between groups 1 and 2, while having very little influence on
the difference between group 3 and the first two groups taken
together. Inferential disparity can remain constant even as
the response of the second difference becomes arbitrarily weak.
If exact compensation requires eliminating that difference,
the necessary premium adjustments, and hence their risk
inadequacy, can become arbitrarily large.

The following proposition establishes this possibility.
We keep the population and individual benchmark risk prices
fixed and vary only the classification, while preserving the
protected-group probabilities, tariff-class probabilities,
and inferential disparity. Each classification uses
conditional-mean aggregation. Exact compensation remains
attainable throughout the family, but its minimum risk
inadequacy becomes unbounded.
\begin{proposition}[Unbounded compensation cost at fixed inferential disparity]
\label{prop:transfer-fixed-disparity}
There exist a fixed population with
$\bbp^\star=\bbp^{\mathrm{reg}}=\bbp$, a fixed benchmark risk-price
map $\pi$ defining $R=\pi(\Theta)$, and three equally likely
protected groups, together with a family of classifications
$C$ into three equally likely tariff classes.
For each classification, use conditional-mean aggregation,
$A^{\mathbf C}(c)=\bbe_{\bbp}[R\mid C(\Theta)=c]$.
Throughout the family,
$\Gamma(C)=1/24$ and $\operatorname{rank}Q=3$.
Exact compensation is attainable for every classification,
but the minimum risk inadequacy required to attain it
is unbounded over the family.
\end{proposition}

The proof, including the construction of the classifications
and the calculation of the compensation costs, is given in
Appendix~\ref{app:transfers-fixed-disparity}.

Conditional-mean aggregation gives $b=0$. Since $Q$ has full
row rank, every protected-group total premium-effect vector
with zero $\rho^\star$-weighted mean is attainable, including
the exact compensation target. Thus the attainable targets
remain unchanged throughout the family. The unbounded cost
arises despite exact compensation remaining feasible.

The proposition establishes that even joint knowledge of
the inferential disparity $\Gamma(C)$ and the distribution
of benchmark risk prices across protected groups, represented
by the joint distribution of $(R,D)$, does not provide a
finite upper bound on the cost of exact compensation.
As characterized by Theorem~\ref{thm:transfer-target-cost},
the exact cost also depends on the conditional distribution
matrix $Q$, the portfolio tariff-class probabilities $q$,
and the tariff-class conditional mean benchmark risk
prices $\bar r$.

The binary construction in
Example~\ref{ex:transfer-dispersion-sharpness} and the
three-group construction underlying
Proposition~\ref{prop:transfer-fixed-disparity}
both allow arbitrary real-valued tariff-class premiums
subject to transfer balance. Additional restrictions can
affect attainability. In particular, requiring nonnegative
premiums may make exact compensation infeasible even when
it is attainable under transfer balance alone.

\subsection{Portfolio coverage at fixed protected-group total premium effects}
\label{subsec:transfer-coverage}

We now ask whether tariff-class premiums can be adjusted to improve
portfolio coverage while preserving a prescribed vector of
protected-group total premium effects, and what additional risk
inadequacy such an improvement requires. The portfolio law $\bbp$
and the selected reference law $\bbp^\star$ may differ. Both laws
remain fixed during the optimization, but determine different
quantities:

\begin{center}
\small
\begin{tabular}{@{}p{0.14\linewidth}p{0.43\linewidth}p{0.32\linewidth}@{}}
\toprule
Law & Relevant quantities & Role in the optimization \\
\midrule
Portfolio $\bbp$
& Tariff-class probabilities $q$, tariff-class conditional mean
benchmark risk prices $\bar r$, classification-induced risk
inadequacy $L(C)$, and the joint distribution of aggregate
claims $S_n$ and portfolio tariff-class memberships
& Risk inadequacy $f_{\mathrm{RA}}$ and portfolio coverage
$f_{\mathrm{cov}}$ \\
\addlinespace
Reference $\bbp^\star$
& Tariff-class probabilities $q^\star$, protected-group
probabilities $\rho^\star$, conditional tariff-class probabilities
$Q$ given protected-group membership, and protected-group
conditional mean benchmark risk prices $\mu^\star$
& Transfer balance and protected-group total premium effects
$M_\Delta^{\mathbf C,\star}$, including their criteria
$R_\Delta^\star$ and $D_\Delta^\star$ \\
\bottomrule
\end{tabular}
\end{center}

Fix the classification $C$, aggregation component $\mathcal A$,
and an attainable protected-group total premium-effect target
$y\in\mathcal T_C$. By Theorem~\ref{thm:transfer-target-cost},
$p_y$ is the unique class-premium vector attaining $y$ with
minimum risk inadequacy. The reference law determines the
constraints defining the premium vectors attaining $y$,
while the portfolio law determines which of these vectors
minimizes risk inadequacy. The inferential disparity $\Gamma(C)$,
evaluated under the regulatory law $\bbp^{\mathrm{reg}}$,
remains fixed because $C$ does not change.

Theorem~\ref{thm:transfer-target-cost} also gives
$\mathcal F_y=p_y+\ker Q$: every class-premium vector attaining
$y$ is of the form $p_y+v$ with $Qv=0$. Such an adjustment
preserves protected-group total premium effects and hence the
criteria $R_\Delta^\star(\mathbf C)$ and
$D_\Delta^\star(\mathbf C)$; it also preserves transfer balance.
By \eqref{eq:transfer-target-decomposition}, the additional
risk inadequacy under $\bbp$ is exactly $\|v\|_q^2$.
An allowance $\delta\geq0$ therefore restricts the adjustments
to the $q$-ball of radius $\sqrt{\delta}$ within $\ker Q$.

The best portfolio coverage attainable within this allowance
is the minimum of the coverage criterion $f_{\mathrm{cov}}$:
\begin{equation}
K_y(\delta)
=
\min_{\substack{v\in\ker Q\\\|v\|_q^2\leq\delta}}
f_{\mathrm{cov}}(p_y+v).
\label{eq:transfer-conditional-coverage}
\end{equation}
The feasible set contains $v=0$ and is compact because
$q_c>0$ for every tariff class $c$. Hence the minimum
exists whenever $f_{\mathrm{cov}}$ is continuous. A total
risk-inadequacy bound $f_{\mathrm{RA}}\leq B$ permits an
additional allowance $\delta=B-L(C)-c_C(y)$ if
$B\geq L(C)+c_C(y)$. Otherwise, no class-premium vector
attaining $y$ satisfies the bound.

To determine to what extent admissible premium adjustments can improve
portfolio coverage, we examine the first-order behavior of
$f_{\mathrm{cov}}$ at the minimum-risk-inadequacy premium vector
$p_y$. When the coverage criterion is differentiable at $p_y$,
its derivative describes the first-order change in the coverage
criterion as a linear functional of the premium adjustment.
We represent the negative of this derivative using the
$q$-inner product, consistent with the $q$-norm used to
measure additional risk inadequacy. The following theorem identifies how admissible premium
adjustments can improve portfolio coverage and quantifies
the optimal improvement attainable under a small allowance
for additional risk inadequacy. The proof is given in
Appendix~\ref{app:transfers-coverage}.

\begin{theorem}[Coverage improvement at fixed protected-group total premium effects]
\label{thm:transfer-coverage}
Suppose $f_{\mathrm{cov}}$ is finite and convex on $\bbr^m$
and differentiable at $p_y$.
Define the negative $q$-gradient $a\in\bbr^m$ of the
coverage criterion at $p_y$ by
\[
Df_{\mathrm{cov}}(p_y)[v]
=
-\langle a,v\rangle_q,
\qquad v\in\bbr^m,
\]
and let $a_{\ker Q}$ be its $q$-orthogonal projection
onto the space $\ker Q$ of admissible premium adjustments.

For every admissible adjustment $v\in\ker Q$, the
first-order change in the coverage criterion is
\[
Df_{\mathrm{cov}}(p_y)[v]
=
-\langle a_{\ker Q},v\rangle_q.
\]
If $a_{\ker Q}\ne0$, then, for sufficiently small $t>0$,
the premium vector $p_y+t a_{\ker Q}$ improves portfolio
coverage, whereas $p_y-t a_{\ker Q}$ worsens it.
Both adjustments incur the same additional risk inadequacy.

More precisely, as $t\downarrow0$,
\begin{equation}
\begin{aligned}
f_{\mathrm{cov}}(p_y+t a_{\ker Q})
&=
f_{\mathrm{cov}}(p_y)
-t\|a_{\ker Q}\|_q^2+o(t),\\
f_{\mathrm{RA}}(p_y+t a_{\ker Q})
&=
f_{\mathrm{RA}}(p_y)
+t^2\|a_{\ker Q}\|_q^2.
\end{aligned}
\label{eq:transfer-coverage-improvement}
\end{equation}
Consequently, the minimum of the coverage criterion under an
allowance $\delta$ for additional risk inadequacy satisfies
\begin{equation}
K_y(\delta)
=
f_{\mathrm{cov}}(p_y)
-\sqrt{\delta}\,\|a_{\ker Q}\|_q
+o(\sqrt{\delta})
\qquad(\delta\downarrow0).
\label{eq:transfer-coverage-frontier}
\end{equation}

If $a_{\ker Q}=0$, convexity implies that $p_y$
already minimizes the coverage criterion over
$\mathcal F_y$, so
$K_y(\delta)=f_{\mathrm{cov}}(p_y)$ for every
$\delta\geq0$. Conversely, this global optimality
implies $a_{\ker Q}=0$.
The condition holds automatically when $\ker Q=\{0\}$,
so the protected-group target determines the
class-premium vector uniquely. When $\ker Q\ne\{0\}$,
it requires the negative $q$-gradient $a$ to be
orthogonal to every admissible premium adjustment.
\end{theorem}

For portfolio coverage measured by AVaR, the effect of premium
adjustments depends on the tariff-class composition of portfolio
outcomes, particularly those with large residual losses.
Let $N_c=\sum_{i=1}^n\mathbf1_{\{C(\Theta_i)=c\}}$
denote the number of policies in tariff class $c$.
For a class-premium vector $p$, the residual loss defined in
\eqref{eq:residual-loss} is
$Z^{\mathbf C}=S_n-\sum_{c=1}^mN_cp_c$.
A premium adjustment $v$ therefore changes residual loss by
$-\sum_{c=1}^mN_cv_c$ in each portfolio outcome.

We use the upper-tail convention for average value at risk
(AVaR), following \citet{RockafellarUryasev2000,RockafellarUryasev2002}:
for an integrable loss $Z$,
$
\operatorname{AVaR}^{\bbp}_\alpha(Z)
=
\inf_{t\in\bbr}
\left\{
t+\frac{1}{1-\alpha}\bbe_{\bbp}[(Z-t)_+]
\right\},
 0<\alpha<1.
$
Thus $\alpha$ denotes the loss-quantile level, while
$1-\alpha$ is the upper-tail probability. Any probability
mass at the quantile boundary is included only to the
extent required by this tail probability.

Suppose portfolio coverage is measured by
$f_{\mathrm{cov}}(p)=\operatorname{AVaR}^{\bbp}_\alpha
(Z^{\mathbf C})$.
At the baseline class-premium vector $p_y$, let
$t_\alpha$ denote an $\alpha$-quantile of the residual
loss $Z^{\mathbf C}$. If
$\bbp(Z^{\mathbf C}=t_\alpha)=0$, then
$f_{\mathrm{cov}}$ is differentiable at $p_y$,
and its negative $q$-gradient has components
\[
a_c=
\frac{\bbe_{\bbp}[N_c\mid Z^{\mathbf C}>t_\alpha]}{q_c},
\qquad c=1,\ldots,m.
\]
Thus the premium sensitivity of AVaR is determined by the
expected tariff-class counts among portfolio outcomes in
the upper loss tail, relative to the portfolio class
probabilities.

A sufficient condition for differentiability is that the
conditional claim distributions
$\kappa_{Y\mid\Theta}(\vartheta,\cdot)$ are atomless for
$\bbp_\Theta$-almost every type $\vartheta$.
Under the conditional-independence structure of
Subsection~\ref{subsec:structural-kernels}, the residual
portfolio loss then has an atomless distribution for every
fixed class-premium vector.
The finite-scenario AVaR representation and its premium
sensitivities are derived in
Appendix~\ref{app:transfers-coverage-AVaR};
Appendix~\ref{app:transfers-coverage-nonsmooth} treats
the nondifferentiable case.

\section{A motor-insurance model}
\label{sec:finite-model}

We illustrate the results of
Sections~\ref{sec:tariff-framework}--\ref{sec:balanced-transfers}
in a motor-insurance model with eight benchmark-price levels
and three protected groups. We fix the distribution of
policyholder characteristics and the benchmark risk-pricing rule,
and compare the benchmark-price classification with three
coarsenings that pool neighboring benchmark-price levels.
The conditional benchmark-price distributions of all three
protected groups have the same support, but assign different
probabilities to the individual levels. For the main analysis, the portfolio, regulatory, and reference
laws coincide, $\bbp=\bbp^{\mathrm{reg}}=\bbp^\star$.
We use conditional-mean aggregation and allow class-dependent
transfers subject to balance. All class premiums in the reported
numerical tariffs remain positive, although nonnegativity is
not imposed as an admissibility condition.

The case study follows the theoretical analysis step by step.
We first examine how classification affects inferential disparity,
risk inadequacy, and protected-group pooling effects.
We then study which protected-group total premium-effect targets
can be attained through balanced transfers, the associated
risk inadequacy costs, and the consequences of choosing different
protected-group criteria. Next, we specify a claim model and
examine to what extent portfolio coverage can be improved
while preserving the protected-group total premium effects.
Finally, we change the reference law $\bbp^\star$ while keeping
the portfolio law $\bbp$ and the classification fixed, to examine
the consequences for transfer balance and attainability.
The model serves to illustrate the mathematical relationships,
rather than to provide an empirical calibration.
Detailed calculations are collected in
Appendix~\ref{app:finite-model}.

\subsection{Risk structure and tariff classifications}
\label{subsec:finite-risk-structure}

A policyholder is characterized by $\Theta=(R,D)$, where
$R\in\{88,92,96,100,104,108,112,116\}$ is the benchmark
risk price and $D\in\{1,2,3\}$ is the protected characteristic,
with $D=d$ identifying protected group $d$.
The benchmark pricing rule is $\pi(r,d)=r$.
The eight benchmark-price levels may represent combinations of
legitimate risk characteristics, such as driving-risk and
mileage categories, which need not be modeled separately here.
The classification and transfer calculations below use the
joint distribution of $(R,D)$.
A conditional claim model consistent with this benchmark
pricing rule is introduced in
Subsection~\ref{subsec:finite-coverage-reference}
for the analysis of portfolio coverage.

\begin{table}[t]
\centering
\small
\caption{Conditional benchmark-price distributions of the protected groups.}
\label{tab:finite-population}
\begin{tabular}{c|rrrrrrrr|r}
\hline
 & 88 & 92 & 96 & 100 & 104 & 108 & 112 & 116
 & $\bbe_{\bbp}[R\mid D]$\\
\hline
$D=1$ & .20 & .20 & .16 & .13 & .11 & .09 & .07 & .04 & 98.00\\
$D=2$ & .08 & .12 & .17 & .20 & .18 & .13 & .08 & .04 & 100.76\\
$D=3$ & .04 & .06 & .09 & .12 & .15 & .18 & .19 & .17 & 105.72\\
\hline
\end{tabular}
\end{table}

The protected groups have equal probabilities $1/3$. Their conditional
distributions over the benchmark-price levels are given in
Table~\ref{tab:finite-population}.
Every protected group occurs throughout the risk spectrum, but
group~1 is more concentrated among lower benchmark risk prices, group~2
around the middle, and group~3 among higher benchmark risk prices. Thus, although the benchmark pricing rule does not depend on $D$,
the conditional distribution of $R$ varies across protected groups.
Consequently, a classification that retains information about
benchmark risk prices can also contain information about $D$.

For the main part of the case study, let
$\bbp=\bbp^{\mathrm{reg}}=\bbp^\star$. We use squared Euclidean dissimilarity for the inferential
disparity $\Gamma(C)$, as in \eqref{eq:MR1-Gamma-variance}. For each classification,
we use conditional-mean aggregation,
$A^{\mathbf C}(c)=\bbe_{\bbp}[R\mid C(\Theta)=c]=\bar r_c$,
and allow class-dependent $\bbp$-balanced transfers. Hence
$\bbe_{\bbp}[W^{\mathbf C}]=0$, so the quantity $b$ from
Section~\ref{subsec:transfer-outcomes-cost} is zero.

We compare four classifications of decreasing granularity. The
benchmark classification $C^8=C^\pi$ retains all eight benchmark-price
levels. The classification $C^4$ has classes
$\{88,92\}$, $\{96,100\}$, $\{104,108\}$, and $\{112,116\}$;
$C^3$ has classes $\{88,92,96\}$, $\{100,104\}$, and
$\{108,112,116\}$; and $C^2$ has classes
$\{88,92,96,100\}$ and $\{104,108,112,116\}$. Thus only neighboring
benchmark-price levels are pooled. The maximum within-class
benchmark-price spreads are, respectively, $0$, $4$, $8$, and $12$.
The three coarser classifications are all coarsenings of $C^8$,
but they do not form a nested sequence: $C^2$ is a coarsening
of $C^4$, whereas $C^3$ is not nested with either.

\subsection{Classification, inferential criterion, and pooling}
\label{subsec:finite-classification}

Table~\ref{tab:finite-classification} summarizes the consequences
of the four classifications for the two objectives studied in
Section~\ref{sec:main-result-I} and for the protected-group
pooling effects introduced in Section~\ref{sec:tariff-framework}.
Standard deviation in the last column is taken under the reference law,
under which the three protected groups are equiprobable.

\begin{table}[t]
\centering
\small
\caption{Classification, risk inadequacy, inferential criterion,
and protected-group pooling effects.}
\label{tab:finite-classification}
\begin{tabular}{Sc|Sr Sr|Sr Sr Sr|Sr}
\hline
$C$ & $L(C)$ & $\Gamma(C)$
& \multicolumn{3}{Sc|}{\shortstack{Protected-group pooling effects\\
$M_W^{\mathbf C,\star}$}}
& \shortstack{Standard\\deviation of\\$M_W^{\mathbf C,\star}$}\\
& & & $d=1$ & $d=2$ & $d=3$ &\\
\hline
$C^8$ & 0.000  & .116 & 0.000 & 0.000 &  0.000 & .000\\
$C^4$ & 3.971  & .107 & 0.196 & 0.034 & -0.230 & .175\\
$C^3$ & 8.431  & .094 & 0.473 & 0.012 & -0.485 & .392\\
$C^2$ & 19.046 & .067 & 1.065 & 0.053 & -1.118 & .892\\
\hline
\end{tabular}
\end{table}

The benchmark classification $C^8=C^\pi$ is perfectly risk
adequate, so $L(C^8)=0$. For this classification, the inferential criterion has value
$\varepsilon_0=\Gamma(C^8)\approx0.116$, which is therefore the
inferential threshold from Theorem~\ref{thm:inferential-threshold}.
Each of the strict coarsenings $C^4$, $C^3$, and $C^2$ has
$\Gamma(C)<\varepsilon_0$ and positive $L(C)$, illustrating the
risk inadequacy cost of moving below the threshold characterized in
Section~\ref{subsec:finite-class-approximation}.

For the coarser classifications, conditional-mean aggregation
assigns the same tariff-class premium to policyholders with
different benchmark risk prices. Since the protected groups
have different distributions over the benchmark-price levels,
pooling produces different average premium deviations across
protected groups.

Table~\ref{tab:finite-classification} shows that, across the
four classifications selected for the case study, inferential
disparity $\Gamma(C)$ decreases while the standard deviation
of protected-group pooling effects increases.
The opposing numerical trends are not a general monotonicity
result: $C^4,C^3,C^2$ do not form a nested sequence, and
the coarsening lemma constrains $\Gamma(C)$, not the dispersion
of protected-group pooling effects.
The comparison shows that lower inferential disparity can be
accompanied by greater differences in the economic effects
of pooling across protected groups.

The classifications are deliberately constructed from consecutive
benchmark-price levels. Hence the increase in $L(C)$ is generated by
pooling policyholders with nearby, rather than radically different,
benchmark risk prices. Proposition~\ref{prop:pooling-effects} describes
how each such merge changes risk inadequacy and the inferential
criterion through, respectively, the benchmark-price differences and
the protected-group profiles of the pooled classes.

\subsection{Balanced transfers: attainability and cost}
\label{subsec:finite-transfers}

We next ask to what extent balanced transfers can change the
protected-group total premium effects created by pooling. Since conditional-mean aggregation
is used and $\bbp^\star=\bbp$, balance implies that the mean of
$M_\Delta^{\mathbf C,\star}$ is zero. Exact compensation therefore
coincides here with group neutrality.

For each classification, the matrix $Q$ from
Section~\ref{subsec:transfer-outcomes-cost} records how the three
protected groups are distributed across the tariff classes.
Theorem~\ref{thm:transfer-target-cost} determines which protected-group
total premium-effect vectors are attainable and the minimum
risk inadequacy required to attain each of them.
Table~\ref{tab:finite-transfers} summarizes attainability,
minimum risk inadequacy, and minimum residual protected-group
dispersion for the four classifications.

\begin{table}[t]
\centering
\scriptsize
\setlength{\tabcolsep}{3.5pt}
\caption{Attainability and cost of exact compensation, and minimum
residual protected-group dispersion under balanced transfers.
Values are rounded to three decimals; dashes indicate
unattainable exact compensation.}
\label{tab:finite-transfers}
\begin{tabular}{Sc|Sc|Sr|Sr|Sr|Sr}
\hline
$C$
& \shortstack{Exact\\compensation\\attainable}
& \shortstack{Additional\\risk inadequacy\\$c_C(0)$}
& \shortstack{Minimum\\$f_{\mathrm{RA}}$\\at exact\\compensation}
& \shortstack{Minimum\\standard deviation\\of $M_\Delta^{\mathbf C,\star}$}
& \shortstack{Minimum\\$f_{\mathrm{RA}}$\\among standard\\deviation minimizers}\\
\hline
$C^8$ & yes & .000 & .000 & .000 & .000\\
$C^4$ & yes & .230 & 4.201 & .000 & 4.201\\
$C^3$ & yes & 1.465 & 9.896 & .000 & 9.896\\
$C^2$ & no  & ---  & ---   & .149 & 26.718\\
\hline
\end{tabular}
\end{table}

Under $C^4$, exact compensation is attainable and requires only a
small increase in risk inadequacy beyond $L(C^4)=3.971$:
the minimum additional cost is $c_{C^4}(0)=0.230$. Exact compensation
remains attainable under $C^3$, but its minimum additional cost
increases to $c_{C^3}(0)=1.465$. Thus attainability and cost are
separate questions. Two classifications may permit the same
protected-group target while requiring substantially different
departures from the portfolio conditional mean benchmark risk prices.

Under $C^2$, exact compensation would require two tariff-class
premiums $p_1,p_2$ satisfying the three protected-group
equations $Qp=\mu^\star$, as characterized by
Theorem~\ref{thm:transfer-target-cost}.
For the population in Table~\ref{tab:finite-population},
the first two equations imply $p_1-p_2=-23$, whereas
the second and third imply $p_1-p_2\approx-19.077$.
Thus the equations are inconsistent, and
$\mathcal F_0=\varnothing$: exact compensation is unattainable. The balanced transfer
minimizing residual standard deviation reduces it from $.892$ before
transfers to $.149$, but the corresponding risk inadequacy
is $26.718$. At this minimum attainable dispersion,
Proposition~\ref{prop:transfer-inferential-bound} gives a
lower bound of approximately $24.528$ on risk inadequacy.
The bound is therefore not attained in this example.

The comparison distinguishes the initial protected-group pooling
effects from the changes subsequently attainable through balanced
transfers. The attainable targets depend on the conditional
tariff-class distribution matrix $Q$, not on the scalar inferential
criterion $\Gamma(C)$ alone. Their minimum risk-inadequacy costs
also depend on the portfolio class probabilities $q$ and the
tariff-class conditional mean benchmark risk prices $\bar r$,
as characterized by Theorem~\ref{thm:transfer-target-cost}.

\subsection{Different protected-group criteria}
\label{subsec:finite-group-criteria}

When exact compensation is unavailable, the choice of protected-group
criterion becomes consequential. Let the deviation measure
$\mathfrak D^\star$ be standard deviation and let the monetary risk
measure $\mathfrak R^\star$ be $\operatorname{AVaR}_{2/3}$ applied
to the protected-group total premium effect. Under the upper-tail convention of
Subsection~\ref{subsec:transfer-coverage},
$\alpha=2/3$ selects the largest one-third of
protected-group effects. Since the three protected groups are equiprobable,
$\operatorname{AVaR}_{2/3}$ is a worst-group criterion:
it equals the largest of the three protected-group
total premium effects.
\begin{table}[htbp]
\centering
\scriptsize
\setlength{\tabcolsep}{5pt}
\caption{Balanced transfers under $C^2$: comparison of the
protected-group standard-deviation and adverse-effect criteria.
Values are rounded to three decimals.}
\label{tab:finite-group-criteria}
\begin{tabular}{Sl|Sc|Sr|Sr|Sr}
\hline
\shortstack{Criterion\\minimized}
& \shortstack{Protected-group total\\premium effects\\$M_\Delta^{\mathbf C,\star}$}
& \shortstack{Standard\\deviation of\\$M_\Delta^{\mathbf C,\star}$}
& \shortstack{$\operatorname{AVaR}_{2/3}$ of\\$M_\Delta^{\mathbf C,\star}$}
& \shortstack{Risk inadequacy\\$f_{\mathrm{RA}}$}\\
\hline
Standard deviation
& $(.141,-.205,.065)$
& .149 & .141 & 26.718\\
$\operatorname{AVaR}_{2/3}$
& $(.107,-.215,.107)$
& .152 & .107 & 27.279\\
\hline
\end{tabular}
\end{table}

Table~\ref{tab:finite-group-criteria} compares the two
criterion-optimal balanced transfers under $C^2$. Minimizing standard
deviation produces the effect vector $(.141,-.205,.065)$, whereas
minimizing $\operatorname{AVaR}_{2/3}$ produces
$(.107,-.215,.107)$. The latter accepts slightly greater dispersion,
increasing the standard deviation from $.149$ to $.152$, in order to
reduce the largest adverse group effect from $.141$ to $.107$. Its
risk inadequacy is correspondingly slightly larger.

Thus the two protected-group criteria from Section~\ref{sec:tariff-framework}
answer different questions even within the same classification and the same
attainable set. A deviation criterion evaluates differences among
protected-group total premium effects, whereas a monetary risk measure
evaluates their adverse level. Once exact compensation is unavailable, the two criteria
need not select the same balanced transfer.

\subsection{Portfolio coverage and the reference population}
\label{subsec:finite-coverage-reference}

We now consider two further aspects of the model.
For the portfolio coverage analysis, we retain the benchmark
risk prices and the joint portfolio distribution of $(R,D)$
specified in Subsection~\ref{subsec:finite-risk-structure},
and introduce a conditional claim distribution.
We examine how portfolio coverage can be improved while
preserving protected-group total premium effects.
We then keep the portfolio law, classification $C^4$,
and aggregation fixed, but change the reference law by
assigning different probabilities to the protected groups,
while retaining their conditional benchmark-price distributions.

\paragraph{Coverage at fixed protected-group total premium effects.}
We illustrate the coverage--risk-inadequacy trade-off from
Section~\ref{subsec:transfer-coverage}. Keep the benchmark risk
prices fixed and use expectation pricing. Conditional on
$\Theta=(R,D)$, let the claim $Y$ equal $0$ or $2R$ with
equal probability, so
$\bbe_{\bbp}[Y\mid\Theta]=R=\pi(\Theta)$.
For a one-policyholder portfolio, evaluate coverage by
$f_{\mathrm{cov}}(p)=\operatorname{AVaR}^{\bbp}_{0.95}
(Y-p_{C(\Theta)})$, corresponding to the largest
$5\%$ of residual losses.

Under $C^4$, let $p_0$ be the class-premium vector
minimizing risk inadequacy subject to exact compensation.
It satisfies $f_{\mathrm{RA}}(p_0)=4.201$ and
$f_{\mathrm{cov}}(p_0)=116.322$.
The matrix $Q$ has rank three and four columns, so
$\ker Q$ is one-dimensional. Adjustments within this kernel
preserve exact compensation and transfer balance.
We choose an additional risk-inadequacy allowance
$\delta=1$, permitting the total risk inadequacy to increase
by at most one unit above its minimum at $p_0$.
The resulting optimization
\eqref{eq:transfer-conditional-coverage} gives
$K_0(1)=115.955$, attained with risk inadequacy $5.201$.
Thus the specified allowance permits a reduction of approximately
$.367$ in the coverage criterion while preserving exact
compensation.

Under $C^3$, by contrast, $Q$ is a nonsingular
$3\times3$ matrix and $\ker Q=\{0\}$.
Exact compensation therefore determines the class-premium
vector uniquely, leaving no admissible adjustment to improve
portfolio coverage. The comparison illustrates how the
remaining freedom in tariff-class premiums determines the
coverage--risk-inadequacy trade-off studied in
Section~\ref{subsec:transfer-coverage}.

\paragraph{Changing the reference population.}
We next examine how the reference law affects transfer balance and
attainability. Keep the portfolio law $\bbp$, classification $C^4$,
and conditional-mean aggregation fixed. Change only the reference law
$\bbp^\star$: retain the conditional distribution of benchmark risk
prices within each protected group, but replace the equal
protected-group probabilities by $\rho^\star=(.5,.3,.2)^\top$.
Consequently, the conditional tariff-class distribution matrix $Q$
and the protected-group conditional mean benchmark risk prices
$\mu^\star$ remain unchanged, while the reference tariff-class
probabilities $q^\star=Q^\top\rho^\star$ change.

Under the new reference law, the mean pooling effect is
$b=\bbe_{\bbp^\star}[W^{\mathbf C}]\approx.062$.
Every $\bbp^\star$-balanced transfer preserves this mean, so
$(\rho^\star)^\top M_\Delta^{\mathbf C,\star}=b$.
Exact compensation is therefore unattainable, since it requires
zero protected-group total premium effects. Under $C^4$, however,
$Q$ has full row rank, so group neutrality remains attainable:
all protected groups can receive the same total premium effect
$b\approx.062$. Thus the choice of reference population changes
which protected-group targets satisfy transfer balance even when
the portfolio and classification are unchanged.

\section{Conclusion and outlook}
\label{sec:conclusion}

The analysis suggests two main conclusions. First, questions of risk adequacy
and anti-discrimination in insurance pricing cannot generally be reduced to a
single criterion. Classification, aggregation, and transfers affect different
properties of a tariff, while portfolio coverage, risk adequacy,
the protected-group composition of tariff classes, and protected-group premium
effects describe different consequences of those choices. Keeping these
objects separate makes it possible to identify where a trade-off arises rather
than attributing all tensions to an undifferentiated notion of fairness.

Second, the effects of classification and subsequent premium adjustment are
closely connected but remain distinct. Making tariff classes less informative
about protected-group membership can require pooling policyholders with
different benchmark risk prices and thereby reduce risk adequacy.
Once a classification and aggregation component have been chosen, balanced
transfers can modify the resulting protected-group premium effects, but the
available corrections and their costs depend on how protected groups are
distributed across tariff classes. A scalar inferential criterion therefore
does not by itself determine protected-group premium effects or the cost of
changing them. These consequences depend on the architecture through which
policyholders are grouped and premiums are adjusted.

Several questions remain for future research. A first extension
concerns the information available for tariff construction.
The present framework defines the classification $C$ as a
function of policyholder type $\Theta$, which need not be
observable. An implementable classification must instead
assign tariff classes through a rule based on available
and admissible observable information. Since observables $X$ may be random
conditional on $\Theta$, this requires extending the type-level
framework. An important question is how such information
restrictions affect the attainable combinations of risk adequacy,
portfolio coverage, and protected-group criteria.

A second direction concerns the algorithmic solution of the
general multi-objective tariff-design problem formulated in
Subsection~\ref{subsec:pareto-formulation}. While
Proposition~\ref{prop:classification-DP} provides a
dynamic-programming solution for a restricted classification
problem, the joint optimization of classification, aggregation,
and balanced transfers remains to be addressed. This calls for
algorithms that compute attainable objective sets and Pareto
frontiers under operational restrictions.

Statistical tariff construction introduces a further problem.
When the relevant probability laws and benchmark risk prices
are unknown, they must be estimated from data, and admissible
tariffs must be learned. This raises questions about statistical
consistency, estimation error, and the convergence of empirical
solutions to their population counterparts.

Further questions arise from uncertainty about the underlying
risk model and population composition. Benchmark risk prices,
portfolio distributions, and regulatory reference populations
may be subject to estimation and model uncertainty.
Robust tariff design would examine how such uncertainty affects
attainable protected-group targets, risk-inadequacy costs,
and Pareto-efficient tariffs, and how tariff decisions can
account for these uncertainties.

Finally, tariff decisions may influence future policyholder
behavior and portfolio composition. A dynamic extension would
allow classifications, premiums, and population laws to evolve,
with current decisions affecting subsequent risks and pricing
opportunities. The resulting problem would concern the evolution
of attainable trade-offs under repeated tariff decisions.

\bibliographystyle{plainnat}
\bibliography{references}

\subsection*{Declaration of generative AI in the manuscript preparation process}

During the preparation of this work, the authors used large language models (multiple model versions of ChatGPT) interactively to assist with literature searches, the discussion and articulation of author-developed ideas, working through local details, improvements to wording and exposition, the identification of possible errors or omissions for subsequent review, and the preparation of LaTeX and code used for plotting and formatting. All AI-assisted output was critically reviewed and edited by the authors, and all mathematical statements, proofs, references, numerical results, tables, and figures were independently verified. The authors retained full control over the scientific content, interpretations, and conclusions and take full responsibility for the content of the manuscript.

\subsection*{Funding information}

Wiebke Hansen's position at Leibniz Universit\"at Hannover was fully financed through the dual PhD programme of the House of Insurance, which is supported by the insurance industry. This financing entailed no obligations regarding the research, and the supporting institutions had no role in the design or conduct of the research, its analysis or interpretation, or the preparation of this manuscript.

\subsection*{Competing interests}

The authors declare that they have no competing interests related to the preparation or publication of this article.

\newpage

\appendix

\section{Proofs and supplementary results for classification}
\label{app:main-result-I}

This appendix provides the proofs and supplementary results for the
classification analysis. It establishes the inferential threshold and the
positive risk inadequacy cost of imposing a bound below this threshold. It also
derives the pooling results, analyzes optimization over the specified family
of admissible classifications, and verifies the implementation condition for
balanced transfers.

Throughout, $C$ denotes a finite measurable classification,
$R=\pi(\Theta)\in L^2(\bbp)$, and
$\bbp_\Theta^{\mathrm{reg}}\ll\bbp_\Theta$.
Conditional expectations and conditional distributions are taken under
the law explicitly indicated. Values on classes of zero probability
under that law may be chosen arbitrarily.

\subsection{Proof of Lemma~\ref{lem:inferential-coarsening}}
\label{app:proof-inferential-coarsening}

For $u\in\{1,\ldots,\widetilde m\}$, write
$\widetilde s_u=\bbp^{\mathrm{reg}}(\widetilde C(\Theta)=u)$ and, whenever
$\widetilde s_u>0$, let
$\widetilde\eta_u
=\bbp^{\mathrm{reg}}(D\in\cdot\mid\widetilde C(\Theta)=u)$.
Since $\widetilde C(\Theta)=\varphi(C(\Theta))$ $\bbp^{\mathrm{reg}}$-a.s.,
we have
$\widetilde s_u=\sum_{\{c:\varphi(c)=u\}}s_c$ and
$\widetilde\eta_u
=\sum_{\{c:\varphi(c)=u\}}(s_c/\widetilde s_u)\eta_c$.

For $u,v$ with $\widetilde s_u,\widetilde s_v>0$, joint convexity of
$\operatorname{dist}_{\dcal}$ therefore gives
\[
\widetilde s_u\widetilde s_v\,
\operatorname{dist}_{\dcal}(\widetilde\eta_u,\widetilde\eta_v)
=
\widetilde s_u\widetilde s_v\,
\operatorname{dist}_{\dcal}\!\left(
\sum_{\varphi(c)=u}\frac{s_c}{\widetilde s_u}\eta_c,
\sum_{\varphi(c')=v}\frac{s_{c'}}{\widetilde s_v}\eta_{c'}
\right)
\leq
\sum_{\substack{c:\varphi(c)=u\\c':\varphi(c')=v}}
s_cs_{c'}\,
\operatorname{dist}_{\dcal}(\eta_c,\eta_{c'}).
\]
Summing over $u,v$ with $\widetilde s_u,\widetilde s_v>0$ yields
$\Gamma(\widetilde C)\leq\Gamma(C)$.


\subsection{Proof of Theorem~\ref{thm:inferential-threshold}}
\label{app:proof-inferential-threshold}


Let $C\in\mathfrak C_{\mathrm{cl}}$ satisfy $L(C)=0$. Then $R$ is
$\sigma(C(\Theta))$-measurable $\bbp$-a.s. By the Doob--Dynkin lemma, there
is a function $h$ such that $R=h(C(\Theta))$ $\bbp$-a.s. For every class $c$ of positive portfolio probability,
$h(c)=r_j$ for a unique benchmark-price level $r_j$.
Define a map $\varphi$ between the class-label sets of $C$ and $C^\pi$
by setting $\varphi(c)=j$ on each such class and assigning arbitrary
labels on the remaining classes. By the definition of $C^\pi$,
$C^\pi(\Theta)=\varphi(C(\Theta))$ $\bbp$-a.s. Because both sides are
measurable functions of $\Theta$ and
$\bbp_\Theta^{\mathrm{reg}}\ll\bbp_\Theta$, this identity also holds
$\bbp^{\mathrm{reg}}$-a.s. Lemma~\ref{lem:inferential-coarsening} therefore
gives $\Gamma(C^\pi)\leq\Gamma(C)$.

Conversely, assigning to each class of $C^\pi$ its corresponding benchmark
risk price gives $L(C^\pi)=0$. Since
$C^\pi\in\mathfrak C_{\mathrm{cl}}$, it follows that
$
\min\{\Gamma(C):C\in\mathfrak C_{\mathrm{cl}},\ L(C)=0\}
=
\Gamma(C^\pi)
=
\varepsilon_0.
$
Hence, within the classification problem, zero risk inadequacy
under the constraint $\Gamma(C)\leq\varepsilon$ is possible if and only if
$\varepsilon\geq\varepsilon_0$.

\subsection{Proof of Corollary~\ref{cor:exact-inferential-neutrality}}
\label{app:proof-exact-inferential-neutrality}

Since $\varepsilon_0=\Gamma(C^\pi)$ and
$\operatorname{dist}_{\dcal}$ is nonnegative and vanishes only for identical
distributions, $\varepsilon_0=0$ if and only if
$\bbp^{\mathrm{reg}}(D\in\cdot\mid R=r_j)
=\bbp^{\mathrm{reg}}(D\in\cdot\mid R=r_\ell)$
for all benchmark-price levels $r_j,r_\ell$ of positive regulatory
probability. If these conditional distributions are identical across all benchmark-price
levels of positive regulatory probability, then their probability-weighted
mixture, which equals the marginal distribution of $D$, is identical to each
of them. Hence $\varepsilon_0=0$ if and only if
$\bbp^{\mathrm{reg}}(D\in\cdot\mid R=r_j)
=\bbp^{\mathrm{reg}}(D\in\cdot)$
for every benchmark-price level $r_j$ of positive regulatory probability,
which is equivalent to independence of $R$ and $D$ under
$\bbp^{\mathrm{reg}}$.

\subsection{Proof of the variance representation of the inferential criterion}
\label{app:proof-Gamma-variance}

For this subsection and the remainder of the appendix, recall the
assumptions $D\in\{1,\ldots,k\}$ and
$\operatorname{dist}_{\dcal}(\mu,\nu)=\|\mu-\nu\|_2^2$
from Section~\ref{sec:main-result-I}. Distributions of $D$ are identified
with probability vectors in $\bbr^k$.

We first record two weighted-variance identities used below.
They hold in every Euclidean dimension $d\geq1$, including the scalar
case $d=1$. Let $w_1,\ldots,w_\ell\geq0$ satisfy
$\sum_jw_j=1$, let $x_1,\ldots,x_\ell\in\bbr^d$, and write
$\bar x=\sum_jw_jx_j$. For every $a\in\bbr^d$,
\begin{align}
\sum_jw_j\|x_j-a\|_2^2
&=
\sum_jw_j\|x_j-\bar x\|_2^2+\|\bar x-a\|_2^2,
\label{eq:app-weighted-variance-decomposition}\\
\frac12\sum_{i,j}w_iw_j\|x_i-x_j\|_2^2
&=
\sum_jw_j\|x_j-\bar x\|_2^2.
\label{eq:app-weighted-pairwise-variance}
\end{align}
The first identity follows by expanding
$x_j-a=(x_j-\bar x)+(\bar x-a)$ and using
$\sum_jw_j(x_j-\bar x)=0$.
Expanding the pairwise squared distances gives
$\frac12\sum_{i,j}w_iw_j\|x_i-x_j\|_2^2
=\sum_jw_j\|x_j\|_2^2-\|\bar x\|_2^2$.
The first identity with $a=0$ identifies this expression with
$\sum_jw_j\|x_j-\bar x\|_2^2$, proving the second identity.

For the inferential criterion, take $w_c=s_c$ and $x_c=\eta_c$,
restricting to classes with $s_c>0$.
Their weighted mean is $\bar x = \eta=\sum_{c:s_c>0}s_c\eta_c$.
Equation~\eqref{eq:app-weighted-pairwise-variance} therefore proves \eqref{eq:MR1-Gamma-variance} by $\Gamma(C)= \sum_{\substack{c,c'\\s_c,s_{c'}>0}} s_c s_{c'}\|\eta_c - \eta_{c'}\|^2_2 = 2\sum_{c:s_c>0}s_c\|\eta_c-\eta\|_2^2$.


\subsection{Excess value of the inferential criterion among perfectly risk-adequate classifications}
\label{app:proof-inferential-excess}

Recall that $R$ has finite essential range $\{r_1,\ldots,r_K\}$
under $\bbp$, and that $C^\pi$ is the corresponding benchmark-price
classification. Let $C$ satisfy $L(C)=0$. As established in the proof of
Theorem~\ref{thm:inferential-threshold}, $C^\pi$ is a coarsening of $C$
under $\bbp^{\mathrm{reg}}$. Thus, for each benchmark-price level $r$
with $\bbp^{\mathrm{reg}}(R=r)>0$,
$\widetilde\eta_r
=\sum_c\bbp^{\mathrm{reg}}(C(\Theta)=c\mid R=r)\eta_c$.

Apply \eqref{eq:app-weighted-variance-decomposition} to the profiles
$x_c=\eta_c$ with weights
$w_c = \bbp^{\mathrm{reg}}(C(\Theta)=c\mid R=r)$, weighted mean
$\bar x = \widetilde\eta_r$, and $a=\eta$.
Averaging over $r$ gives
$\bbe_{\bbp^{\mathrm{reg}}}[\|\eta_{C(\Theta)}-\eta\|_2^2]
=\bbe_{\bbp^{\mathrm{reg}}}
[\|\eta_{C(\Theta)}-\widetilde\eta_R\|_2^2]
+\bbe_{\bbp^{\mathrm{reg}}}[\|\widetilde\eta_R-\eta\|_2^2]$.
By \eqref{eq:MR1-Gamma-variance}, the left-hand side is
$\Gamma(C)/2$, and the last term is
$\Gamma(C^\pi)/2=\varepsilon_0/2$.
Rearranging proves \eqref{eq:MR1-inferential-excess}.
Its right-hand side vanishes if and only if
$\eta_{C(\Theta)}=\widetilde\eta_R$
$\bbp^{\mathrm{reg}}$-almost surely, which gives the equality
characterization.

\subsection{Proof of Theorem~\ref{thm:inferential-cost}}
\label{app:proof-inferential-cost}

We use the weighted-variance identities
\eqref{eq:app-weighted-variance-decomposition}
and \eqref{eq:app-weighted-pairwise-variance} for protected-group
probability vectors and benchmark risk prices. Write the distinct essential values of $R$ as
$r_1,\ldots,r_K$ and set
$\delta_R=\min_{i\ne j}|r_i-r_j|>0$. Fix a finite classification $C$.

\noindent\textnormal{(i)} We first derive an auxiliary bound from the
classification obtained by jointly recording $C$ and $C^\pi$. Define
$J:\widetilde\Theta\to\{1,\ldots,m\}\times\{1,\ldots,K\}$ by
$J(\vartheta)=(C(\vartheta),C^\pi(\vartheta))$. Thus both $C$ and $C^\pi$
are coarsenings of $J$.
For every pair $(c,j)$ with
$\bbp^{\mathrm{reg}}(C(\Theta)=c,C^\pi(\Theta)=j)>0$, let
\begin{align*}
    \nu_{c,j}
=
\bbp^{\mathrm{reg}}
(D\in\cdot\mid C(\Theta)=c,C^\pi(\Theta)=j).
\end{align*}
For every $c$ with $s_c>0$, define
\begin{align*}
q_j(c)
=
\bbp^{\mathrm{reg}}
(C^\pi(\Theta)=j\mid C(\Theta)=c)
=
\bbp^{\mathrm{reg}}
(R=r_j\mid C(\Theta)=c),
\end{align*}
In sums involving $\nu_{c,j}$, only indices with
$s_c>0$ and $q_j(c)>0$ are included. In particular,
$\eta_c
= \bbp^{\mathrm{reg}}
(D \in \cdot \mid C(\Theta)=c)
=
\sum_j
q_j(c) \nu_{c,j}$. Recall that
$\eta=\bbp^{\mathrm{reg}}(D\in\cdot)$
is the unconditional protected-group probability vector.
Applying \eqref{eq:app-weighted-variance-decomposition} conditionally on
$C(\Theta)=c$, with $x_j=\nu_{c,j}$, weights $w_j=q_j(c)$,
weighted mean $\eta_c$, and $a=\eta$, gives $\bbe_{\bbp^{\mathrm{reg}}}
[\|\nu_{C(\Theta),C^\pi(\Theta)}-\eta\|_2^2\mid C(\Theta)=c]
=
\bbe_{\bbp^{\mathrm{reg}}}
[\|\nu_{C(\Theta),C^\pi(\Theta)}-\eta_c\|_2^2\mid C(\Theta)=c]
+\|\eta_c-\eta\|_2^2$. 
Taking expectations in the preceding identity gives
$\bbe_{\bbp^{\mathrm{reg}}}
[\|\nu_{C(\Theta),C^\pi(\Theta)}-\eta\|_2^2]
=
\bbe_{\bbp^{\mathrm{reg}}}
[\|\nu_{C(\Theta),C^\pi(\Theta)}-\eta_{C(\Theta)}\|_2^2]
+
\bbe_{\bbp^{\mathrm{reg}}}
[\|\eta_{C(\Theta)}-\eta\|_2^2]$.
By \eqref{eq:MR1-Gamma-variance}, twice the left-hand side equals
$\Gamma(J)$, while twice the last term equals $\Gamma(C)$. Hence
$\Gamma(J)-\Gamma(C)
=
2\bbe_{\bbp^{\mathrm{reg}}}
[\|\nu_{C(\Theta),C^\pi(\Theta)}-\eta_{C(\Theta)}\|_2^2]$.
Since $C^\pi$ is a coarsening of $J$,
Lemma~\ref{lem:inferential-coarsening} gives
$\Gamma(J)\geq\Gamma(C^\pi)=\varepsilon_0$. Therefore
\begin{equation}
\bigl(\varepsilon_0-\Gamma(C)\bigr)_+
\leq
2\bbe_{\bbp^{\mathrm{reg}}}
\!\left[
\|\nu_{C(\Theta),C^\pi(\Theta)}-\eta_{C(\Theta)}\|_2^2
\right].
\label{eq:app-MR1-refinement-bound}
\end{equation}

\noindent\textnormal{(ii)} We next prove
\begin{equation}
\bigl(\varepsilon_0-\Gamma(C)\bigr)_+
\leq
\frac{4}{\delta_R^2}
\bbe_{\bbp^{\mathrm{reg}}}
\!\left[
\operatorname{Var}_{\bbp^{\mathrm{reg}}}(R\mid C(\Theta))
\right].
\label{eq:app-MR1-regulatory-cost}
\end{equation}
Since $\eta_c=\sum_jq_j(c)\nu_{c,j}$ and, conditional on
$C(\Theta)=c$, the vector
$\nu_{C(\Theta),C^\pi(\Theta)}$ takes the value $\nu_{c,j}$
with probability $q_j(c)$, \eqref{eq:app-weighted-pairwise-variance} gives
$
\bbe_{\bbp^{\mathrm{reg}}}
[\|\nu_{C(\Theta),C^\pi(\Theta)}-\eta_c\|_2^2\mid C(\Theta)=c]
=
\frac12\sum_{i,j}q_i(c)q_j(c)\|\nu_{c,i}-\nu_{c,j}\|_2^2.
$
Since $\nu_{c,i}$ and $\nu_{c,j}$ are probability vectors,
$\|\nu_{c,i}-\nu_{c,j}\|_2^2\leq2$, while the terms with $i=j$ vanish.
Hence
$
\frac12\sum_{i,j}q_i(c)q_j(c)\|\nu_{c,i}-\nu_{c,j}\|_2^2
\leq
\sum_{i\ne j}q_i(c)q_j(c)$.
Conditional on $C(\Theta)=c$, the benchmark risk price $R$ takes the value $r_j$
with probability $q_j(c)$. Applying \eqref{eq:app-weighted-pairwise-variance} to
the values $r_1,\ldots,r_K$ with weights $q_1(c),\ldots,q_K(c)$ therefore gives
$\operatorname{Var}_{\bbp^{\mathrm{reg}}}(R\mid C(\Theta)=c)
=
\frac12\sum_{i,j}q_i(c)q_j(c)(r_i-r_j)^2$.
Since $(r_i-r_j)^2\geq\delta_R^2$ whenever $i\ne j$, while the terms with
$i=j$ vanish, it follows that
$\operatorname{Var}_{\bbp^{\mathrm{reg}}}(R\mid C(\Theta)=c)
\geq
\frac{\delta_R^2}{2}\sum_{i\ne j}q_i(c)q_j(c)
$. Therefore
$\bbe_{\bbp^{\mathrm{reg}}}
[\|\nu_{C(\Theta),C^\pi(\Theta)}-\eta_c\|_2^2\mid C(\Theta)=c]
\leq
\frac{2}{\delta_R^2}
\operatorname{Var}_{\bbp^{\mathrm{reg}}}(R\mid C(\Theta)=c)$.
Taking expectations and using \eqref{eq:app-MR1-refinement-bound} proves
\eqref{eq:app-MR1-regulatory-cost}.

\noindent\textnormal{(iii)} Recall that
$\bar r_{C(\Theta)}=\bbe_{\bbp}[R\mid C(\Theta)]$ is the
portfolio conditional-mean predictor. Write
$\zeta=d\bbp_\Theta^{\mathrm{reg}}/d\bbp_\Theta$.
We next prove the auxiliary estimate
\begin{equation}
\bbe_{\bbp^{\mathrm{reg}}}
[\operatorname{Var}_{\bbp^{\mathrm{reg}}}(R\mid C(\Theta))]
\leq
\bbe_{\bbp}[\zeta(\Theta)(R-\bar r_{C(\Theta)})^2].
\label{eq:app-MR1-law-transfer}
\end{equation}
Since the conditional mean under $\bbp^{\mathrm{reg}}$ minimizes squared
error among $C(\Theta)$-measurable predictors,
$\bbe_{\bbp^{\mathrm{reg}}}
[\operatorname{Var}_{\bbp^{\mathrm{reg}}}(R\mid C(\Theta))]
\leq
\bbe_{\bbp^{\mathrm{reg}}}[(R-\bar r_{C(\Theta)})^2]$.
Moreover, $(R-\bar r_{C(\Theta)})^2$ is a measurable function of
$\Theta$, so the definition of $\zeta$ gives
$\bbe_{\bbp^{\mathrm{reg}}}[(R-\bar r_{C(\Theta)})^2]
=\bbe_{\bbp}[\zeta(\Theta)(R-\bar r_{C(\Theta)})^2]$.
Together, these two relations prove \eqref{eq:app-MR1-law-transfer}.

\noindent\textnormal{(iv)} We now prove the first assertion of the theorem:
for every $0\leq\varepsilon<\varepsilon_0$, there exists
$c_\varepsilon>0$ such that every finite classification $C$ satisfies
$\Gamma(C)\leq\varepsilon
\Longrightarrow
L(C)\geq c_\varepsilon$.
Set $B=\max_jr_j-\min_jr_j$.
Since both $R$ and $\bar r_{C(\Theta)}$ lie in
$[\min_jr_j,\max_jr_j]$ $\bbp$-almost surely,
$|R-\bar r_{C(\Theta)}|\leq B$. Recall that the optimal classwise constant predictor of $R$ is
$\bar r_{C(\Theta)}=\bbe_{\bbp}[R\mid C(\Theta)]$, so that
$L(C)=\bbe_{\bbp}[(R-\bar r_{C(\Theta)})^2]$.
Splitting according to $\{\zeta(\Theta)\leq t\}$ and
$\{\zeta(\Theta)>t\}$, for every $t>0$,
$\bbe_{\bbp}[\zeta(\Theta)(R-\bar r_{C(\Theta)})^2]
\leq
tL(C)
+
B^2\bbe_{\bbp}
[\zeta(\Theta)\mathbf 1_{\{\zeta(\Theta)>t\}}]$.
Fix $0\leq\varepsilon<\varepsilon_0$ and set
$g=\varepsilon_0-\varepsilon>0$.
Since $\bbe_{\bbp}[\zeta(\Theta)]=1$,
$\bbe_{\bbp}
[\zeta(\Theta)\mathbf 1_{\{\zeta(\Theta)>t\}}]
\longrightarrow0
\text{ as }t\to\infty$.
Choose a finite $t>0$, independently of $C$, such that
$B^2\bbe_{\bbp}
[\zeta(\Theta)\mathbf 1_{\{\zeta(\Theta)>t\}}]
\leq
g\delta_R^2/8$.
If $\Gamma(C)\leq\varepsilon$, then
$(\varepsilon_0-\Gamma(C))_+\geq g$.
By \eqref{eq:app-MR1-regulatory-cost},
\eqref{eq:app-MR1-law-transfer}, and the preceding bound,
$g
\leq
\frac{4}{\delta_R^2}
\bbe_{\bbp}[\zeta(\Theta)(R-\bar r_{C(\Theta)})^2]
\leq
\frac{4t}{\delta_R^2}L(C)+\frac{g}{2}$.
Hence
$L(C)\geq\frac{g\delta_R^2}{8t}=:c_\varepsilon>0$.

\noindent\textnormal{(v)} Under the additional assumption
$\zeta\leq M$ $\bbp_\Theta$-almost surely, we prove the second assertion,
\eqref{eq:MR1-inferential-cost}. Indeed,
\eqref{eq:app-MR1-law-transfer} gives
\[
\bbe_{\bbp^{\mathrm{reg}}}
[\operatorname{Var}_{\bbp^{\mathrm{reg}}}(R\mid C(\Theta))]
\leq
\bbe_{\bbp}[\zeta(\Theta)(R-\bar r_{C(\Theta)})^2]
\leq
M\bbe_{\bbp}[(R-\bar r_{C(\Theta)})^2]
=
ML(C).
\]
Together with \eqref{eq:app-MR1-regulatory-cost}, this yields
$
\bigl(\varepsilon_0-\Gamma(C)\bigr)_+
\leq
\frac{4M}{\delta_R^2}L(C),
$
which proves \eqref{eq:MR1-inferential-cost}.

\noindent\textnormal{(vi)} Finally, we prove that the numerical constant
$1/4$ in \eqref{eq:MR1-inferential-cost} is sharp. Let the portfolio and
regulatory laws coincide, let $D\in\{1,2\}$ satisfy
$\bbp(D=1)=\bbp(D=2)=1/2$, and set $R=r_D$, with $r_1\ne r_2$.
Then $M=1$ and $\delta_R=|r_2-r_1|$. Recall that $\varepsilon_0=\Gamma(C^\pi)$, where $C^\pi$ is the
benchmark-price classification, which in the present example equals $D$.
Hence
$\varepsilon_0
=
2(1/2)(1/2)\|(1,0)-(0,1)\|_2^2
=
1$.
For the constant classification $C$, we have $\Gamma(C)=0$, while the constant premium minimizing the squared-error criterion in
$L(C)$ is $(r_1+r_2)/2$ and therefore
$
L(C)
=
\frac12\left(r_1-\frac{r_1+r_2}{2}\right)^2
+
\frac12\left(r_2-\frac{r_1+r_2}{2}\right)^2
=
\frac{\delta_R^2}{4}$. Since $\varepsilon_0=1$, $\Gamma(C)=0$, and
$M=1$, we have
$L(C)=\delta_R^2(\varepsilon_0-\Gamma(C))/(4M)$, so equality holds in
\eqref{eq:MR1-inferential-cost}. Hence no constant larger than $1/4$ can
satisfy \eqref{eq:MR1-inferential-cost} under the assumptions of the theorem.

\subsection{Proof of Proposition~\ref{prop:pooling-effects}}
\label{app:proof-pooling-effects}

We compute separately the changes in $L(C)$ and $\Gamma(C)$ caused by
merging classes $a$ and $b$.

Recall that 
$
L(C)
=
\bbe_{\bbp}
[(R-\bar r_{C(\Theta)})^2].
$
The conditional mean benchmark risk price of the merged class is
$
\bar r_{a,b}
=
\frac{q_a\bar r_a+q_b\bar r_b}{q_a+q_b}.
$
All classes other than $a$ and $b$ are unchanged. For $c\in\{a,b\}$, write
$R-\bar r_{a,b}=(R-\bar r_c)+(\bar r_c-\bar r_{a,b})$. Using
$\bbe_{\bbp}[R-\bar r_c\mid C(\Theta)=c]=0$ gives
$
\bbe_{\bbp}[(R-\bar r_{a,b})^2\mid C(\Theta)=c]
=
\bbe_{\bbp}[(R-\bar r_c)^2\mid C(\Theta)=c]
+
(\bar r_c-\bar r_{a,b})^2.
$
Thus the within-class squared errors for $a$ and $b$ remain as in $L(C)$,
while pooling adds the two squared deviations of their conditional means
from the merged conditional mean. Since the two classes have probabilities
$q_a$ and $q_b$, respectively, $L(C^{a,b})-L(C)
=
q_a(\bar r_a-\bar r_{a,b})^2
+
q_b(\bar r_b-\bar r_{a,b})^2$. 
We obtain
$
\bar r_a-\bar r_{a,b}
=
\bar r_a-\frac{q_a\bar r_a+q_b\bar r_b}{q_a+q_b}
=
\frac{q_b}{q_a+q_b}(\bar r_a-\bar r_b),
$
and similarly
$
\bar r_b-\bar r_{a,b}
=
\frac{q_a}{q_a+q_b}(\bar r_b-\bar r_a).
$
Hence $L(C^{a,b})-L(C)
=
q_a\frac{q_b^2}{(q_a+q_b)^2}(\bar r_a-\bar r_b)^2
+
q_b\frac{q_a^2}{(q_a+q_b)^2}(\bar r_a-\bar r_b)^2 
=
\frac{q_aq_b(q_a+q_b)}{(q_a+q_b)^2}(\bar r_a-\bar r_b)^2
=
\frac{q_aq_b}{q_a+q_b}(\bar r_a-\bar r_b)^2$.

Recall that \(\Gamma(C)=2\sum_{c:s_c>0}s_c\|\eta_c-\eta\|_2^2\). The protected-group profile of the merged class is
$
\eta_{a,b}
=
\frac{s_a\eta_a+s_b\eta_b}{s_a+s_b},
$
while all other profiles are unchanged. Therefore,
$\frac12\bigl(\Gamma(C)-\Gamma(C^{a,b})\bigr)
=s_a\|\eta_a-\eta\|_2^2+s_b\|\eta_b-\eta\|_2^2
-(s_a+s_b)\|\eta_{a,b}-\eta\|_2^2$.
Apply \eqref{eq:app-weighted-variance-decomposition} to
$\eta_a,\eta_b$ with weights $s_a/(s_a+s_b)$ and $s_b/(s_a+s_b)$,
weighted mean $\eta_{a,b}$, and reference vector $\eta$.
Multiplying by $s_a+s_b$ and rearranging gives
$\frac12\bigl(\Gamma(C)-\Gamma(C^{a,b})\bigr)
=s_a\|\eta_a-\eta_{a,b}\|_2^2
+s_b\|\eta_b-\eta_{a,b}\|_2^2$.
Applying \eqref{eq:app-weighted-pairwise-variance} with the same
vectors and weights then yields
$s_a\|\eta_a-\eta_{a,b}\|_2^2+s_b\|\eta_b-\eta_{a,b}\|_2^2
=(s_a+s_b)\frac{s_a}{s_a+s_b}\frac{s_b}{s_a+s_b}
\|\eta_a-\eta_b\|_2^2
=\frac{s_as_b}{s_a+s_b}\|\eta_a-\eta_b\|_2^2$.
Multiplying by $2$ proves the second identity in
\eqref{eq:MR1-pooling-effects}.

\subsection{Proof and implementation of
Proposition~\ref{prop:classification-DP}}
\label{app:proof-classification-DP}

The argument has two steps. We first show that, for a coarsening $C$ of
$C^\pi$, both $L(C)$ and $\Gamma(C)$ can be computed by adding the
contributions of the individual tariff classes. We then restrict tariff
classes to consecutive benchmark-price levels. A classification can then be
viewed as a segmentation of the ordered benchmark-price levels, and the
additive structure leads to a dynamic-programming recursion.

Recall that $R$ has essential range $\{r_1, \hdots, r_K\}$ under $\mathbb{P}$ with distinct benchmark-price levels $r_j$. Order $r_1<\cdots<r_K$. For all $j \in \{1, \hdots, K\}$, write
$q_j=\bbp(R=r_j)>0$ and
$s_j=\bbp^{\mathrm{reg}}(R=r_j)$.
For a nonempty set $B\subseteq\{1,\ldots,K\}$, define
$q_B=\sum_{j\in B}q_j$,
$s_B=\sum_{j\in B}s_j$, and
$\bar r_B=q_B^{-1}\sum_{j\in B}q_jr_j$.
If $s_B>0$, let
$\eta_B=\bbp^{\mathrm{reg}}
(D\in\cdot\mid R\in\{r_j:j\in B\})$.

Every coarsening $C$ of $C^\pi$ corresponds to a partition
$\mathcal P_C$ of $\{1,\ldots,K\}$: a set $B\in\mathcal P_C$ contains
exactly the indices of the benchmark-price levels merged into one tariff class $c_B$. We now compute the contribution of $c_B$ to each of the two
criteria.

Under the portfolio law, $\bbp(C(\Theta)=c_B) = q_B$ and $\bbe_{\bbp}[R \mid C(\Theta)=c_B] = \bar r_B$. Since
$L(C)=\bbe_{\bbp}[(R-\bar r_{C(\Theta)})^2]$, the contribution of $c_B$ to $L(C)$ is $q_B\bbe_{\bbp}[(R-\bar r_B)^2\mid R\in\{r_j:j\in B\}]
=\sum_{j\in B}q_j(r_j-\bar r_B)^2$. 
Summing these contributions over all tariff classes of $C$ gives
\begin{equation*}
L(C)
=
\sum_{B\in\mathcal P_C}\sum_{j\in B}q_j(r_j-\bar r_B)^2.
\end{equation*}

Under the regulatory law, $\bbp^{\mathrm{reg}}(C(\Theta)=c_B)=s_B$ and, when $s_B>0$, $\bbp^{\mathrm{reg}}(D \in \cdot \mid C(\Theta)=c_B)=\eta_B$.
Hence \eqref{eq:MR1-Gamma-variance} gives
\begin{equation*}
\Gamma(C)
=
2\sum_{\substack{B\in\mathcal P_C\\ s_B>0}}
s_B\|\eta_B-\eta\|_2^2.
\end{equation*}
Thus the contribution of each tariff class to $L(C)+\lambda\Gamma(C)$
depends only on the benchmark-price levels contained in that class. This is the additive structure needed for the optimization.

We now impose the consecutive-level restriction from
Proposition~\ref{prop:classification-DP}. For
$1\leq i\leq j\leq K$, let $B(i,j)=\{i,\ldots,j\}$.
For $\lambda\geq0$, define the contribution of the tariff class $c_{B(i,j)}$
to the scalarized objective $L(C)+\lambda\Gamma(C)$ by
\begin{equation*}
w_\lambda(i,j)
=
\sum_{h=i}^jq_h(r_h-\bar r_{B(i,j)})^2
+
2\lambda s_{B(i,j)}\|\eta_{B(i,j)}-\eta\|_2^2,
\end{equation*}
where the second term is defined to be zero when $s_{B(i,j)}=0$.

The classification problem has therefore become a segmentation problem:
partition the ordered indices $\{1,\ldots,K\}$ into at most $m$ consecutive
sets and minimize the sum of their costs $w_\lambda(i,j)$. Dynamic
programming applies because, once the last tariff class is fixed, the
preceding tariff classes constitute a problem of the same form on an initial
segment of the benchmark-price levels.

To make the recursion explicit, let $V_\lambda(j,\ell)$ be the minimum
total cost for partitioning the first $j$ benchmark-price levels
$\{1,\ldots,j\}$ into exactly $\ell$ nonempty consecutive tariff classes.
Its cost is the sum of the block costs $w_\lambda$ defined above.
All subproblems use the original portfolio and regulatory masses and
the fixed full-population profile $\eta$; these quantities are not
renormalized or recomputed for the initial segment.
For $\ell=1$, there is only one possible tariff class, so
$V_\lambda(j,1)=w_\lambda(1,j)$.
For $2\leq\ell\leq j$, suppose that the last tariff class begins at level
$i+1$. It then has cost $w_\lambda(i+1,j)$, while the preceding levels
$\{1,\ldots,i\}$ must be partitioned into $\ell-1$ nonempty consecutive
tariff classes, with minimum total cost $V_\lambda(i,\ell-1)$. Since
$\ell-1\leq i<j$, minimizing over the possible positions of this last
boundary gives
\begin{equation}
V_\lambda(j,\ell)
=
\min_{\ell-1\leq i<j}
\left\{
V_\lambda(i,\ell-1)+w_\lambda(i+1,j)
\right\}.
\label{eq:MR1-classification-DP}
\end{equation}
This is the standard dynamic-programming recursion obtained from Bellman's
principle of optimality; see, for example, \cite{Bertsekas2017}.

Since at most $m$ tariff classes are allowed, the minimum in
Proposition~\ref{prop:classification-DP} is
\[
\min_{1\leq\ell\leq\min\{m,K\}}V_\lambda(K,\ell).
\]
An optimal classification is recovered by recording a minimizing boundary
$i$ in \eqref{eq:MR1-classification-DP} for each state $(j,\ell)$ and
tracing these boundaries backwards from a minimizing value of $\ell$ at
$j=K$.

It remains to describe how the block costs can be computed efficiently.
For the portfolio contribution, define $P_j=\sum_{h=1}^j q_h$, $H_j=\sum_{h=1}^j q_hr_h$, and $T_j=\sum_{h=1}^jq_hr_h^2$, with $P_0=H_0=T_0=0$.
For $B(i,j)$, we have $q_{B(i,j)}=P_j-P_{i-1}$.
The corresponding mean benchmark price is $\bar r_{B(i,j)}=\sum_{h=i}^j q_hr_h / \sum_{h=i}^j q_h=(H_j-H_{i-1})/(P_j-P_{i-1})$.
The contribution of $c_{B(i,j)}$ to $L(C)$ is
$\sum_{h=i}^j q_h(r_h-\bar r_{B(i,j)})^2=T_j-T_{i-1}-(H_j-H_{i-1})^2/(P_j-P_{i-1})$.

For the regulatory contribution, define $U_{j,d}=\sum_{h=1}^j \bbp^{\mathrm{reg}}(R=r_h,D=d)$ for $d=1,\ldots,k$, with $U_{0,d}=0$.
The regulatory mass of $B(i,j)$ is $s_{B(i,j)}=\sum_{d=1}^k(U_{j,d}-U_{i-1,d})$.
When $s_{B(i,j)}>0$, the conditional group probabilities satisfy
$s_{B(i,j)}\bbp^{\mathrm{reg}}(D=d\mid R\in\{r_h:h\in B(i,j)\})=U_{j,d}-U_{i-1,d}$ for $d=1,\ldots,k$.
Thus all quantities entering $w_\lambda(i,j)$ can be obtained from
cumulative sums. Computing the regulatory quantities for all $O(K^2)$
intervals requires $O(kK^2)$ arithmetic operations, and the
dynamic-programming recursion requires $O(mK^2)$ operations when $m\leq K$.

\subsection{Aggregation and implementation by balanced transfers}
\label{app:classification-implementation}

Fix a finite classification $C$ with $L(C)=0$ and a prescribed
aggregation component $\mathcal A$. Let $\Lambda$ vary and, for each
transfer, write $\mathbf C=(C,\mathcal A,\Lambda)$. Since
$A^{\mathbf C}(c)=\mathcal A_c(\pi|_{C^{-1}(\{c\})})$ depends only on
$C$ and $\mathcal A$, the aggregate benchmark prices remain fixed as
$\Lambda$ varies. Assume
$\bbp_\Theta^\star\ll\bbp_\Theta$ and
$R,A^{\mathbf C}(C(\Theta))\in L^1(\bbp^\star)$, and suppose that transfers
are unrestricted apart from class dependence and the balance condition
$\bbe_{\bbp^\star}[\Lambda(C(\Theta))]=0$.

Suppose first that a $\bbp^\star$-balanced transfer produces a perfectly
risk-adequate tariff. Then
$
A^{\mathbf C}(C(\Theta))+\Lambda(C(\Theta))=R
$
$\bbp$-almost surely. Since each term is a measurable function of $\Theta$,
the set of types on which this equality fails has $\bbp_\Theta$-measure
zero. The assumption $\bbp_\Theta^\star\ll\bbp_\Theta$ therefore implies
that the equality also holds $\bbp^\star$-almost surely. Taking
$\bbp^\star$-expectations and using the balance condition gives
\[
\bbe_{\bbp^\star}[A^{\mathbf C}(C(\Theta))]
=
\bbe_{\bbp^\star}[R].
\]

Conversely, suppose that this equality holds. Recall that
$\bar r_{C(\Theta)}=\bbe_{\bbp}[R\mid C(\Theta)]$ is the minimizing class premium in
the definition of $L(C)$ and that $L(C)=\bbe_{\bbp}[(R-\bar r_{C(\Theta)})^2]$. Hence $L(C)=0$ implies
$R=\bar r_{C(\Theta)}$ $\bbp$-almost surely. Since $\bar r_{C(\Theta)}$ is a function of the finite
classification $C$, there exists
$h:\{1,\ldots,m\}\to\bbr$ such that
$R=h(C(\Theta))$ $\bbp$-almost surely. As above,
$\bbp_\Theta^\star\ll\bbp_\Theta$ implies that the same identity holds
$\bbp^\star$-almost surely.

Define
$
\Lambda(c)=h(c)-A^{\mathbf C}(c)
$
for $c\in\{1,\ldots,m\}$. Under $\bbp$,
$
A^{\mathbf C}(C(\Theta))+\Lambda(C(\Theta))
=h(C(\Theta))=R
$
almost surely, so the resulting tariff is perfectly risk adequate. Under
$\bbp^\star$,
$
\Lambda(C(\Theta))
=
R-A^{\mathbf C}(C(\Theta))
$
almost surely. The assumed integrability of the two terms on the right
therefore makes the transfer $\bbp^\star$-integrable, and
\[
\bbe_{\bbp^\star}[\Lambda(C(\Theta))]
=
\bbe_{\bbp^\star}[R]
-
\bbe_{\bbp^\star}[A^{\mathbf C}(C(\Theta))]
=
0.
\]
Thus the transfer is $\bbp^\star$-balanced, proving sufficiency.

The absolute-continuity assumption here concerns the reference law
$\bbp^\star$ used for transfer balance, whereas the threshold results use
$\bbp^{\mathrm{reg}}$. Without
$\bbp_\Theta^\star\ll\bbp_\Theta$, perfect risk adequacy under the portfolio
law need not determine the required transfer on types outside the portfolio
support.

\section{Proofs and supplementary results for balanced transfers}
\label{app:balanced-transfers}

This appendix supplies the proofs and supplementary calculations for
Section~\ref{sec:balanced-transfers}. We first determine which
protected-group total premium effects balanced transfers can
attain and the minimum risk inadequacy required for each
attainable target. The balance and target constraints define affine
sets of class-premium vectors, so the minimum-cost problem reduces
to orthogonal projection. We then derive the lower bound involving
the inferential criterion and construct examples showing both its
sharpness and why that criterion does not determine the exact cost.
Finally, we examine whether premium changes preserving a given
target can improve portfolio coverage, first for a differentiable
convex coverage criterion and then for finite-scenario AVaR,
including the nondifferentiable case.

We retain the notation and assumptions of
Section~\ref{sec:balanced-transfers}. Risk inadequacy
is measured by squared premium deviation. Unless explicitly
specialized, risk inadequacy and portfolio coverage
are evaluated under the portfolio law $\bbp$, while transfer
balance and protected-group total premium effects are
evaluated under the reference law $\bbp^\star$.
In particular, $R\in L^2(\bbp)\cap L^1(\bbp^\star)$,
$q_c=\bbp(C(\Theta)=c)>0$, and
$\rho_d^\star=\bbp^\star(D=d)>0$.
Within each optimization over transfers, the classification,
aggregation component, population laws, and benchmark risk prices
remain fixed; class premiums are otherwise unrestricted real
numbers.
\subsection{Quadratic formulation for balanced transfers}
\label{app:transfers-quadratic-formulation}

We express the transfer problem in terms of class premiums.
Balance and a prescribed protected-group target impose affine
constraints on these premiums, while risk inadequacy
is a quadratic function of their distance from the conditional
mean benchmark risk prices. This formulation provides the
projection problem used in the next subsection.

For fixed $C$ and $\mathcal A$, a class-premium vector
$p=(p_1,\ldots,p_m)^\top\in\bbr^m$ specifies the transfer
$\Lambda(c)=p_c-A^{\mathbf C}(c)$. Consequently,
$\bbe_{\bbp^\star}[\Lambda(C(\Theta))]
=(q^\star)^\top p
-\bbe_{\bbp^\star}[A^{\mathbf C}(C(\Theta))]$.
Thus the balanced class-premium vectors form the affine hyperplane
$\mathcal H$ defined in
Subsection~\ref{subsec:transfer-outcomes-cost}.
It is nonempty because the vector
$(A^{\mathbf C}(1),\ldots,A^{\mathbf C}(m))^\top$
corresponds to zero transfer.
A premium change $v\in\bbr^m$ preserves balance if and only if
$(q^\star)^\top v=0$, so the direction space of $\mathcal H$ is
$\{v\in\bbr^m:(q^\star)^\top v=0\}$.

To express the protected-group constraints, recall that
$Q_{dc}=\bbp^\star(C(\Theta)=c\mid D=d)$ and
$\mu_d^\star=\bbe_{\bbp^\star}[R\mid D=d]$.
Each row of $Q$ sums to one, giving
$Q\mathbf1_m=\mathbf1_k$.
The law of total probability gives
$\sum_d\rho_d^\star Q_{dc}=q_c^\star$ for every $c$,
and hence $(\rho^\star)^\top Q=(q^\star)^\top$.
Likewise,
$(\rho^\star)^\top\mu^\star=\bbe_{\bbp^\star}[R]$.
Conditioning the total premium effect $\Delta^{\mathbf C}$
on $D=d$ yields
$\bbe_{\bbp^\star}[p_{C(\Theta)}-R\mid D=d]
=\sum_cQ_{dc}p_c-\mu_d^\star$.
Therefore $M_\Delta^{\mathbf C,\star}=Qp-\mu^\star$, and
the attainable set of protected-group total premium-effect vectors
is $\mathcal T_C=Q\mathcal H-\mu^\star$.

These identities also explain why an attainable target
$y\in\mathcal T_C$ already encodes balance. If
$Qp-\mu^\star=y$, then
$(\rho^\star)^\top y
=(q^\star)^\top p-\bbe_{\bbp^\star}[R]$.
Balance means
$(q^\star)^\top p=\bbe_{\bbp^\star}[A^{\mathbf C}(C(\Theta))]$.
Recalling $b=\bbe_{\bbp^\star}[A^{\mathbf C}(C(\Theta))-R]$,
such a vector $p$ belongs to $\mathcal H$ if and only if
$(\rho^\star)^\top y=b$.
Thus, for an attainable target $y$,
$\mathcal F_y=\{p\in\bbr^m:Qp-\mu^\star=y\}$.
Choosing any $p_0\in\mathcal F_y$, the target equation is
equivalent to $Q(p-p_0)=0$, so
$\mathcal F_y=p_0+\ker Q$.
The vectors in $\ker Q$ are therefore the premium changes
that preserve every protected-group total premium effect.
They also preserve balance, since
$(q^\star)^\top v=(\rho^\star)^\top Qv=0$.

It remains to express risk inadequacy on this affine
set. Recall that
$\bar r_c=\bbe_{\bbp}[R\mid C(\Theta)=c]$.
For every class-premium vector $p$, decompose its premium deviation as
$p_{C(\Theta)}-R
=(p_{C(\Theta)}-\bar r_{C(\Theta)})
+(\bar r_{C(\Theta)}-R)$.
The first term is measurable with respect to
$\sigma(C(\Theta))$, and the second has conditional mean zero
given $C(\Theta)$ under $\bbp$.
Their product therefore has expectation zero.
Taking the expectation of the squared sum gives
$f_{\mathrm{RA}}(p)
=\bbe_{\bbp}[(p_{C(\Theta)}-R)^2]
=L(C)+\sum_c q_c(p_c-\bar r_c)^2
=L(C)+\|p-\bar r\|_q^2$.

Since $L(C)$ is fixed, minimizing risk inadequacy
for an attainable target amounts to finding the point of
$\mathcal F_y$ closest to $\bar r$ in the $q$-norm.
The reference law $\bbp^\star$ determines the affine constraints;
the portfolio law $\bbp$ determines $\bar r$ and the distance
being minimized. The two laws need not coincide.

\subsection{Attainable protected-group total premium effects and minimum risk inadequacy}
\label{app:transfers-minimum-cost}

We first prove the attainability and minimum-cost statements in
Theorem~\ref{thm:transfer-target-cost}. We then give matrix formulas
for computing the minimizing class-premium vector and its cost,
including when $Q$ has linearly dependent rows.

\begin{proof}[Proof of Theorem~\ref{thm:transfer-target-cost}]
\emph{Attainability.}
A proposed target $y\in\bbr^k$ must satisfy the mean restriction
imposed by balance and be reachable through the map
$p\mapsto Qp-\mu^\star$.
Indeed, if a balanced class-premium vector $p$ attains $y$, the
probability identities from the preceding subsection give
$(\rho^\star)^\top y
=\bbe_{\bbp^\star}[A^{\mathbf C}(C(\Theta))-R]=b$.
Since $w_C=Q\bar r-\mu^\star$, we also have
$y-w_C=Q(p-\bar r)\in\operatorname{im}Q$.

Conversely, suppose that
$(\rho^\star)^\top y=b$ and
$y-w_C\in\operatorname{im}Q$.
Choose $z\in\bbr^m$ satisfying $Qz=y-w_C$ and put $p=\bar r+z$.
Then $Qp-\mu^\star=y$, and
$(q^\star)^\top p
=(\rho^\star)^\top(\mu^\star+y)
=\bbe_{\bbp^\star}[R]+b
=\bbe_{\bbp^\star}[A^{\mathbf C}(C(\Theta))]$.
Thus $p$ is balanced and attains $y$, proving
\eqref{eq:transfer-target-feasibility}.

\emph{Minimum cost and remaining freedom.}
For an attainable target $y\in\mathcal T_C$, the set
$\mathcal F_y$ is a nonempty closed affine subspace with
direction space $\ker Q$.
Since every $q_c$ is positive, the formula
$\langle x,z\rangle_q=\sum_cq_cx_cz_c$
defines an inner product on $\bbr^m$.
The orthogonal-projection theorem therefore gives a unique
$p_y\in\mathcal F_y$ minimizing $\|p-\bar r\|_q^2$.
The displacement from $\bar r$ to this projection satisfies
$\langle p_y-\bar r,v\rangle_q=0$ for every $v\in\ker Q$.

Every $p\in\mathcal F_y$ is uniquely of the form $p=p_y+v$
with $v\in\ker Q$.
Expanding the squared norm, the cross term vanishes by this
orthogonality, and hence
$\|p-\bar r\|_q^2
=\|p_y-\bar r\|_q^2+\|v\|_q^2
=c_C(y)+\|v\|_q^2$.
Together with
$f_{\mathrm{RA}}(p)=L(C)+\|p-\bar r\|_q^2$,
this proves \eqref{eq:transfer-minimum-risk-inadequacy} and
\eqref{eq:transfer-target-decomposition}.
Any nonzero $v$ strictly increases risk inadequacy,
so $p_y$ is its unique minimizer on $\mathcal F_y$.
\end{proof}

\smallskip
\noindent
\emph{Explicit computation of the minimum cost.}
The projection characterization reduces computation to finding
a vector that attains the target and whose displacement from
$\bar r$ is $q$-orthogonal to $\ker Q$.
For this purpose, let
$D_q=\operatorname{diag}(q_1,\ldots,q_m)$ and
$H_C=QD_q^{-1}Q^\top$.
The matrix $H_C$ is symmetric and positive semidefinite, since
$z^\top H_Cz=\|D_q^{-1/2}Q^\top z\|_2^2$ for every
$z\in\bbr^k$.
This identity gives $\ker H_C=\ker Q^\top$.
Taking Euclidean orthogonal complements therefore yields
$\operatorname{im}H_C=\operatorname{im}Q$.

The matrix $H_C$ may be singular, so we use its Moore--Penrose
inverse $H_C^\dagger$.
For a symmetric positive semidefinite matrix, this inverse is
obtained by replacing each positive eigenvalue by its reciprocal
and leaving each zero eigenvalue equal to zero.
Thus $H_CH_C^\dagger$ is the Euclidean orthogonal projection
onto $\operatorname{im}H_C$, and
$H_C^\dagger H_CH_C^\dagger=H_C^\dagger$.

For an attainable target $y$, consider
$p=\bar r+D_q^{-1}Q^\top H_C^\dagger(y-w_C)$.
Because $y-w_C\in\operatorname{im}Q=\operatorname{im}H_C$,
we have
$Q(p-\bar r)=H_CH_C^\dagger(y-w_C)=y-w_C$.
Hence $p\in\mathcal F_y$.
Moreover, for every $v\in\ker Q$,
$\langle p-\bar r,v\rangle_q
=(y-w_C)^\top H_C^\dagger Qv=0$.
These two properties identify $p$ as the projection $p_y$.

Substituting this displacement into the squared $q$-norm gives
$c_C(y)
=(y-w_C)^\top H_C^\dagger H_CH_C^\dagger(y-w_C)
=(y-w_C)^\top H_C^\dagger(y-w_C)$.
If $Q$ has full row rank, $H_C$ is positive definite and
$H_C^\dagger=H_C^{-1}$.
The formula using $H_C^\dagger$ also covers dependent rows of
$Q$, provided the target satisfies the attainability conditions.

Consequently, a target $y$ can be attained under a risk inadequacy bound $B$ if and only if it satisfies
\eqref{eq:transfer-target-feasibility} and
$L(C)+c_C(y)\leq B$:
the first condition ensures that implementing premiums exist,
and the second ensures that their minimum cost meets the bound.

\subsubsection{Balanced changes in protected-group total premium effects}

We now describe how balanced premium changes can move the
protected-group total premium-effect vector within
$\mathcal T_C$. This gives the full-row-rank criterion stated
in Subsection~\ref{subsec:transfer-outcomes-cost} and identifies
situations in which some or all group differences cannot be
changed.

\smallskip
\noindent
\emph{Available changes.}
A premium change $v\in\bbr^m$ preserves balance when
$(q^\star)^\top v=0$ and changes protected-group means by $Qv$.
The available changes therefore form the linear subspace
$\mathcal G_C
=Q\{v\in\bbr^m:(q^\star)^\top v=0\}$.
Since $(\rho^\star)^\top Q=(q^\star)^\top$, this space satisfies
$\mathcal G_C
=\operatorname{im}Q
\cap\{z\in\bbr^k:(\rho^\star)^\top z=0\}$.
Indeed, for $z=Qv$,
$(\rho^\star)^\top z=(\rho^\star)^\top Qv=(q^\star)^\top v$.
Thus a premium change is balanced if and only if its image
has zero $\rho^\star$-weighted mean, proving the equality
of the two descriptions of $\mathcal G_C$.

For any attainable target $y_0$, choose an implementing
$p_0\in\mathcal H$.
Every other balanced premium vector differs from $p_0$ by a
vector with zero $q^\star$-weighted mean.
Consequently, $\mathcal T_C=y_0+\mathcal G_C$:
the subspace $\mathcal G_C$ describes all movements from an
attainable target.

\smallskip
\noindent
\emph{Attainability of every target with the required mean.}
Since $(\rho^\star)^\top y_0=b$, every target satisfying
$(\rho^\star)^\top y=b$ is attainable if and only if
$\mathcal G_C=\{z\in\bbr^k:(\rho^\star)^\top z=0\}$.
If $\operatorname{rank}Q=k$, this equality follows from
$\operatorname{im}Q=\bbr^k$ and the preceding characterization
of $\mathcal G_C$.

For the converse, suppose $\mathcal G_C$ contains every vector
with zero $\rho^\star$-weighted mean.
Any $z\in\bbr^k$ has the decomposition
$z=((\rho^\star)^\top z)\mathbf1_k
+[z-((\rho^\star)^\top z)\mathbf1_k]$.
The constant term belongs to $\operatorname{im}Q$ because
$Q\mathbf1_m=\mathbf1_k$.
The remaining term has zero $\rho^\star$-weighted mean and
therefore belongs to $\mathcal G_C\subseteq\operatorname{im}Q$.
Thus $\operatorname{im}Q=\bbr^k$, proving
$\operatorname{rank}Q=k$.

\smallskip
\noindent
\emph{Restrictions arising from identical rows of $Q$.}
If rows $d$ and $e$ of $Q$ coincide, the two protected groups
have the same distribution across tariff classes.
Consequently, $(Qp)_d=(Qp)_e$ for every class-premium vector
$p$: their mean charged premiums are identical.
The corresponding protected-group total premium effects satisfy
$(Qp-\mu^\star)_d-(Qp-\mu^\star)_e
=\mu_e^\star-\mu_d^\star$.
No class-dependent transfer can change this difference.

At the level of all groups, balanced transfers leave the entire
protected-group total premium-effect vector unchanged if
and only if $C(\Theta)$ and $D$ are independent under
$\bbp^\star$.
To prove this, independence gives
$Q=\mathbf1_k(q^\star)^\top$, so $Qv=0$ whenever
$(q^\star)^\top v=0$.
Hence $\mathcal G_C=\{0\}$.

Conversely, suppose $\mathcal G_C=\{0\}$.
For any $v\in\bbr^m$, decompose
$v=((q^\star)^\top v)\mathbf1_m
+[v-((q^\star)^\top v)\mathbf1_m]$.
The second term has zero $q^\star$-weighted mean, so its image
under $Q$ vanishes.
Using $Q\mathbf1_m=\mathbf1_k$, we obtain
$Qv=((q^\star)^\top v)\mathbf1_k$ for every $v$.
Thus $Q=\mathbf1_k(q^\star)^\top$: the conditional distribution
of tariff classes is the same in every protected group,
which proves independence.

\subsubsection{Minimum protected-group standard deviation}

The group-neutrality question in
Subsection~\ref{subsec:transfer-outcomes-cost} also leads to a
quantitative question: if balanced transfers cannot equalize
protected-group total premium effects, what is their
smallest attainable dispersion?
Choose standard deviation for $\mathfrak D^\star$ and equip
$\bbr^k$ with the inner product
$\langle x,z\rangle_{\rho^\star}
=\sum_d\rho_d^\star x_dz_d$ and its norm
$\|\cdot\|_{\rho^\star}$.
Since every attainable target has $\rho^\star$-weighted mean $b$,
the minimum dispersion over balanced transfers is
$d_C=\min_{y\in\mathcal T_C}\|y-b\mathbf1_k\|_{\rho^\star}$.

To identify the centered attainable vectors
$\{y-b\mathbf1_k:y\in\mathcal T_C\}$, define
$\mu_0=\mu^\star
-((\rho^\star)^\top\mu^\star)\mathbf1_k$,
the centered vector of protected-group conditional mean
benchmark risk prices.
Every balanced premium vector can be written as
$p=\bbe_{\bbp^\star}[A^{\mathbf C}(C(\Theta))]\mathbf1_m+z$
with $(q^\star)^\top z=0$.
Using $Q\mathbf1_m=\mathbf1_k$ and the definition of $b$ gives
$Qp-\mu^\star-b\mathbf1_k=Qz-\mu_0$.
As $z$ ranges over the balanced premium changes, $Qz$ ranges
over $\mathcal G_C$. Hence
$\{y-b\mathbf1_k:y\in\mathcal T_C\}
=\mathcal G_C-\mu_0$.

Minimizing dispersion therefore amounts to finding the point
of $\mathcal G_C$ closest to $\mu_0$ in the
$\rho^\star$-norm.
Let $\operatorname{proj}_C$ denote orthogonal projection onto
$\mathcal G_C$ for this inner product.
The projection theorem gives
$d_C=\|\mu_0-\operatorname{proj}_C\mu_0\|_{\rho^\star}$,
with unique minimizing protected-group total premium-effect
vector
$b\mathbf1_k+\operatorname{proj}_C\mu_0-\mu_0$.
The corresponding class-premium vector need not be unique:
its implementing vectors differ by elements of $\ker Q$.
Thus the minimum dispersion is the $\rho^\star$-norm of the
component of $\mu_0$ orthogonal to $\mathcal G_C$, which balanced
premium changes cannot offset.
It vanishes if and only if $\mu_0\in\mathcal G_C$,
which characterizes attainability of group neutrality.
Exact compensation $y=0$ is attainable if and only if
$\mu_0\in\mathcal G_C$ and $b=0$.

\subsubsection{Dependence on the aggregation component}

We finally verify the aggregation statement in
Subsection~\ref{subsec:transfer-outcomes-cost}.
Fix the classification, population laws, and benchmark risk
prices, and consider two aggregation components.
With unrestricted transfers, either component can generate
every premium vector satisfying its balance condition.
If their aggregate benchmark prices have the same expectation
under $\bbp^\star$, these balance conditions define the same
hyperplane $\mathcal H$.

The two components consequently generate the same available
class-premium vectors, the same attainable set $\mathcal T_C$,
and the same implementing sets $\mathcal F_y$.
The portfolio conditional mean vector $\bar r$ and the
$q$-inner product are also unchanged.
Their projections $p_y$ and minimum costs $c_C(y)$ therefore
coincide for every attainable target.
The implementing transfers may differ, since each is obtained
by subtracting the chosen aggregate benchmark prices from the
class premiums.

\subsection{The inferential criterion and risk inadequacy}
\label{app:transfers-inferential-cost}

We prove the bound in
Proposition~\ref{prop:transfer-inferential-bound} by comparing
the change in protected-group means required to meet a dispersion
constraint with the change that a balanced premium adjustment
can produce at a given cost. We then construct a binary example
in which the bound is attained for every dispersion allowance.

Throughout this subsection,
$\bbp^\star=\bbp^{\mathrm{reg}}=\bbp$,
the inferential criterion uses squared Euclidean dissimilarity,
and $\mathfrak D^\star$ is standard deviation.
The protected-group profile $\eta_c$ from
Section~\ref{sec:main-result-I} has components
$\eta_{c,d}=\bbp(D=d\mid C(\Theta)=c)$.
By \eqref{eq:MR1-Gamma-variance},
$\Gamma(C)=2\bbe_{\bbp}[
\sum_{d=1}^k(\eta_{C(\Theta),d}-\rho_d^\star)^2]$.
Recall also that
$\rho_{\min}^\star=\min_d\rho_d^\star>0$.

\begin{proof}[Proof of Proposition~\ref{prop:transfer-inferential-bound}]
\emph{Cost and effect of a balanced adjustment.}
Let $p$ be a balanced class-premium vector and write
$u=p-\bar r-b\mathbf1_m$.
Since the population laws coincide,
$q^\top\bar r=\bbe_{\bbp}[R]$, while balance gives
$q^\top p
=\bbe_{\bbp}[A^{\mathbf C}(C(\Theta))]
=\bbe_{\bbp}[R]+b$.
Thus $q^\top u=0$: the vector $u$ is a balanced adjustment
to the baseline $\bar r+b\mathbf1_m$.
The vectors $b\mathbf1_m$ and $u$ are $q$-orthogonal, so
$f_{\mathrm{RA}}(p)
=L(C)+\|p-\bar r\|_q^2
=L(C)+\|b\mathbf1_m+u\|_q^2
=L(C)+b^2+\|u\|_q^2$.
In particular, the baseline minimizes risk inadequacy
over balanced premiums, and $\|u\|_q^2$ is the additional cost.

The resulting protected-group total premium-effect vector is
$M_\Delta^{\mathbf C,\star}
=Qp-\mu^\star=b\mathbf1_k+w_C+Qu$.
The probability identities give
$(\rho^\star)^\top w_C
=q^\top\bar r-(\rho^\star)^\top\mu^\star=0$
and $(\rho^\star)^\top Qu=q^\top u=0$.
Consequently,
$D_\Delta^\star(\mathbf C)=\|w_C+Qu\|_{\rho^\star}$:
the adjustment changes the centered group-effect vector from
$w_C$ to $w_C+Qu$.

\emph{Bounding the change in group means.}
We now relate $Qu$ to the adjustment cost and the inferential
criterion.
For each protected group,
$\rho_d^\star(Qu)_d
=\bbe_{\bbp}[u_{C(\Theta)}\mathbf1_{\{D=d\}}]
=\bbe_{\bbp}[u_{C(\Theta)}\eta_{C(\Theta),d}]$,
where the second equality follows by conditioning on $C(\Theta)$.
Since $\bbe_{\bbp}[u_{C(\Theta)}]=q^\top u=0$, this becomes
$\rho_d^\star(Qu)_d
=\bbe_{\bbp}[u_{C(\Theta)}
(\eta_{C(\Theta),d}-\rho_d^\star)]$.
Thus the group mean adjustment depends on how the premium
adjustment varies with the class protected-group profile.

Cauchy--Schwarz gives
$(\rho_d^\star)^2(Qu)_d^2
\leq\|u\|_q^2
\bbe_{\bbp}[(\eta_{C(\Theta),d}-\rho_d^\star)^2]$.
Summing over $d$ yields
$\sum_d(\rho_d^\star)^2(Qu)_d^2
\leq\Gamma(C)\|u\|_q^2/2$.
Since $\rho_d^\star\geq\rho_{\min}^\star$,
the left-hand side is at least
$\rho_{\min}^\star\sum_d\rho_d^\star(Qu)_d^2
=\rho_{\min}^\star\|Qu\|_{\rho^\star}^2$.
Hence
$\|Qu\|_{\rho^\star}^2
\leq\Gamma(C)\|u\|_q^2/(2\rho_{\min}^\star)$.
This estimate holds for every $u$ with $q^\top u=0$.

\emph{Cost of meeting the dispersion constraint.}
If $D_\Delta^\star(\mathbf C)\leq s$, the triangle inequality gives
$\|w_C\|_{\rho^\star}
\leq\|w_C+Qu\|_{\rho^\star}+\|Qu\|_{\rho^\star}
\leq s+\|Qu\|_{\rho^\star}$.
The adjustment must therefore satisfy
$\|Qu\|_{\rho^\star}
\geq(\|w_C\|_{\rho^\star}-s)_+$.
When $\Gamma(C)>0$, combining this requirement with the norm
estimate yields
$\|u\|_q^2
\geq 2\rho_{\min}^\star
(\|w_C\|_{\rho^\star}-s)_+^2/\Gamma(C)$.
Substitution into
$f_{\mathrm{RA}}(p)=L(C)+b^2+\|u\|_q^2$
proves \eqref{eq:transfer-budget-bound}.

If $\Gamma(C)=0$, the norm estimate instead gives $Qu=0$
for every balanced adjustment $u$.
All balanced class-premium vectors then generate
$b\mathbf1_k+w_C$.
Since $\bar r+b\mathbf1_m$ is itself balanced, the dispersion
constraint is feasible if and only if
$\|w_C\|_{\rho^\star}\leq s$.
\end{proof}

\subsubsection{The binary example and partial compensation}
\label{app:transfers-binary-example}

We now verify the sharpness claim in
Subsection~\ref{subsec:transfer-inferential-cost}.
With two equally likely tariff classes, balance fixes the
average of their class premiums, leaving one free parameter.
Both its risk inadequacy cost and its effect on protected-group
dispersion can therefore be computed directly. The construction attains the bound for every
$s\geq0$, including exact compensation at $s=0$.

\smallskip
\noindent
\emph{Construction and baseline quantities.}
Let both tariff-class labels and protected-group labels be
$\{1,2\}$, and define $\sigma_1=-1$ and $\sigma_2=1$.
For $0<\theta\leq1$, specify
$\bbp(C(\Theta)=c,D=d)=(1+\theta\sigma_c\sigma_d)/4$.
This distribution can be realized on the type space
$\{1,2\}^2$ with $C$ and $D$ as the coordinate maps.
Both marginals are uniform, and
$Q_{dc}=(1+\theta\sigma_d\sigma_c)/2$.
In particular,
$\bbe_{\bbp}[\sigma_D\mid C(\Theta)=c]=\theta\sigma_c$
and
$\bbe_{\bbp}[\sigma_{C(\Theta)}\mid D=d]=\theta\sigma_d$.

Take $R=r_0+a\sigma_D$, where $r_0\geq a>0$, and use
conditional-mean aggregation,
$A^{\mathbf C}(c)=\bar r_c=r_0+a\theta\sigma_c$.
Since $\bbe_{\bbp}[R]=r_0$, this aggregation gives $b=0$.
The conditional variance of $\sigma_D$ given $C(\Theta)$
is $1-\theta^2$, so $L(C)=a^2(1-\theta^2)$.

The protected-group profile in class $c$ is
$\eta_c=((1-\theta\sigma_c)/2,(1+\theta\sigma_c)/2)^\top$.
Its squared distance from the population profile
$(1/2,1/2)^\top$ is $\theta^2/2$; hence
$\Gamma(C)=\theta^2$.
Moreover, $(Q\bar r)_d=r_0+a\theta^2\sigma_d$, giving
$(w_C)_d=-a(1-\theta^2)\sigma_d$ and
$\|w_C\|_{\rho^\star}=a(1-\theta^2)$.

\smallskip
\noindent
\emph{Balanced premiums and exact compensation.}
Balance requires $(p_1+p_2)/2=r_0$.
Thus every balanced class-premium vector
$p=(p_1,p_2)^\top$ has components
$p_c=r_0+x\sigma_c$ for a unique $x\in\bbr$.
The protected-group total premium effect for group $d$ is
$\bbe_{\bbp^\star}[p_{C(\Theta)}-R\mid D=d]
=(\theta x-a)\sigma_d$.
Since $p_c-\bar r_c=(x-a\theta)\sigma_c$, its risk inadequacy is
$f_{\mathrm{RA}}(p)=a^2(1-\theta^2)+(x-a\theta)^2$.

Exact compensation requires $\theta x=a$.
Its unique implementing premium vector therefore has components
$p_c=r_0+(a/\theta)\sigma_c$, and its risk inadequacy is
$f_{\mathrm{RA}}(p)
=a^2(1-\theta^2)+a^2(\theta^{-1}-\theta)^2
=a^2(\theta^{-2}-1)$.
Together with $\Gamma(C)=\theta^2$, this proves
\eqref{eq:transfer-binary-cost}.
The cost diverges as $\theta\downarrow0$, although exact
compensation remains attainable for every $\theta>0$.

\smallskip
\noindent
\emph{Partial compensation and attainment of the bound.}
The two group effects have opposite signs and equal probabilities,
so their standard deviation is $|\theta x-a|$.
Consequently, $D_\Delta^\star(\mathbf C)\leq s$ is equivalent to
$x\in[(a-s)/\theta,(a+s)/\theta]$.
Minimizing risk inadequacy amounts to projecting its
unconstrained minimizer $a\theta$ onto this interval.

For $s\geq0$ and $0<\theta\leq1$,
$a\theta\leq(a+s)/\theta$, so the minimizer never lies above
the upper endpoint.
Its distance from the interval is therefore
$((a-s)/\theta-a\theta)_+
=(a(1-\theta^2)-s)_+/\theta$.
The minimum risk inadequacy under the dispersion
constraint is
$a^2(1-\theta^2)
+(a(1-\theta^2)-s)_+^2/\theta^2$.

In this family, $b=0$, $\rho_{\min}^\star=1/2$,
$\Gamma(C)=\theta^2$, and
$\|w_C\|_{\rho^\star}=a(1-\theta^2)$.
Substitution into \eqref{eq:transfer-budget-bound} gives this
minimum, proving attainment of the bound for every $s\geq0$.

\subsection{Risk inadequacy at a fixed value of the inferential criterion}
\label{app:transfers-fixed-disparity}

We prove Proposition~\ref{prop:transfer-fixed-disparity} by
constructing classifications for which premium adjustments become
arbitrarily weak in one direction of protected-group differences.
The benchmark risk prices are chosen so that exact compensation
requires a change in this direction. A second, orthogonal direction
keeps the inferential criterion constant as the first response
weakens. The population and benchmark risk-price map remain fixed,
and $\bbp^\star=\bbp^{\mathrm{reg}}=\bbp$ throughout.

\begin{proof}[Proof of Proposition~\ref{prop:transfer-fixed-disparity}]
\emph{Construction on a fixed population.}
Let $e=(1,-1,0)^\top/\sqrt2$ and
$f=(1,1,-2)^\top/\sqrt6$.
These vectors are Euclidean orthonormal and orthogonal to
$\mathbf1_3$, so they describe two independent directions of
differences among three equally likely groups.

Let $D$ be uniform on $\{1,2,3\}$ and let $U$ be uniform on
$[0,1]$, independently of $D$.
Take $\Theta=(D,U)$.
Fix $a>0$ and $r_0\geq a\sqrt{2/3}$, and define the benchmark
risk-price map by $\pi(d,u)=r_0+a f_d$.
Then $R=\pi(\Theta)=r_0+a f_D\geq0$.
This fixes the type distribution and the joint distribution of
$(R,D)$ before any classification is chosen.

For $0<\tau<1/4$, put
$s_\tau=\sqrt{1/16-\tau^2}$ and define
$Q_\tau
=\frac13\mathbf1_3\mathbf1_3^\top
+s_\tau ee^\top+\tau ff^\top$.
The row and column sums of $Q_\tau$ equal one because
$e^\top\mathbf1_3=f^\top\mathbf1_3=0$.
Moreover, $|e_de_c|\leq1/2$ and $|f_df_c|\leq2/3$.
Since $s_\tau<1/4$ and $\tau<1/4$, every entry satisfies
$(Q_\tau)_{dc}>1/3-1/8-1/6=1/24$.
Thus each row specifies a strictly positive distribution of
tariff classes within a protected group.

To realize these conditional probabilities, partition $[0,1]$,
for each $d$ and $\tau$, into three consecutive intervals of
lengths $(Q_\tau)_{d1}$, $(Q_\tau)_{d2}$, and $(Q_\tau)_{d3}$.
Define $C_\tau(d,u)$ as the corresponding interval label,
assigning endpoints consistently.
Then
$\bbp(C_\tau(\Theta)=c\mid D=d)=(Q_\tau)_{dc}$ and
$\bbp(C_\tau(\Theta)=c,D=d)=(Q_\tau)_{dc}/3$.
The column sums imply that the tariff-class probabilities are
also uniform.
Only the classification varies with $\tau$.

\emph{Fixed inferential disparity and attainable targets.}
The matrix $Q_\tau$ acts on the three mutually orthogonal
directions as
$Q_\tau\mathbf1_3=\mathbf1_3$,
$Q_\tau e=s_\tau e$, and $Q_\tau f=\tau f$.
Its eigenvalues are therefore $1$, $s_\tau$, and $\tau$,
all positive, so $\operatorname{rank}Q_\tau=3$.
The response in direction $f$ tends to zero as $\tau\downarrow0$,
while the definition of $s_\tau$ keeps
$s_\tau^2+\tau^2=1/16$.

Because both marginals are uniform, the protected-group profile
in class $c$ is the $c$th column of $Q_\tau$.
Its difference from the population profile
$\frac13\mathbf1_3$ is $s_\tau e e_c+\tau f f_c$.
Orthonormality and \eqref{eq:MR1-Gamma-variance} give
$\Gamma(C_\tau)
=\frac23\sum_{c=1}^3
\|s_\tau e e_c+\tau f f_c\|_2^2
=\frac23(s_\tau^2+\tau^2)=1/24$.

For each classification, use conditional-mean aggregation.
Suppressing the dependence of $\bar r$ on $\tau$, its components
are
$\bar r_c
=\bbe_{\bbp}[R\mid C_\tau(\Theta)=c]
=r_0+a(Q_\tau^\top f)_c
=r_0+a\tau f_c$.
Thus $A^{\mathbf C}(c)=\bar r_c$ and $b=0$.
By the full-row-rank characterization, every protected-group
total premium-effect vector with zero mean is attainable
for every $\tau$, including exact compensation.

\emph{Diverging compensation cost.}
Since $R=r_0+a f_D$, exact compensation requires
$Q_\tau p=r_0\mathbf1_3+a f$.
The identities above and nonsingularity of $Q_\tau$ give the
unique solution $p=r_0\mathbf1_3+(a/\tau)f$.
It is balanced because its mean under the uniform tariff-class
probabilities is $r_0$.
The factor $1/\tau$ expresses the premium adjustment needed to
offset the weakening response in direction $f$.

To compute its risk inadequacy, uniform group
probabilities and $\|f\|_2=1$ give
$\operatorname{Var}_{\bbp}(R)=a^2/3$.
Likewise,
$\operatorname{Var}_{\bbp}(\bar r_{C_\tau(\Theta)})
=a^2\tau^2/3$.
The conditional-variance decomposition therefore yields
$L(C_\tau)
=\bbe_{\bbp}[\operatorname{Var}_{\bbp}(R\mid C_\tau(\Theta))]
=a^2(1-\tau^2)/3$.
Furthermore,
$p-\bar r=a(\tau^{-1}-\tau)f$, so
$\|p-\bar r\|_q^2=a^2(\tau^{-1}-\tau)^2/3$.
Consequently, the minimum risk inadequacy for exact
compensation is
$f_{\mathrm{RA}}(p)
=L(C_\tau)+\|p-\bar r\|_q^2
=\frac{a^2}{3}[(1-\tau^2)+(\tau^{-1}-\tau)^2]
=\frac{a^2}{3}(\tau^{-2}-1)$.
This tends to infinity as $\tau\downarrow0$, while the population,
benchmark risk-price map, inferential criterion, and attainable
protected-group targets remain unchanged.
\end{proof}

\subsection{Portfolio coverage at fixed protected-group total premium effects}
\label{app:transfers-coverage}

We return to possibly distinct portfolio and reference laws.
Fix an attainable target $y\in\mathcal T_C$, and let $p_y$
minimize $f_{\mathrm{RA}}(p)$ over $p\in\mathcal F_y$.
We ask whether the remaining premium freedom can improve
portfolio coverage and how the best attainable coverage depends
on an allowance for additional risk inadequacy.

Theorem~\ref{thm:transfer-target-cost} gives
$\mathcal F_y=p_y+\ker Q$ and
$f_{\mathrm{RA}}(p_y+v)
=f_{\mathrm{RA}}(p_y)+\|v\|_q^2$ for $v\in\ker Q$.
Thus these directions preserve the protected-group target and
balance, and an allowance $\delta$ for additional risk inadequacy
permits precisely the changes satisfying $\|v\|_q^2\leq\delta$.
The minimum coverage criterion over these changes is
$K_y(\delta)$ from \eqref{eq:transfer-conditional-coverage}.
We first prove its local expansion for a differentiable convex
coverage criterion, then derive the finite-scenario AVaR
representation needed to treat both differentiable and
nondifferentiable cases.

\begin{proof}[Proof of Theorem~\ref{thm:transfer-coverage}]
Under the theorem's assumptions, $f_{\mathrm{cov}}$ is finite
and convex on $\bbr^m$, hence continuous.
For every $\delta\geq0$, the feasible set
$\{v\in\ker Q:\|v\|_q^2\leq\delta\}$ is nonempty and compact,
so the minimum defining $K_y(\delta)$ is attained.

\emph{Optimality at the minimum-cost premium vector.}
Differentiability at $p_y$ gives a linear functional
$Df_{\mathrm{cov}}(p_y)$.
Represent its negative by the unique vector $a\in\bbr^m$
satisfying
$Df_{\mathrm{cov}}(p_y)[v]=-\langle a,v\rangle_q$
for every $v\in\bbr^m$.
Let $a_{\ker Q}$ be its $q$-orthogonal projection onto $\ker Q$.
For feasible directions $v\in\ker Q$,
$\langle a,v\rangle_q=\langle a_{\ker Q},v\rangle_q$,
so this projection determines the first-order coverage effect
of every target-preserving premium change.

If $a_{\ker Q}=0$, convexity gives
$f_{\mathrm{cov}}(p_y+v)
\geq f_{\mathrm{cov}}(p_y)+Df_{\mathrm{cov}}(p_y)[v]
=f_{\mathrm{cov}}(p_y)$
for every $v\in\ker Q$.
Hence $p_y$ minimizes the coverage criterion on $\mathcal F_y$.
Since $v=0$ is feasible for every allowance,
$K_y(\delta)=f_{\mathrm{cov}}(p_y)$ for all $\delta\geq0$.

Conversely, if $p_y$ minimizes the coverage criterion on
$\mathcal F_y$, then, for every $v\in\ker Q$, the differentiable
function $t\mapsto f_{\mathrm{cov}}(p_y+tv)$ has a minimum at
$t=0$.
Its derivative there is zero, so
$\langle a,v\rangle_q=0$ for every $v\in\ker Q$.
This is equivalent to $a_{\ker Q}=0$.

\emph{A feasible improvement direction.}
Suppose $a_{\ker Q}\ne0$.
Moving in this direction preserves the target and balance.
Differentiability gives
$f_{\mathrm{cov}}(p_y+t a_{\ker Q})
=f_{\mathrm{cov}}(p_y)
-t\langle a,a_{\ker Q}\rangle_q+o(t)$
as $t\downarrow0$.
Since orthogonal projection implies
$\langle a,a_{\ker Q}\rangle_q=\|a_{\ker Q}\|_q^2$,
coverage improves to first order.
The additional risk inadequacy is exactly
$t^2\|a_{\ker Q}\|_q^2$, by
\eqref{eq:transfer-target-decomposition}.
This proves both identities in
\eqref{eq:transfer-coverage-improvement}.

\emph{The minimum under an additional cost allowance.}
To obtain the expansion of $K_y$, the first-order approximation
must hold uniformly over the shrinking feasible ball.
Write
$f_{\mathrm{cov}}(p_y+v)
=f_{\mathrm{cov}}(p_y)-\langle a,v\rangle_q+\varepsilon(v)$,
where $\varepsilon(0)=0$ and
$\varepsilon(v)/\|v\|_q\to0$ as $v\to0$, $v\ne0$.
For every $\epsilon>0$, differentiability gives
$|\varepsilon(v)|\leq\epsilon\|v\|_q$
throughout a sufficiently small ball.
Consequently,
$\sup_{\|v\|_q\leq r}|\varepsilon(v)|=o(r)$
as $r\downarrow0$.

For $v\in\ker Q$ with $\|v\|_q\leq r$,
Cauchy--Schwarz gives
$-\langle a,v\rangle_q
=-\langle a_{\ker Q},v\rangle_q
\geq-r\|a_{\ker Q}\|_q$.
When $a_{\ker Q}\ne0$, equality is attained at
$v=r a_{\ker Q}/\|a_{\ker Q}\|_q$.
When $a_{\ker Q}=0$, the linear term vanishes throughout
the feasible ball.
In either case, its minimum is $-r\|a_{\ker Q}\|_q$.
The uniform remainder changes the minimum by at most $o(r)$.
Taking $r=\sqrt{\delta}$ proves
\eqref{eq:transfer-coverage-frontier}.
\end{proof}

\subsubsection{Finite-scenario AVaR}
\label{app:transfers-coverage-AVaR}

For finite-scenario AVaR, we can identify the coverage effect
of a premium change through tariff-class counts in the scenarios
receiving weight in the risk calculation.
We first express AVaR as a maximum of affine functions of the
class premiums. Their slopes then provide the negative gradient
used in Theorem~\ref{thm:transfer-coverage}, whenever it exists,
and the collection of slopes needed in the nondifferentiable case.

Let $j=1,\ldots,J$ index portfolio scenarios under $\bbp$,
with probabilities $\gamma_j>0$ satisfying $\sum_j\gamma_j=1$.
Write $S_{n,j}$ for aggregate claims generated by $n$ policies
in scenario $j$, and $N_{jc}$ for the number of policies in
tariff class $c$ in that scenario.
For a class-premium vector $p\in\bbr^m$, the residual loss is
$z_j(p)=S_{n,j}-\sum_{c=1}^mN_{jc}p_c$.
For $0\leq\alpha<1$, take
$f_{\mathrm{cov}}(p)
=\operatorname{AVaR}_\alpha^\bbp(Z^{\mathbf C})$,
where $Z^{\mathbf C}$ takes value $z_j(p)$ in scenario $j$.

\smallskip
\noindent
\emph{Representation by scenario weights.}
The finite-distribution AVaR formula is
$f_{\mathrm{cov}}(p)
=\min_{t\in\bbr}
\{t+(1-\alpha)^{-1}\sum_j\gamma_j(z_j(p)-t)_+\}$.
Define
$\mathcal U_\alpha
=\{\xi\in\bbr^J:
0\leq\xi_j\leq(1-\alpha)^{-1},
\ \sum_j\gamma_j\xi_j=1\}$.
Here $\xi_j$ multiplies the original scenario probability,
so the weights $\gamma_j\xi_j$ form a probability distribution.
The set $\mathcal U_\alpha$ is a nonempty compact polytope;
the constant vector $\xi_j=1$ is feasible.
We show that
$f_{\mathrm{cov}}(p)
=\max_{\xi\in\mathcal U_\alpha}
\sum_j\gamma_j\xi_jz_j(p)$.

For any $\xi\in\mathcal U_\alpha$ and $t\in\bbr$,
normalization and the bounds on $\xi_j$ give
$\sum_j\gamma_j\xi_jz_j(p)
=t+\sum_j\gamma_j\xi_j(z_j(p)-t)
\leq t+(1-\alpha)^{-1}\sum_j\gamma_j(z_j(p)-t)_+$.
Thus the maximum over $\xi$ is no larger than the minimum
in the AVaR formula.

To obtain equality, choose a support value $t$ satisfying
$\bbp(Z^{\mathbf C}>t)\leq1-\alpha
\leq\bbp(Z^{\mathbf C}\geq t)$.
Such a value exists by ordering the finitely many residual
loss values.
Set $\xi_j=(1-\alpha)^{-1}$ when $z_j(p)>t$ and
$\xi_j=0$ when $z_j(p)<t$.
For scenarios with $z_j(p)=t$, choose a common value in
$[0,(1-\alpha)^{-1}]$ to make $\sum_j\gamma_j\xi_j=1$.
The two probability inequalities ensure that the remaining
weight lies between zero and the available capacity at $t$.
For this choice,
$\sum_j\gamma_j\xi_jz_j(p)
=t+(1-\alpha)^{-1}\sum_j\gamma_j(z_j(p)-t)_+$.
Together with the preceding inequality, this proves the
maximization representation.

Each $z_j(p)$ is affine in $p$, and a linear function on
$\mathcal U_\alpha$ attains its maximum at a vertex.
Consequently, $f_{\mathrm{cov}}$ is the maximum of finitely
many affine functions of $p$.
In particular, it is finite, convex, and continuous.

\smallskip
\noindent
\emph{Premium sensitivities and the projected negative gradient.}
For $\xi\in\mathcal U_\alpha$, define
$a_c(\xi)=q_c^{-1}\sum_j\gamma_j\xi_jN_{jc}$.
For every premium change $v\in\bbr^m$,
$\sum_j\gamma_j\xi_jz_j(p+v)
=\sum_j\gamma_j\xi_jz_j(p)-\langle a(\xi),v\rangle_q$.
Thus $a(\xi)$ is the negative gradient, in the $q$-inner
product, of the affine function associated with $\xi$.
Its components are the tariff-class counts averaged under
the weights $\gamma_j\xi_j$, divided by the portfolio
class probabilities.

At $p_y$, let
$\mathcal U_y
=\arg\max_{\xi\in\mathcal U_\alpha}
\sum_j\gamma_j\xi_jz_j(p_y)$.
This set is nonempty, compact, and convex.
For $\xi\in\mathcal U_y$, the associated affine function
equals $f_{\mathrm{cov}}$ at $p_y$ and is bounded above by it
everywhere.
Hence, for every $v\in\bbr^m$,
$f_{\mathrm{cov}}(p_y+v)
\geq f_{\mathrm{cov}}(p_y)-\langle a(\xi),v\rangle_q$.

If $f_{\mathrm{cov}}$ is differentiable at $p_y$, applying
this inequality to $tv$ and $-tv$, dividing by $t>0$,
and letting $t\downarrow0$ gives
$Df_{\mathrm{cov}}(p_y)[v]=-\langle a(\xi),v\rangle_q$.
Thus every maximizing weight yields the same vector
$a(\xi)=a$ from Theorem~\ref{thm:transfer-coverage},
even if the maximizing weights themselves are not unique.
Projecting $a$ onto $\ker Q$ gives $a_{\ker Q}$, which
determines the first-order coverage effect of changes
preserving the protected-group target.

\smallskip
\noindent
\emph{Fixed tariff-class counts.}
If $N_{jc}=N_c$ is independent of the scenario,
normalization gives $a_c(\xi)=N_c/q_c$ for every $\xi$.
A premium change then shifts every residual loss by the same
amount, namely $-\sum_cN_cv_c$.
The AVaR representation consequently gives
$f_{\mathrm{cov}}(p+v)
=f_{\mathrm{cov}}(p)-\sum_cN_cv_c$.
In particular, if $\sum_cN_cv_c=0$, both residual losses and
the coverage criterion remain unchanged.

\subsubsection{Finite-scenario AVaR without differentiability}
\label{app:transfers-coverage-nonsmooth}

When AVaR is not differentiable at $p_y$, maximizing scenario
weights can yield different negative gradients.
A feasible improvement direction must therefore account for
all affine functions attaining the maximum there.
We collect their projected negative gradients and show that
the element of smallest norm determines both whether improvement
is possible and the exact local dependence of coverage on the
allowance for additional risk inadequacy.

For the fixed attainable target $y\in\mathcal T_C$, retain
$\mathcal U_y$ from the preceding subsection, and let
$P_{\ker Q}^q$ denote $q$-orthogonal projection onto $\ker Q$.
Define
$\mathcal V_y
=\{P_{\ker Q}^q a(\xi):\xi\in\mathcal U_y\}$.
Since this is the image of a nonempty compact convex set
under a linear map, it is a nonempty compact convex subset
of $\ker Q$.
Let $a_0$ be its element of minimum $q$-norm.
Existence follows from compactness, and uniqueness follows
from strict convexity of the squared norm.

\begin{proposition}[Finite-scenario AVaR under additional risk inadequacy]
\label{prop:transfer-AVaR-local}
The class-premium vector $p_y$ minimizes the coverage criterion
over $\mathcal F_y$ if and only if $a_0=0$.
In that case,
$K_y(\delta)=f_{\mathrm{cov}}(p_y)$ for every $\delta\geq0$.

If $a_0\ne0$, then, for all sufficiently small $t>0$,
$f_{\mathrm{cov}}(p_y+t a_0)
=f_{\mathrm{cov}}(p_y)-t\|a_0\|_q^2$ and
$f_{\mathrm{RA}}(p_y+t a_0)
=f_{\mathrm{RA}}(p_y)+t^2\|a_0\|_q^2$.
Moreover, for all sufficiently small $\delta\geq0$,
$K_y(\delta)
=f_{\mathrm{cov}}(p_y)-\sqrt{\delta}\,\|a_0\|_q$.
\end{proposition}

\begin{proof}
\emph{The minimum-norm property.}
We first establish the relation between $a_0$ and all other
projected negative gradients.
For $a\in\mathcal V_y$ and $0\leq t\leq1$, convexity gives
$a_0+t(a-a_0)\in\mathcal V_y$.
Minimality of $a_0$ therefore implies
$\|a_0+t(a-a_0)\|_q^2\geq\|a_0\|_q^2$.
Expanding, dividing by $t>0$, and letting $t\downarrow0$
gives $\langle a-a_0,a_0\rangle_q\geq0$.
Since $a_0$ itself belongs to $\mathcal V_y$, it follows that
$\min_{a\in\mathcal V_y}\langle a,a_0\rangle_q
=\|a_0\|_q^2$.

\emph{A global lower bound.}
For every $\xi\in\mathcal U_y$ and $v\in\ker Q$,
the affine function associated with $\xi$ gives
$f_{\mathrm{cov}}(p_y+v)
\geq f_{\mathrm{cov}}(p_y)
-\langle P_{\ker Q}^q a(\xi),v\rangle_q$.
By the definition of $\mathcal V_y$, some maximizing weight
has projected negative gradient $a_0$.
Consequently,
$f_{\mathrm{cov}}(p_y+v)
\geq f_{\mathrm{cov}}(p_y)-\langle a_0,v\rangle_q
\geq f_{\mathrm{cov}}(p_y)-\|a_0\|_q\|v\|_q$.
For every $\delta\geq0$, this implies
$K_y(\delta)
\geq f_{\mathrm{cov}}(p_y)-\sqrt{\delta}\,\|a_0\|_q$.
If $a_0=0$, the same inequality proves global minimality
of $p_y$ on $\mathcal F_y$.
Since $v=0$ is feasible, it also gives
$K_y(\delta)=f_{\mathrm{cov}}(p_y)$ for every $\delta\geq0$.

\emph{Local attainment of the lower bound.}
The finite affine representation allows us to determine
coverage exactly near $p_y$.
Every vertex of $\mathcal U_\alpha$ that does not maximize
at $p_y$ has a strictly positive gap from the maximum there.
Fix a vertex that does maximize at $p_y$.
By continuity, its affine function remains above each of
the initially nonmaximizing vertex functions in a sufficiently
small neighborhood.
Because there are only finitely many vertices, one neighborhood
works for all of them.
Thus only vertices maximizing at $p_y$ need be considered
within that neighborhood.

These maximizing vertices generate the face $\mathcal U_y$.
Their affine functions all have value $f_{\mathrm{cov}}(p_y)$
at $p_y$, so, for every sufficiently small $v\in\ker Q$,
$f_{\mathrm{cov}}(p_y+v)
=f_{\mathrm{cov}}(p_y)
-\min_{a\in\mathcal V_y}\langle a,v\rangle_q$.
This local identity explains why an exact improvement formula
is available.

Suppose $a_0\ne0$ and take $v=t a_0$.
For all sufficiently small $t>0$, the minimum-norm property
then yields
$f_{\mathrm{cov}}(p_y+t a_0)
=f_{\mathrm{cov}}(p_y)-t\|a_0\|_q^2$.
The decrease is strict, so $p_y$ does not minimize coverage
on $\mathcal F_y$.
Together with the case $a_0=0$, this proves the stated
optimality equivalence.
Since $a_0\in\ker Q$,
\eqref{eq:transfer-target-decomposition} also gives
$f_{\mathrm{RA}}(p_y+t a_0)
=f_{\mathrm{RA}}(p_y)+t^2\|a_0\|_q^2$.

Finally, for sufficiently small $\delta>0$, choose
$v=\sqrt{\delta}\,a_0/\|a_0\|_q$.
This vector satisfies $\|v\|_q^2=\delta$, and the local
identity gives
$f_{\mathrm{cov}}(p_y+v)
=f_{\mathrm{cov}}(p_y)-\sqrt{\delta}\,\|a_0\|_q$.
It therefore attains the global lower bound for $K_y(\delta)$.
The case $\delta=0$ follows from $v=0$.
\end{proof}

If $f_{\mathrm{cov}}$ is differentiable at $p_y$, every
$\xi\in\mathcal U_y$ gives the same negative gradient $a$.
Hence $\mathcal V_y=\{a_{\ker Q}\}$ and $a_0=a_{\ker Q}$.
For finite-scenario AVaR,
Proposition~\ref{prop:transfer-AVaR-local} thus strengthens
Theorem~\ref{thm:transfer-coverage}: along the stated direction
the local coverage change is exactly linear, and the formula
for $K_y(\delta)$ is exact for sufficiently small additional
risk inadequacy.

\section{Calculations for the motor-insurance model}
\label{app:finite-model}

This appendix provides the calculations underlying
Section~\ref{sec:finite-model}. We first derive the population quantities
used throughout the case study, then calculate the classification and
pooling criteria, the attainable protected-group effects and their
risk-inadequacy costs, and the resulting portfolio-coverage effects.
Finally, we examine the consequences of changing the reference law.

Numerical values reported as decimals are rounded for presentation;
the calculations use the exact quantities defined by the model.
Except for the reference-law perturbation in
Appendix~\ref{app:finite-reference-law}, the portfolio, regulatory,
and reference laws coincide. The three protected groups each have
probability $1/3$, and their conditional distributions over the eight
benchmark-price levels are given in Table~\ref{tab:finite-population}.

Write $r=(88,92,96,100,104,108,112,116)^\top$. Since the protected
groups have equal probability, the unconditional probability of each
benchmark-price level $r_j$ is the average
$\bbp(R=r_j)=\frac13\sum_{d=1}^3\bbp(R=r_j\mid D=d)$.
Using Table~\ref{tab:finite-population} gives
\begin{equation}
\bigl(\bbp(R=88),\ldots,\bbp(R=116)\bigr)^\top
=
\frac1{300}(32,38,42,45,44,40,34,25)^\top.
\label{eq:app-finite-risk-distribution}
\end{equation}
The unconditional mean benchmark price is consequently
$\bbe_{\bbp}[R]=\sum_{j=1}^8 r_j\bbp(R=r_j)
=7612/75=101.493\overline3$.

For every classification considered in the case study, aggregation
is defined by the conditional benchmark-price mean,
$A^{\mathbf C}(c)=\bbe_{\bbp}[R\mid C(\Theta)=c]=\bar r_c$.
Transfers are $\bbp$-balanced, so
$\bbe_{\bbp}[\Lambda(C(\Theta))]=0$.
Thus, the mean total premium remains equal to
$\bbe_{\bbp}[R]$, independently of the classification and the
particular balanced transfer.

Since aggregation uses portfolio conditional means, its risk
inadequacy equals the pooling loss $L(C)$ defined in
\eqref{eq:MR1-RA}. Applying the decomposition
$f_{\mathrm{RA}}(p)=L(C)+\|p-\bar r\|_q^2$ from
Section~\ref{subsec:transfer-outcomes-cost} with
$p_c=\bar r_c+\Lambda(c)$ gives
\begin{equation}
f_{\mathrm{RA}}(\mathbf C)
=
\bbe_{\bbp}\!\left[
\bigl(A^{\mathbf C}(C(\Theta))+\Lambda(C(\Theta))-R\bigr)^2
\right]
=
L(C)+\bbe_{\bbp}[\Lambda(C(\Theta))^2].
\label{eq:app-finite-risk-decomposition}
\end{equation}
This decomposition allows the subsequent calculations to treat
the risk-inadequacy cost of classification and the additional
cost of balanced transfers separately.

\subsection{Classification, inferential criterion, and pooling}
\label{app:finite-classification}

We calculate the quantities reported in
Table~\ref{tab:finite-classification} in three steps. First, we determine
the class probabilities, conditional mean benchmark risk prices, and
pooling losses for each classification. Second, we calculate the
inferential criterion from the protected-group distributions within
classes. Third, we determine the protected-group premium effects
generated by aggregation without transfers and their dispersion.

For the three coarsenings $C^4,C^3,C^2$, write $q^j$ and $\bar r^j$
for the vectors of class probabilities and portfolio conditional mean
benchmark risk prices under $C^j$. Both are obtained by aggregating
the benchmark-price levels belonging to each class. In particular,
if class $c$ contains the benchmark-price levels indexed by $I_c$,
then $q_c=\sum_{i\in I_c}\bbp(R=r_i)$ and
$\bar r_c=q_c^{-1}\sum_{i\in I_c}r_i\bbp(R=r_i)$.

For example, the first class of $C^4$ contains benchmark-price levels
$88$ and $92$. Its probability is
$q^4_1=(32+38)/300=7/30$, and its conditional mean benchmark risk
price is $\bar r^4_1=(32\cdot88+38\cdot92)/(32+38)=90.1714$
after rounding. Applying the same calculation to the remaining
classes gives
\begin{equation}
\begin{array}{c|c|c}
C^j & q^j & \bar r^j\\
\hline
C^4
&( .23333,.29000,.28000,.19667 )
&( 90.1714,98.0690,105.9048,113.6949 )\\
C^3
&( .37333,.29667,.33000 )
&( 92.3571,101.9775,111.3939 )\\
C^2
&( .52333,.47667 )
&( 94.5478,109.1189 )
\end{array}
\label{eq:app-finite-class-probabilities}
\end{equation}
where entries are rounded for presentation.

Evaluating the conditional-variance representation of $L(C)$
from Section~\ref{subsec:exact-inferential-threshold} gives
$L(C)=\sum_c\sum_{i\in I_c}
\bbp(R=r_i)(r_i-\bar r_c)^2$. For the first class of $C^4$,
the contribution is
$(32/300)(88-\bar r^4_1)^2+
(38/300)(92-\bar r^4_1)^2=0.92648$ after rounding.
Summing over all classes gives $L(C^4)=3.97092$,
$L(C^3)=8.43102$, and $L(C^2)=19.04624$.
For $C^8=C^\pi$, each class has a single benchmark risk price,
so $\bar r_{C^8(\Theta)}=R$ and $L(C^8)=0$.

We next calculate the inferential criterion. The relevant quantities
are the conditional class-probability matrices
$Q^j_{dc}=\bbp(C^j(\Theta)=c\mid D=d)$, obtained by summing the
conditional benchmark-price probabilities in
Table~\ref{tab:finite-population} over the levels belonging to
each class. This gives
\begin{equation}
Q^4=
\begin{pmatrix}
.40&.29&.20&.11\\
.20&.37&.31&.12\\
.10&.21&.33&.36
\end{pmatrix},
\qquad
Q^3=
\begin{pmatrix}
.56&.24&.20\\
.37&.38&.25\\
.19&.27&.54
\end{pmatrix},
\qquad
Q^2=
\begin{pmatrix}
.69&.31\\
.57&.43\\
.31&.69
\end{pmatrix}.
\label{eq:app-finite-Q}
\end{equation}

Since the portfolio and regulatory laws coincide, the
protected-group probability vector in the notation of
Section~\ref{sec:main-result-I} is
$\eta=(1/3,1/3,1/3)^\top$. For each class $c$, Bayes' formula
gives the posterior protected-group probabilities
$\eta_c(d)=\bbp(D=d\mid C(\Theta)=c)
=\eta_d Q_{dc}/q_c$.
Equation~\eqref{eq:MR1-Gamma-variance} therefore becomes
\begin{equation}
\Gamma(C)=2\sum_c q_c\|\eta_c-\eta\|_2^2.
\label{eq:app-finite-Gamma}
\end{equation}
For example, the first column of $Q^4$, together with
$q^4_1=7/30$, gives
$\eta_1=(4/7,2/7,1/7)^\top$. Its contribution to
$\Gamma(C^4)$ is
$2(7/30)\|\eta_1-\eta\|_2^2=2/45$.
Calculating the remaining contributions in the same way,
and using the conditional distributions in
Table~\ref{tab:finite-population} for $C^8$, gives
\begin{equation}
\Gamma(C^8)=0.1158054,
\qquad
\Gamma(C^4)=0.1073037,
\qquad
\Gamma(C^3)=0.0942810,
\qquad
\Gamma(C^2)=0.0672279.
\label{eq:app-finite-Gamma-values}
\end{equation}

Finally, we calculate the protected-group premium effects
generated by aggregation without transfers. In the notation of
Section~\ref{subsec:transfer-outcomes-cost}, the reference
protected-group probabilities and conditional mean benchmark risk
prices are
$\rho^\star=(1/3,1/3,1/3)^\top$ and
$\mu^\star=(98,100.76,105.72)^\top$.
The matrices above satisfy
$q^j=(Q^j)^\top\rho^\star$ for $j\in\{4,3,2\}$.

Under classification $C^j$, the vector of protected-group mean
premiums without transfers is $Q^j\bar r^j$. Subtracting the
corresponding protected-group mean benchmark risk prices gives
$w_{C^j}=Q^j\bar r^j-\mu^\star$.
For example, the first component of $w_{C^4}$ is
$.40\bar r^4_1+.29\bar r^4_2+.20\bar r^4_3+
.11\bar r^4_4-98=.19596$ after rounding.
The resulting vectors are
\begin{equation}
\begin{aligned}
w_{C^4}&=(.19596,.03367,-.22963)^\top,\\
w_{C^3}&=(.47339,.01209,-.48548)^\top,\\
w_{C^2}&=(1.06481,.05335,-1.11816)^\top.
\end{aligned}
\label{eq:app-finite-w}
\end{equation}

Since $\bbp^\star=\bbp^{\mathrm{reg}}=\bbp$, each
$w_{C^j}$ has zero $\rho^\star$-weighted mean, as established in
Section~\ref{subsec:transfer-inferential-cost}.
Its standard deviation is therefore
$\|w_{C^j}\|_{\rho^\star}
=\bigl(\frac13\sum_{d=1}^3(w_{C^j,d})^2\bigr)^{1/2}$.
This gives $.17537$, $.39155$, and $.89200$ for
$C^4,C^3,C^2$, respectively.

Together, the pooling losses, inferential criterion values,
and standard deviations calculated above establish
Table~\ref{tab:finite-classification}.

\subsection{Attainability and minimum risk inadequacy}
\label{app:finite-transfers}

We now examine how balanced transfers can change the protected-group
premium effects calculated above. For each classification, we
first determine which effects are attainable. Where exact compensation
is attainable, we calculate the balanced transfer with the smallest
additional risk inadequacy. Where it is not attainable,
we determine the transfer that minimizes the standard deviation of
the remaining protected-group effects.

Because aggregation is by portfolio conditional means and the
reference law is the portfolio law, a class-premium vector has the
form $p=\bar r+\lambda$, where
$\lambda=(\Lambda(1),\ldots,\Lambda(m))^\top$.
Balance requires $q^\top\lambda=0$. For $C=C^j$,
$j\in\{4,3,2\}$, the resulting protected-group total
premium-effect vector is
\begin{equation}
M_\Delta^{\mathbf C,\star}
=
w_{C^j}+Q^j\lambda.
\label{eq:app-finite-group-effect}
\end{equation}
By \eqref{eq:app-finite-risk-decomposition}, the additional
risk inadequacy generated by the transfer is
$\|\lambda\|_q^2=\sum_cq_c\lambda_c^2$.
Thus, attaining a specified effect vector $y$ requires
$Q^j\lambda=y-w_{C^j}$, and its minimum additional cost is
obtained by minimizing $\|\lambda\|_{q^j}^2$ over all transfers
satisfying this equation and balance.

\smallskip
\noindent
\emph{Four classes.}
For $C^4$, the matrix $Q^4$ has rank three. Here $b=0$,
so Theorem~\ref{thm:transfer-target-cost} implies that every
protected-group total premium-effect vector $y$ satisfying
$(\rho^\star)^\top y=0$ is attainable by a balanced transfer.
In particular, exact compensation is attainable.

The compensating transfers form a one-dimensional affine set.
To compute the unique transfer minimizing risk inadequacy, let
$D_{q^4}=\operatorname{diag}(q^4)$ and set
\[
\lambda_{\mathrm{comp}}^4
=
-D_{q^4}^{-1}(Q^4)^\top
\bigl(Q^4D_{q^4}^{-1}(Q^4)^\top\bigr)^{-1}w_{C^4}.
\]
The inverse exists because $Q^4$ has full row rank and
all portfolio class probabilities are positive.
This vector satisfies $Q^4\lambda_{\mathrm{comp}}^4=-w_{C^4}$
and is $q^4$-orthogonal to $\ker Q^4$:
for every $v\in\ker Q^4$,
\[
\langle\lambda_{\mathrm{comp}}^4,v\rangle_{q^4}
=
-w_{C^4}^\top
\bigl(Q^4D_{q^4}^{-1}(Q^4)^\top\bigr)^{-1}Q^4v
=0.
\]
Thus $\bar r^4+\lambda_{\mathrm{comp}}^4$ is the
$q^4$-orthogonal projection characterized in
Theorem~\ref{thm:transfer-target-cost}.

Using the quantities calculated above gives
\begin{equation}
\lambda_{\mathrm{comp}}^4
=
(-.69760,-.14797,.25222,.68676)^\top.
\label{eq:app-finite-C4-transfer}
\end{equation}
Its additional risk inadequacy is
$\|\lambda_{\mathrm{comp}}^4\|_{q^4}^2
=\sum_{c=1}^4q_c^4(\lambda_{\mathrm{comp},c}^4)^2
=.23047$. Consequently, $c_{C^4}(0)=.23047$ and the
total risk inadequacy is
$f_{\mathrm{RA}}=3.97092+.23047=4.20139$.

\smallskip
\noindent
\emph{Three classes.}
For $C^3$, the matrix $Q^3$ is nonsingular.
Since $b=0$, Theorem~\ref{thm:transfer-target-cost} again
establishes attainability of exact compensation by a balanced
transfer. Here $\ker Q^3=\{0\}$, so the unique solution of
$Q^3\lambda=-w_{C^3}$ is also the unique compensating transfer
minimizing risk inadequacy.

Solving $\lambda=-(Q^3)^{-1}w_{C^3}$ gives
\begin{equation}
\lambda_{\mathrm{comp}}^3
=
(-1.56270,.79957,1.04910)^\top.
\label{eq:app-finite-C3-transfer}
\end{equation}
Its additional risk inadequacy is
$\|\lambda_{\mathrm{comp}}^3\|_{q^3}^2
=\sum_{c=1}^3q_c^3(\lambda_{\mathrm{comp},c}^3)^2
=1.46455$.
Thus, $c_{C^3}(0)=1.46455$ and
$f_{\mathrm{RA}}=8.43102+1.46455=9.89557$.

\smallskip
\noindent
\emph{Two classes.}
For $C^2$, there are only two class transfers, and balance
leaves one degree of freedom. Writing
$q^2=(q_1^2,q_2^2)^\top$, let
$\xi=(q_2^2,-q_1^2)^\top$. Since
$(q^2)^\top\xi=0$, every balanced transfer can be written
as $\lambda=t\xi$ for some $t\in\bbr$.
Its protected-group total premium-effect vector is
therefore $w_{C^2}+ta$, where
$a=Q^2\xi=(.16667,.04667,-.21333)^\top$.
The entries of $a$ are rounded here; the calculations below
use their exact values.

Exact compensation would require $w_{C^2}+ta=0$.
However, the first component requires
$t=-w_{C^2,1}/a_1\approx-6.389$, whereas the second
requires $t=-w_{C^2,2}/a_2\approx-1.143$.
These requirements cannot both hold. Thus, exact compensation
is not attainable under $C^2$.

We instead minimize the standard deviation of the remaining
protected-group total premium effects. Both $w_{C^2}$
and $a$ have zero mean under $\rho^\star$, so this is
equivalent to minimizing
$\|w_{C^2}+ta\|_{\rho^\star}^2
=\sum_{d=1}^3\rho_d^\star(w_{C^2,d}+ta_d)^2$.
This is a strictly convex quadratic function of $t$.
Differentiating and setting the derivative to zero gives
$t_{\mathrm{SD}}
=-\langle w_{C^2},a\rangle_{\rho^\star}/
\|a\|_{\rho^\star}^2$, where
$\langle u,v\rangle_{\rho^\star}
=\sum_{d=1}^3\rho_d^\star u_dv_d$.
Substituting the calculated vectors yields
$t_{\mathrm{SD}}\approx-5.545$ and hence
\begin{equation}
\lambda_{\mathrm{SD}}^2
=
(-2.64335,2.90214)^\top.
\label{eq:app-finite-C2-SD-transfer}
\end{equation}

The resulting protected-group total premium-effect vector is
$w_{C^2}+t_{\mathrm{SD}}a
=(.14057,-.20544,.06488)^\top$.
Its mean remains zero, and its standard deviation is
$\bigl(\frac13(.14057^2+(-.20544)^2+.06488^2)\bigr)^{1/2}
=.14852$, with the calculation using unrounded values.

Finally, the additional risk inadequacy of this
transfer is
$\|\lambda_{\mathrm{SD}}^2\|_{q^2}^2
=t_{\mathrm{SD}}^2
\bigl(q_1^2(q_2^2)^2+q_2^2(q_1^2)^2\bigr)
=t_{\mathrm{SD}}^2q_1^2q_2^2
=7.67139$, where the penultimate equality uses
$q_1^2+q_2^2=1$.
The total risk inadequacy is consequently
$f_{\mathrm{RA}}=19.04624+7.67139=26.71763$.

These calculations establish the results in
Table~\ref{tab:finite-transfers}. They distinguish the
minimum costs of exact compensation under $C^4$ and $C^3$
from the minimum achievable standard deviation under $C^2$,
where exact compensation is infeasible.

\subsection{Standard deviation and adverse protected-group total premium effects}
\label{app:finite-group-criteria}

We now compare the transfer that minimizes the standard deviation of
protected-group total premium effects with the transfer that minimizes
their $\operatorname{AVaR}_{2/3}$. The calculations concern $C^2$, for which
exact compensation is not attainable. Both optimization problems use the
same one-dimensional balanced-transfer space, but they measure the
remaining protected-group effects differently.

Under $C^2$, every balanced transfer has the form $\lambda=t\xi$, where
$\xi=(q_2^2,-q_1^2)^\top$, as established in the preceding subsection.
Writing $w=w_{C^2}$ and $a=Q^2\xi$, the resulting protected-group
total premium effects are $w+ta$. Their $\rho^\star$-weighted mean is
zero for every $t$, since both $w$ and $a$ are centered.

Because the three protected groups are equiprobable,
$\operatorname{AVaR}_{2/3}$ of these effects is their largest component.
Minimizing $\operatorname{AVaR}_{2/3}$ therefore amounts to solving
\begin{equation}
\min_{\lambda:\,(q^2)^\top\lambda=0}
\max_{d=1,2,3}\{(w_{C^2})_d+(Q^2\lambda)_d\}.
\label{eq:app-finite-C2-AVaR-problem}
\end{equation}

To solve this problem, we use the balanced-transfer parameterization
$\lambda=t\xi$. The exact values of $Q^2$ and $q^2$ give
$a=(1/6,7/150,-16/75)^\top$. Thus, the three protected-group effects
are the affine functions
$h_1(t)=w_1+t/6$,
$h_2(t)=w_2+7t/150$, and
$h_3(t)=w_3-16t/75$.
The problem is to minimize
$\max\{h_1(t),h_2(t),h_3(t)\}$ over $t\in\bbr$.

The first function is strictly increasing, whereas the third is
strictly decreasing. Their maximum is therefore minimized at their
unique intersection. Solving $h_1(t)=h_3(t)$ gives
$t_*=(w_3-w_1)/(19/50)\approx-5.7447$.
At this value, the three effects are
$h_1(t_*)=.10737$, $h_2(t_*)=-.21474$, and
$h_3(t_*)=.10737$, after rounding. In particular,
$h_2(t_*)<h_1(t_*)=h_3(t_*)$.

This establishes optimality for the full problem: for every $t$,
the maximum of the three functions is at least
$\max\{h_1(t),h_3(t)\}$, which cannot be smaller than its value
at $t_*$. At $t_*$, the second function lies below the other two,
so this lower bound is attained. The minimizing balanced transfer is
therefore
\begin{equation}
\lambda_{\mathrm{AVaR}}^2
=
(-2.73830,3.00638)^\top.
\label{eq:app-finite-C2-AVaR-transfer}
\end{equation}

The resulting protected-group total premium-effect vector is
$(.10737,-.21474,.10737)^\top$. Its largest component gives
$\operatorname{AVaR}_{2/3}=.10737$. Since its mean is zero,
its standard deviation is
$\bigl(\frac13\sum_{d=1}^3(h_d(t_*))^2\bigr)^{1/2}
=.15184$, using unrounded values.

The additional risk inadequacy generated by the transfer is
$\|\lambda_{\mathrm{AVaR}}^2\|_{q^2}^2
=t_*^2q_1^2q_2^2=8.23237$.
Together with the pooling loss, this gives
$f_{\mathrm{RA}}=19.04624+8.23237=27.27860$,
where the total is calculated before rounding.

For comparison, the standard-deviation-minimizing transfer
$\lambda_{\mathrm{SD}}^2$ from
\eqref{eq:app-finite-C2-SD-transfer} produces
$\operatorname{AVaR}_{2/3}=.14057$ and standard deviation $.14852$,
with additional risk inadequacy $7.67139$.
Thus, minimizing $\operatorname{AVaR}_{2/3}$ reduces the largest
protected-group effect from $.14057$ to $.10737$, while increasing
the standard deviation of the protected-group effects and the
associated risk inadequacy.

These calculations establish the results in
Table~\ref{tab:finite-group-criteria}.

\subsection{Coverage at fixed protected-group total premium effects}
\label{app:finite-coverage}

We calculate the portfolio coverage attainable under $C^4$
while preserving exact compensation and allowing an additional
risk inadequacy of at most $\delta=1$.
We first identify the admissible premium adjustments and their
risk-inadequacy costs, then minimize the coverage criterion
under the prescribed allowance. Finally, we compare the result
with $C^3$, where exact compensation leaves no freedom
for premium adjustments.

For this illustration, we use expectation pricing and assume that,
conditional on $\Theta=(R,D)$, the claim $Y$ equals $0$ or $2R$,
each with probability $1/2$. Consequently,
$\bbe_{\bbp}[Y\mid\Theta]=R$, so $\pi(\Theta)=R$.
For a one-policyholder portfolio with class-premium vector $p$,
coverage is evaluated by
$f_{\mathrm{cov}}(p)=\operatorname{AVaR}^{\bbp}_{.95}
(Y-p_{C(\Theta)})$, corresponding to the largest
$5\%$ of residual losses.

\smallskip
\noindent
\emph{Feasible premium adjustments and their cost.}
Under $C^4$, the minimum-risk-inadequacy premium vector attaining
exact compensation is obtained by adding the compensating transfer
from \eqref{eq:app-finite-C4-transfer} to the conditional mean
benchmark risk prices:
\begin{equation}
p_0^4
=
\bar r^4+\lambda_{\mathrm{comp}}^4
=
(89.47382,97.92100,106.15698,114.38168)^\top.
\label{eq:app-finite-C4-exact-premium}
\end{equation}
Its risk inadequacy is approximately $4.2013889$.

To identify premium adjustments that preserve exact compensation,
we use the kernel of $Q^4$. Direct substitution into the matrix
in \eqref{eq:app-finite-Q} shows that
\begin{equation}
\ker Q^4
=
\operatorname{span}\{(-111,413,-527,273)^\top\}.
\label{eq:app-finite-C4-kernel}
\end{equation}
Write $v=(-111,413,-527,273)^\top$.
By Theorem~\ref{thm:transfer-target-cost}, all class-premium
vectors preserving exact compensation are
$p(t)=p_0^4+tv$, with $t\in\bbr$; these also preserve
transfer balance. Equation~\eqref{eq:transfer-target-decomposition}
gives their additional risk inadequacy as $t^2\|v\|_{q^4}^2$.
Using the exact portfolio class probabilities gives
\[
\|v\|_{q^4}^2
=\sum_{c=1}^4q_c^4v_c^2
=144761.4,
\qquad
f_{\mathrm{RA}}(p(t))
=f_{\mathrm{RA}}(p_0^4)+144761.4\,t^2.
\]
Thus, $t=0$ minimizes risk inadequacy, while the remaining
premium adjustments can be evaluated for portfolio coverage
as in Section~\ref{subsec:transfer-coverage}.

\smallskip
\noindent
\emph{Coverage under the additional risk-inadequacy allowance.}
For the additional risk-inadequacy allowance $\delta=1$ in
\eqref{eq:transfer-conditional-coverage}, the constraint is
$144761.4\,t^2\leq1$. Hence the allowable premium adjustments satisfy
\[
|t|\leq t_{\max}:=\frac{1}{\sqrt{144761.4}}
\approx0.00262829.
\]

For each benchmark-price level $r_i$, let $c_i$ denote
its tariff class under $C^4$.
For each $i$, the residual-loss distribution assigns probability mass
$\bbp(R=r_i)/2$ to each of the values
$-p_{c_i}(t)$ and $2r_i-p_{c_i}(t)$.
Specializing the AVaR representation in
Section~\ref{subsec:transfer-coverage} to $\alpha=.95$, we obtain
\begin{equation}
\operatorname{AVaR}^{\bbp}_{.95}(Z)
=
\min_{z\in\bbr}
\left\{
z+20\bbe_{\bbp}[(Z-z)_+]
\right\}.
\label{eq:app-finite-coverage-AVaR}
\end{equation}

The probabilities in
\eqref{eq:app-finite-risk-distribution}, combined with
the conditional claim distribution, give $600$ probability
units. Thus, the upper $5\%$ tail comprises $30$ units.
At $t=0$, the largest residual loss is
$232-p_{0,4}^4=117.61832$, corresponding to $R=116$
and having $25$ units.
The next largest is
$216-p_{0,3}^4=109.84302$, corresponding to $R=108$
and having $40$ units.
The upper tail therefore contains all $25$ units of the
first outcome and five units of the second, giving
\[
f_{\mathrm{cov}}(p_0^4)
=
\frac{25(232-p_{0,4}^4)+5(216-p_{0,3}^4)}{30}
\approx116.32244.
\]

\smallskip
\noindent
\emph{Minimum coverage criterion under the allowance.}
To determine the optimal adjustment, consider the residual
losses associated with the three largest benchmark-price
levels:
\[
x_{116}(t)=232-p_4(t),\qquad
x_{112}(t)=224-p_4(t),\qquad
x_{108}(t)=216-p_3(t).
\]
Since $p_3(t)=p_{0,3}^4-527t$ and
$p_4(t)=p_{0,4}^4+273t$, we have
\[
\begin{aligned}
x_{116}(t)&=232-p_{0,4}^4-273t,\\
x_{112}(t)&=224-p_{0,4}^4-273t,\\
x_{108}(t)&=216-p_{0,3}^4+527t.
\end{aligned}
\]

Direct comparison of the residual-loss outcomes shows that,
throughout the admissible interval
$[-t_{\max},t_{\max}]$, the largest $25$ probability units
correspond to $R=116$.
The remaining five units in the upper tail come from
whichever of the outcomes for $R=108$ and $R=112$
has the larger residual loss.
Consequently,
\[
f_{\mathrm{cov}}(p(t))
=
\frac{
25x_{116}(t)
+5\max\{x_{108}(t),x_{112}(t)\}
}{30}.
\]

Writing the entire coverage criterion as the maximum of two
affine functions of $t$, their slopes are $-419/3$ and $-273$,
respectively.
Both are strictly negative, so the coverage criterion
decreases throughout the admissible interval.
The minimum is therefore attained at $t=t_{\max}$.

The corresponding class-premium vector is
\begin{equation}
p_{\mathrm{cov}}^4
=
p_0^4+t_{\max}v
\approx
(89.18208,99.00648,104.77187,115.09920)^\top.
\label{eq:app-finite-C4-coverage-premium}
\end{equation}
The minimum coverage criterion and its associated
risk inadequacy are
\[
K_0(1)
=
f_{\mathrm{cov}}(p_{\mathrm{cov}}^4)
\approx115.95535,
\qquad
f_{\mathrm{RA}}(p_{\mathrm{cov}}^4)
\approx5.20139.
\]

Thus, the additional risk-inadequacy allowance $\delta=1$
reduces portfolio AVaR from $116.32244$ to $115.95535$,
an improvement of approximately $0.36708$.
Exact compensation and transfer balance are preserved
throughout the adjustment.

For $C^3$, the matrix $Q^3$ is nonsingular, so
$\ker Q^3=\{0\}$.
Exact compensation therefore determines the class-premium
vector uniquely, leaving no freedom to improve portfolio
coverage while preserving the target.
This establishes the comparison in
Subsection~\ref{subsec:finite-coverage-reference}.

\subsection{Reference-law balance, group neutrality, and exact compensation}
\label{app:finite-reference-law}

We finally examine how changing the reference population affects
the attainable protected-group total premium effects. The
purpose is to distinguish exact compensation, which eliminates
these effects for every protected group, from group neutrality,
which makes them equal across groups. We calculate the effect
of the reference-law change on the balance constraint, show why
exact compensation becomes impossible, and construct a balanced
transfer that still achieves group neutrality.

Keep the portfolio law $\bbp$, the classification $C^4$, and the
conditional-mean aggregation component fixed. Change only the
reference law used for transfer balance and protected-group
evaluation. Specifically, retain the conditional distribution
of benchmark risk prices within each protected group, but replace
the original protected-group probabilities $(1/3,1/3,1/3)^\top$
by $\rho^\star=(.5,.3,.2)^\top$.

Because the conditional distributions within protected groups
are unchanged, the conditional class-probability matrix $Q^4$
and the protected-group mean benchmark risk prices
$\mu^\star=(98,100.76,105.72)^\top$ remain unchanged.
The aggregation component also remains fixed at the portfolio
conditional means $\bar r^4$. Consequently, the protected-group
pooling-effect vector $w_{C^4}=Q^4\bar r^4-\mu^\star$
is the same as in \eqref{eq:app-finite-w}.

The reference-law class probabilities, however, change because
the protected groups now have different weights. For each class
$c$, they are obtained as
$q_c^\star=\sum_{d=1}^3\rho_d^\star Q^4_{dc}$.
Using the matrix in \eqref{eq:app-finite-Q} gives
\begin{equation}
q^\star=(Q^4)^\top\rho^\star
=(.280,.298,.259,.163)^\top.
\label{eq:app-finite-reference-class-probabilities}
\end{equation}

We first determine the mean protected-group effect that remains
under reference-law balance. Its value is
\begin{equation}
b
=
(\rho^\star)^\top w_{C^4}
=
0.0621562.
\label{eq:app-finite-reference-mean}
\end{equation}
Unlike under the original reference law, this mean is not zero.
The reason is that the aggregation component still uses the
portfolio conditional means, whereas the reference population
assigns different weights to the protected groups.

The fixed-mean restriction in
Section~\ref{subsec:transfer-outcomes-cost} applies under the
new reference law: every $\bbp^\star$-balanced transfer
generates a protected-group total premium-effect vector with
$\rho^\star$-weighted mean $b$.
Since $b\ne0$, Theorem~\ref{thm:transfer-target-cost}
excludes exact compensation. The same theorem establishes
attainability of the group-neutral target $b\mathbf1_3$,
because it has the required reference-law mean and $Q^4$
has full row rank.

An explicit implementation uses the transfer
$\lambda_{\mathrm{comp}}^4$ in
\eqref{eq:app-finite-C4-transfer}. Define
\[
\lambda_{\mathrm{neutral}}^4
=\lambda_{\mathrm{comp}}^4+b\mathbf1_4.
\]
Using $Q^4\lambda_{\mathrm{comp}}^4=-w_{C^4}$,
$(q^\star)^\top=(\rho^\star)^\top Q^4$, and
$Q^4\mathbf1_4=\mathbf1_3$, we obtain
\[
(q^\star)^\top\lambda_{\mathrm{neutral}}^4=-b+b=0,
\qquad
w_{C^4}+Q^4\lambda_{\mathrm{neutral}}^4=b\mathbf1_3.
\]
Thus this transfer implements group neutrality while satisfying
balance under the new reference law.

The reference-law change therefore leaves group neutrality
attainable but makes exact compensation impossible. The
distinction arises from the nonzero mean pooling effect $b$,
which balanced transfers cannot change. This completes the
reference-population calculation used in
Section~\ref{subsec:finite-coverage-reference}.

\end{document}